\documentclass[paper=a4, fontsize=10pt,
  DIV=10,
  headheight=13.6pt,
]{scrartcl}
\usepackage[utf8]{inputenc}
\usepackage{graphicx}
\usepackage{subcaption}
\usepackage{epstopdf, epsfig}
\usepackage{amsmath}
\usepackage{nicematrix}
\usepackage{xcolor}
\usepackage{multirow}
\usepackage{rotating, graphicx}
\usepackage{xcolor}
\usepackage{float}
\usepackage[utf8]{inputenc}
\usepackage[T1]{fontenc}
\usepackage{xcolor} 
\usepackage[utf8]{inputenc}

\usepackage[hyperpageref]{backref}

\usepackage{apacite}

\usepackage{natbib}
\usepackage[hidelinks,colorlinks=true,linkcolor=blue,citecolor=blue]{hyperref}
\setcitestyle{authoryear,open={(},close={)}}
\newcommand*{\doi}[1]{\href{http://dx.doi.org/#1}{#1}}

\begin{document}
\title{FRM Financial Risk Meter for Emerging Markets}

\author{Souhir Ben Amor\footnote{Blockchain Research Center, Humboldt-Universität zu Berlin, Germany. International Research Training Group 1792, Humboldt-Universität zu Berlin, Germany. Alexander von Humboldt Stiftung.  benamors@hu-berlin.de.} \and Michael Althof \footnote {Blockchain Research Center, Humboldt-Universität zu Berlin, Germany. International Research Training Group 1792, Humboldt-Universität zu Berlin, Germany. michael.althof@hu-berlin.de} \and Wolfgang Karl Härdle \footnote{Blockchain Research Center, Humboldt-Universität zu Berlin, Germany.
Wang Yanan Institute for Studies in Economics, Xiamen University, China.
Sim Kee Boon Institute for Financial Economics, Singapore Management University, Singapore. Faculty of Mathematics and Physics, Charles University, Czech Republic. National Chiao Tung University, Taiwan. haerdle@wiwi.hu-berlin.de.} 
\footnote{
Financial support of the European Union’s Horizon 2020 research and innovation program “FIN- TECH: A Financial supervision and Technology compliance training programme” under the grant agree- ment No 825215 (Topic: ICT-35-2018, Type of action: CSA), the European Cooperation in Science \& Technology COST Action grant CA19130 - Fintech and Artificial Intelligence in Finance - Towards a transparent financial industry, the Deutsche Forschungsgemeinschaft’s IRTG 1792 grant, the Yushan Scholar Program of Taiwan, the Czech Science Foundation’s grant no. 19-28231X / CAS: XDA 23020303}
}

\maketitle
\begin{abstract}

The fast-growing Emerging Market (EM) economies and their improved transparency and liquidity has attracted international investors. However, the external price shocks can result in a higher level of volatility as well as domestic policy instability. Therefore, an efficient risk measure and hedging strategies are needed to help investors protect their investments against this risk. In this paper, a daily systemic risk measure, called FRM (Financial Risk Meter) is proposed. The FRM@ EM is applied to capture systemic risk behavior embedded in the returns of the 25 largest EMs’ FIs, covering the BRIMST (Brazil, Russia, India, Mexico, South Africa, and Turkey), and thereby reflects the financial linkages between these economies. Concerning the Macro factors, in addition to the \cite{tobias2016covar} Macro, we include the EM sovereign yield spread over respective US Treasuries and the above-mentioned countries' currencies. The results indicated that the FRM of EMs’ FIs reached its maximum during the US financial crisis following by COVID 19 crisis and the Macro factors explain the BRIMST' FIs with various degree of sensibility. We then study the relationship between those factors and the tail event network behavior to build our policy recommendations to help the investors to choose the suitable market for investment and tail-event optimized portfolios. For that purpose, an overlapping region between portfolio optimisation strategies and FRM network centrality is developed. We propose a robust and well diversified tail-event and cluster risk sensitive portfolio allocation model and compare it to more classical approaches.
\vspace{\baselineskip}

\vspace{\baselineskip}

\vspace{\baselineskip}

\textbf{Highlights}
\begin{itemize}
    \item{We select the 25 biggest FIs in the BRIMST by market capitalization,}
    \item{Select Macro variables to reflect the state and impact of the developed and emerging economies,}
    \item{The FRM is based on Lasso quantile regression designed to capture tail event co-movements, between the selected EM FIs and the Macro variables,}
    \item{We use different quantile risk levels to check the robustness of our results.}
    \item{We found a high positive spillover effects between FIs of the same country  and often negative spillover effects between FIs in between regions.}
    \item{ We examplify EM FIs illustrating high-CoStress and high network centrality.} 
    \item{We give examples of "risk receivers" during the 2020 COVID-19 crisis.}
    \item{The EMs are strongly influenced by the Macros the Emerging Market Yield spread, followed the VIX, Moody's Baa corporate yield spreads, and the shape of the U.S. Treasury yield curve. However, Emerging market currencies have a lesser impact.}
    \item{We use the FRM technology based results to construct robust tail-event sensitive portfolios based on an uplifted HRP approach and compare them to other classical approaches.}
    \item{We analyze the relationship between the FRM network centralities and the portfolio weights and specifically risky concentrations,}
    \item{ The classical portfolio approach leads to high weight concentration. However, the uplifted HRP approach provides better diversification.}
    \item{The uplifted HRP portfolio overweight low-central FIs and underweight high-central ones.}
    \item{The Inv$\lambda$ is less at risk of spill-over effects across EM regions, FIs, and financial sub-sectors.}
    
    \end{itemize}
\end{abstract}

\section{Introduction}
Emerging markets have been commonly acclaimed for providing robust growth potential and offering investors a higher expected return compared to developed markets. Indeed, due to the possibility of higher profits and the low level of global equity markets integration, EMs have been considered as an investment opportunity for investors, whose aim to build an internationally diversified portfolio. EMs liquidity and transparency have continuously enhanced [\cite{mcguire2006common}; \cite{bunda2009correlations}]. Moreover, the reputation of EMs, in the framework of portfolio diversification, has received the attention of international investors, especially after the financial crisis that affected mostly developed markets. 

However, the investment’s benefits of EMs come with additional risks, which are usually not as prominent in developed markets. In fact, EMs are exposed to additional economic, political, and currency risks. Further, the EMs’ economies fast growth and, as consequence, the quick evolution of their structure result in market information being rapidly outdated. Therefore, the existing more traditional methods of risk evaluation may be misleading in EMs, especially in short and medium investment horizons. Indeed, the existing risk measure methods are not suited to provide the up to date point of view representing current market structures, if they not supplemented with the latest market information. Hence, an efficient systemic risk measure is needed for EM.

For that purpose, it is crucial to understand and measure the spillover risk across EMs financial system network, which is important for financial risk measurement and portfolio diversification;\\
From the perspective of financial risk measurement, the interdependence among FIs becomes more important, especially during periods of distress, when losses spread through institutions, rendering the global financial system more vulnerable. In this regard, a systemic crisis that disturbs the financial system stability can have serious effects and lead to high losses for the entire economy and society. 

From the perspectives of risk management and portfolio diversification, the contagion risk across the FIs from the same market, and across the worldwide markets, leads to a decrease in diversification potential. Hence, understanding the network structure of interdependence among FIs is crucial to risk managers and portfolio investors, as this can help them design investment strategies to reduce dependence risk and thereby increase diversification. Investors are also interested in recognizing the FIs that contribute the most (least) risk to their portfolio so that they are considered with caution in their portfolio design, especially during financial market turmoil. 

To understand the FIs co-movements in EMs, our study examines the effect of Macro factors on the FIs in BRIMST (Brazil, Russia, India, Mexico, South Africa, Turkey) EMs, using daily interval data for the period between 2000–2020, with particular focus on the last two years. It is well known that the economics of the mentioned markets are strongly linked to the US economy [\cite{ozatay2009emerging}], so it is crucial to analyze the effects of US Macro factors on volatility in EMs financial system. It is worth to note that our dataset covers several global crisis periods, allowing us to examine how the EMs Financial systems respond to the different crisis. After determining the interdependencies among FIs and macroeconomic factors, our research aims also to build a robust strategy based on portfolio diversification in EMs. 

In order to achieve the mentioned goals, our paper seeks to answer the following questions: Can the US and EMs Macro factors explain BRIMST financial equity indices? Are some categories of Macro factors more important than others? What FIs from EMs are the largest (smallest) spillover transmitters (receivers)? What FIs contribute the most (least) risk to total portfolio risk? What FIs offer greater diversification benefits? And lastly, how the tail spillover effect and portfolio weights change over time, and how they react to the different tail risk levels?

Answering all these questions is crucial, as international investors  interested in understanding the forces behind the interdependence among macroeconomic factors and FIs, to identify potential risks and rewards and benefit from global diversification. Economic policymakers and regulators in BRIMST are interested in forces behind the co-movement between these markets to further establish market resilience in EMs. 

We tackle these questions by employing the Financial Risk Meter (FRM) technology in EMs. The FRM is based on Lasso quantile regression designed to examine tail event co-movements financial securites. The objective is to understand the FIs interconnectedness and represent them in a network topology. Moreover, the FRM indices summarize the systematic risk at a given area and identify risk factors. Briefly, FRM represents tail event behavior in a financial risk factors network, which allows us (i) to identify the “stress emitters” and “stress receivers” companies, (ii) to measure the tail dependencies among the FIs and the Macro factors (iii) analyze the risk level in EMs over time. Concerning risk management, (iv) the FRM network is adopted in the portfolio selection process. More precisely, by interpreting the correlation coefficients of FIs equity indices in the adjacency matrix of the FRM network, an overlapping region between the portfolio optimisation strategies and FRM network is developed, (v) The FRM adjacency matrix is also adopted to lift the Hierarchical Risk Parity (HRP) approach [propsed by \cite{de2016building}] to the quantile level. 

The remainder of this research is organized as follows. Section 2 presents a brief review of risk measurement and risk management methods. The econometric approach is discussed in Section 3, as well as a discussion on centrality measures. Section 4 illustrates the proposed portfolio optimization strategies. The FRM results are analysed in section 5. Section 6 develops the portfolio construction, while section 7 provides policy recommendations before concluding the general scope.\\ 
The codes are published on \url{www.quantlet.de} indicated by  \includegraphics[scale=0.008]{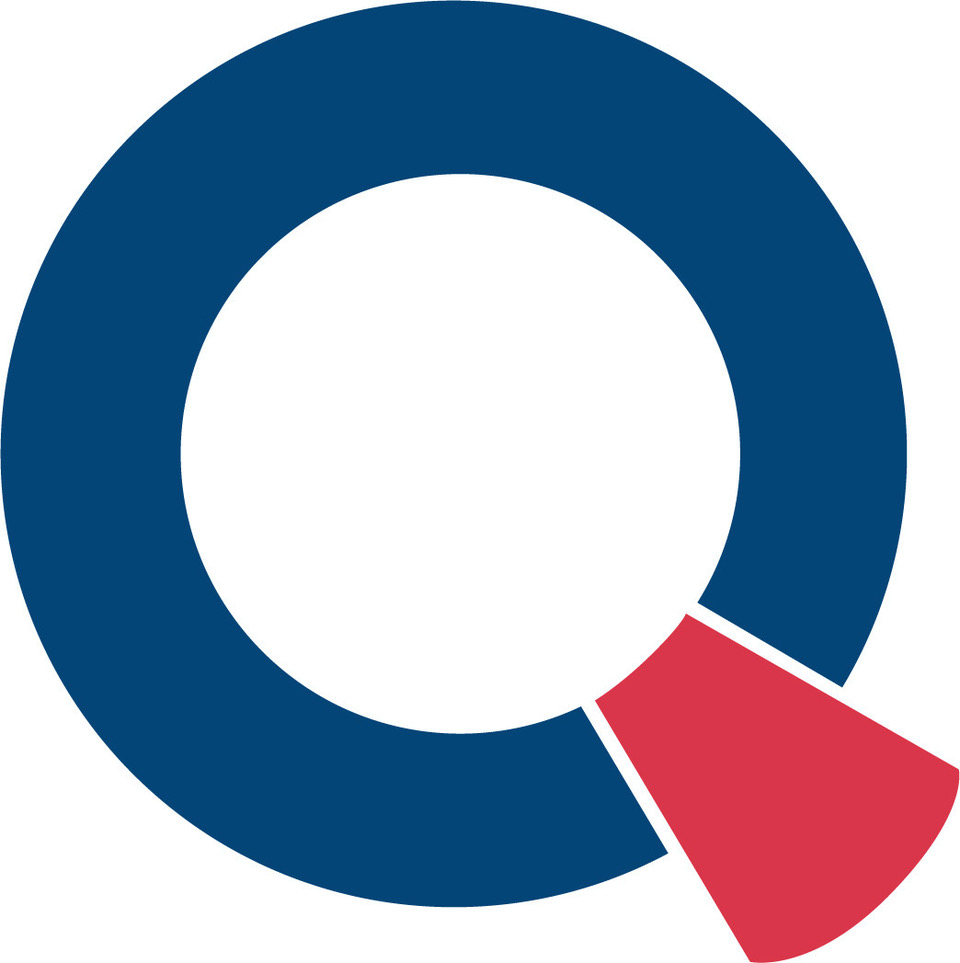} with keyword FRM. The current level of FRM for Emerging markets as well as other channels can be found at \url{http://hu.berlin/frm}.

\section{Literature Review}
The large-scale breakdown of FIs after the global Lehman brother’s crisis in 2008 had caused serious social and economic losses. Therefore, controlling and maintaining financial system stability is one of the principals and the mutual responsibilities of central banks and financial market regulators around the world. Previous researches have been conducted in the framework of FIs interconnectedness. In this context, empirical studies commonly focus on resolving this issue in European countries; \cite{betz2016systemic}, \cite{abbassi2017credit}, and \cite{aldasoro2018multiplex}. Some studies have studied the FIs networks of developed American countries. Among others, \cite{cai2018syndication}; \cite{kreis2018systemic}, and \cite{tonzer2015cross}. Relatively few researchers have investigated bank networks of emerging American countries; \cite{silva2016network}; and \cite{berndsen2018financial}. All of the above studies focused on data from only one country and did not consider networks between FIs from several countries across the globe.

In order to bridge this gap, our study investigates the FIs networks from six among the biggest emerging countries worldwide, as globalization leads to closely linked economic activity among these countries in terms of trade and finance.
It is well known that FIs are interconnected within networks of several types of financial connections and contracts. The complex links among these institutions, which can be considered as risk factors, can cause systemic risk and results in spillover-effects that deteriorate the network stability and its functioning. These observations on the joint network dynamics motivated practitioners and researchers to insert tail events into risk management. 

The Value-at-Risk (VaR) approach \cite{franke2019statistics} is frequently used to measure market risk, by computing the monetary loss of an institution for a given confidence level [\cite{slim2017value}]. However, the VaR measures a tail event probability hosting only one single node, which does not reflect its connection to overall systemic risk. 
Recently, \cite{adrian2011covar} \& (\citeyear{tobias2016covar})  developed CoVaR to measure the systemic risk [e.g.\cite{fang2018identifying}, \cite{souhir2019price}]. The CoVaR measures spillover effects across financial markets by providing the VaR of one market under the condition that the other market is in financial distress as given by its VaR. Consequently, the CoVaR approach can gauge the size of the financial risk spillover. However, it can only capture the extent of risk spillovers for a simple bivariate system and cannot simultaneously measure the risk spillover effects across multiple financial markets. 
\cite{hardle2016tenet} developed the Tail Event NETwork (TENET) risk approach by generalizing the CoVaR to be able to accommodate all system nodes as risk factors. TENET applies the quantile regression method on a set of network nodes stock market information and macroeconomic variables in a rolling window approach. The innovation of TENET model is to employ dimensional reduction (in a semi-parametric setting) by using the Lasso in a quantile regression framework \cite{tibshirani1996regression}. Also, \cite{chen2019tail} propose a Tail Event driven Network Quantile Regression (TENQR) to adress the interdependence, dynamics and riskiness of financial institutions. The TENQR captures risk dynamics within a panel quantile autoregression in a network topology, quantified through a time-varying adjacency matrix. TENQR technique is evaluated using the SIFIs (systemically important financial institutions). To extend the TENET further, \cite{mihoci2020frm} developed an improved systemic risk measure that summarizes the high-dimensional tail stress into a single real value indicator, called the Financial Risk Meter (FRM). The FRM is computed as an average of each node and at each time window selected penalization terms. The FRM level contains fundamental information about the active set of influential neighboring nodes and about the contributors to systemic risk. \\
\cite{yu2019ai} compared the proposed FRM to other measures for systemic risk, such as Google Trends, SRISK, and VIX. They found a Granger causality between the FRM and these measures, which confirmed the validity of the FRM as an efficient systemic risk measure. \cite{mihoci2020frm} applied the FRM to measure the dependencies between FIs and Macro factors exploiting tail event information. They built two FRM indices namely FRM@Americas and FRM@Europe, but also on bond yield and credit default swap spread co-movements. \\ 
So far, the FRM applications took into consideration financial securities from one economic region and its domestic Macro factors. In our study, we extend the application of FRM for EMs, and we will consider the domestic Macro factors from the BRIMST as well as those from the US (proposed by \cite{tobias2016covar} to represent both the domestic and foreign state of the economy.

In many portfolio construction approaches, the correlation between financial assets represents the basis for portfolio selection. To exemplify, \cite{markowitz1959portfolio} developed the modern portfolio theory (MPT). He found that when correlation between assets is not perfect, a diversified portfolio can be constructed. Therefore, to reach efficient diversification, investors should select anti-correlated assets and verify and ensure the satisfaction of this condition over time. But the correlation structure among assets changes over time and evolves especially during crises periods. For that reason, the Markowitz theory is usually oriented to select the most stable assets such as the industrial assets, and hence, the Markowitz optimal portfolio is often composed of a limited and invariant set of assets. 
The second weakness of Markowitz's MPT [\cite{markowitz1959portfolio}] is related to the large estimation errors of the expected returns vector  [\cite{merton1980estimating}], as well as of the covariance matrices [\cite{jobson1980estimation}]. Hence, a robust method for the modelling of the dynamic interconnectedness of assets is needed to support the MPT and guide investors in building an efficient and well diversified portfolio.
In this regard, recent researchers designed financial markets in networks (FMN), where assets are represented by nodes and the correlation of returns are represented by links that relate these nodes [\cite{chi2010network}; \cite{diebold2014network}; \cite{peralta2016network}]. Despite the originality and interesting results provided by this network approach, the majority of its applications are descriptive and lack concrete applications in the portfolio optimization procedure.
Recently, \cite{pozzi2013spread} extracted the dependency structure of financial equities from the network approach to build a diversified portfolio to reduce investment risk. This procedure visualized portfolio selection directly over the FMN design. \cite{peralta2016network} used the FMN as a powerful device to enhance the portfolio optimization procedure by selecting a set of assets according to their network centrality. 

However, the adopted spillover and financial network methods focused on estimating the risk spillover based on the return distributions' first two conditional moments, thereby ignoring higher moments of the distribution (i.e. right and left tails). Indeed, the existing spillover measures concentrate only on the mean and variance of the distributions, which may underestimate the real spillover effects among FIs in tail-events, since they do not take into consideration the extreme risk spillover across financial markets. More specifically, previous researchers investigated mean-to-mean effect, assuming that investors are mean–variance optimizers. But ideally, a portfolio selection decision should be based on the entire return distribution estimation, since investors are more risk averse to the extreme downside risk related to the tails of the distribution. 
To overcome this limitation, we adopt the FRM to measure the joint tail events. In fact, the FRM has the advantage to represent instantaneously the co-movements and the dynamics of high-dimension networks. Furthermore, the FRM can display the hidden interdependency structures between the financial network's nodes. 
Given that the functions of complex FMN are reflected in the risk evaluation and portfolio optimization [\cite{haluszczynski2017linear}; \cite{WANG20181}], we contribute to this line of research by investigating the extent to which the underlying structure of this financial market tail event network can be used as an effective tool in enhancing the portfolio selection process. For that purpose, we establish a bridge between the FRM network and portfolio optimisation strategies. More precisely, we study the relationship between optimal portfolio weights and the FIs’ centrality in the FRM network. \vspace{\baselineskip}

The HRP is a second approach that aims to overcome the shortcomings of MPT. The Hierarchical risk parity (HRP) is firstly proposed by \cite{LopezdePrado59} as a risk parity allocation algorithm. The HRP applies machine learning and graph theory to extrapolate a hierarchical implementation of an inverse-variance allocation with weights computed among the formed correlated asset return clusters.  By substituting the covariance structure with a hierarchical structure of clusters, HRP fully benefits from the covariance matrix information and improves the stability of the weights.  
 \cite{LopezdePrado59} shows that the HRP achieves higher risk-adjusted return and lower out-of-sample volatility than inverse-variance allocations. Despite its appealing features, the application of HRP is still rare, with increased interest \cite{lohre2020hierarchical}, \cite{risks7030074}, and \cite{Raffinot89} test the performance of HRP in a multi-asset allocation. Furthermore, \cite{Raffinot89} and \cite{Alipour2016} evaluate the performance of variant HRP. 
However, the HRP focusses on the covariance matrix in computing the weights, consequently, it offers only a conditional mean view of the assets’ connectedness, which may underestimate the real spillover effects among assets in tail events. Taking into consideration tail spillover effects using quantile regression methods seeks to broaden this view, by providing a complete description of the stochastic relationship between assets and offering more robust and more efficient estimation \cite{Lingjie2008}. In this context, our paper contributes to the growing literature by employing an uplifted HRP based on FRM as a quantile regression method.  
More precisely, our research introduces two modifications to the HRP algorithm of \cite{LopezdePrado59}: (i) Replace the covariance matrix with the FRM-adjacency matrix. Indeed, the HRP assign portfolio weights without the necessity to invert the covariance matrix. This propriety of HRP makes the replacement of the symmetric covariance matrix with the non-symmetric adjacency matrix possible. (ii) The HRP uses the inverse variance of each asset to calculate the optimal weights, and we replace the variance with the FRM's Lasso penalization parameter ($\lambda$). By introducing these two modifications, we build the uplifted HRP based on the FRM technology.
The HRP portfolio is then benchmarked against the classical HRP, the Minimum variance (MinVar) and inverse variance portfolio (IVP) optimization methods. \vspace{\baselineskip}

Our research investigates a comprehensive set of risk measurement and portfolio optimization in EMs, contributing a new dimension to the existing literature as follows:

(1)	The FRM based Lasso quantile regression yields novel insights into the co-movement among FIs and US and EM domestic macroeconomic risk factors. The risk of spillover and its direction are also quantified and visualized, as well as how spillover risk evolves during the financial crisis. Therefore, our research paper is among a few investigating the risk spillover across BRIMST EMs. 
To the best of our knowledge, there is a considerable gap in financial literature on the subject of tail risk spillover among US Macro factors and FIs in EMs. This paper is an attempt to bridge this gap by building our research on the above subject. For that purpose, we cover a sample of the biggest six emerging countries from different areas around the world which offers a reasonable basis for comparisons at country and regional market levels. Most of the earlier studies focus on fewer countries, mostly in one region. In addition, we adopt a sample that contains the largest FIs from each country as well as the domestic and US Macro factors to build our FRM network. Hence, our study investigates a comprehensive set of emerging equity markets, contributing a new dimension to the literature on international equity market co-movement that has traditionally focused on developed markets. 

(2)	Examine the existence of portfolio diversification benefits for foreigners investing in EMs.
Several studies have been conducted to examine the existence of portfolio diversification benefits in less correlated financial markets; \cite{jrfm7020045}; and \cite{SAITI2014196}. However, few other studies investigate this subject from the perspective of an EMs framework [\cite{ARREOLAHERNANDEZ2020101219}].

(3)	This paper sheds light on the connection between the portfolio optimisation approaches and the financial tail event networks. Our strategy aims to simplify the portfolio selection procedure by targeting a set of assets within a certain range of network centrality. As far as we are aware, \cite{ARREOLAHERNANDEZ2020101219} is the only paper that attempts to take advantage of the topology of the tail event financial market network for investment purposes. They applied a directional spillover index, the tail-event driven network (TENET), and nonlinear portfolio optimization methods on the bank returns from emerging and developed America.

(4) While most previous studies focus on portfolio optimization based on mean variance estimation, there is a lack of empirical literature on quantile estimation (both tails of distribution). This study attempts to fill this gap by extending the classical HRP approach to build an optimal portfolio based on tail information provided by the FRM quantile regression.

(5)	Finally, it is worth mentioning that our results are robust to different estimation time windows, market situations (pre-crisis, crisis, and post crisis), and tail risk levels. We argue that our proposed FRM based portfolio optimisation approach benefits from the tail event co-movement and asset clustering making more useful use of fundamental tail-event information, resulting in an efficient risk portfolio selection strategy. This practice is unique to this research and remains an important contribution to the literature on risk measurement and international portfolio diversification. 

\section{Econometric Methodology}
\subsection{FRM Systemic Risk Measure Framework}

This section describes the genesis and framework of FRM. For that purpose, we lay down the fundamental structure and the background of our systemic risk analysis. More precisely, we will state the systemic risk measure models that have been evaluated to lead to an augmented risk measure, the FRM.
\subsubsection{FRM Genesis}

The Value at Risk (VaR) and the expected shortfall (ES) are traditional risk measures. They compute the risk of a given financial institution based either on company characteristics or by introducing macroeconomic variables as a proxy for the state of economy. To exemplify, the VaR of a financial institution $i$ at a quantile level $\tau$ is given by the following equation:
\begin{equation}
    P(X_{i,t}\leq VaR_{i,t,\tau})\overset{def}{=}\tau, \quad\tau \in(0, 1)
\label{eq1}
\end{equation}
Where $X_{i,t}$ represent the log return of financial institution $i$ at time $t$.

\cite{adrian2011covar} proposed the Conditional Value at Risk (CoVaR), the CoVaR considers the spillover effects and the macro economy state. The CoVaR of $FI_j$ given $X_i$ at quantile level $\tau$ at time $t$ is given as follow:
\begin{equation}
P(X_{j,t}\leq CoVaR_{j|i,t,\tau}|R_{i,t})\overset{def}{=}\tau, 
\label{eq2}
\end{equation}
Where $M_{t-1}$ denotes the vector of macro state variables, and 
$R_{i,t}$ indicates the information set that involves the event of $X_{i,t}=VaR_{i,t,\tau}$ and $M_{t-1}$. 
The CoVaR is estimated in two steps of linear quantile regression assuming the following equations:
\begin{equation}
X_{i,t}=\alpha_{i}+\gamma_{i}^{\top}M_{t-1}+\varepsilon_{i,t},
\label{eq3}
\end{equation}
\begin{equation}
X_{j,t}=\alpha_{j|i}+\beta_{j|i}X_{i,t}+\gamma_{j|i}^{\top}M_{t-1}+\varepsilon_{j,t},
\label{eq4}
\end{equation}
\begin{equation}
F^{-1}_{\varepsilon_{i,t}}(\tau|M_{t-1})=0\,\, \mbox{and}\,\, F^{-1}_{\varepsilon_{j,t}}(\tau|M_{t-1},X_{i,t})=0
\label{eq5}
\end{equation}
Where $\beta_{j|i}$ in Equation(\ref{eq4}) defines the sensitivity of log return of a company $j$ to variation in tail event log return of a company $i$.

\textbf{Step 1: Estimate the $VaR$ of an institution $i$ ($VaR_i$)}\\
To estimate the $VaR_i$, \cite{adrian2011covar} apply the quantile regression of log return of company $i$ on macro state variables. The estimated equation is defined as follow:
\begin{equation}
\widehat{VaR}^{\tau}_{i,t}=\widehat\alpha_{i}+\widehat\gamma^{\top}_{i}M_{t-1},
\label{eq6}
\end{equation}

\textbf{Step 2: Estimate the $CoVaR$}\\
The CoVaR is computed by integrating the $VaR_{i,t}^{\tau}$ in Equation (\ref{eq6}) into  the CoVaR equation as follow:
\begin{equation}
\widehat{CoVaR}^{\tau}_{j|i,t}=\widehat\alpha_{j|i}+\widehat\beta_{j|i}\widehat{VaR}^{\tau}_{i,t}+\widehat\gamma^{\top}_{j|i}M_{t-1}.
\label{eq7}
\end{equation}

Where the coefficient $\beta{j|i}$ of Equation (\ref{eq7}) indicates the degree of interconnectedness.

Therefore, the risk of a financial company $j$ is computed through a $VaR$ and macro state of company $i$. 
By setting $j$ equal to the return of a system, and $i$ to be the return of a financial company $i$, we obtain the contribution CoVaR that illustrates how a company $i$ effects the rest of the financial system. Or, by setting $j$ to be a financial company and $i$ to be a financial system, we obtain exposure $CoVaR$, which characterizes how a single institution is exposed to the overall risk of a system.

Recently,  \cite{hardle2016tenet} developed the tail-event driven network (TENET). The TENET is a risk approach generalizes CoVaR by joining systemic interconnectedness between FIs and accommodates all system nodes as risk factors. It is estimated based on quantile regressions and illustrated by three steps. 
\paragraph{Step 1: Estimate the $VaR$ of each FI}\

The $VaR$ for the studied FIs is estimated using the linear quantile regression based on Equations (\ref{eq3} and \ref{eq6}).

\paragraph{Step 2: The network analysis}\

The TENET estimates the non-linear relationship among FIs, and takes more institutions into consideration to compute the tail-driven risk interdependencies. Therefore, we have: 
\begin{equation}
    X_{j,t}=\mbox{g}(\beta_{j|R_{j}}^{\top}R_{j,t})+\varepsilon_{j,t},
    \label{eq8}
\end{equation}
\begin{equation}
\widehat{CoVaR}^{TENET}_{j|\tilde{R}_{j,t,\tau}}\overset{def}{=}\widehat{\mbox{g}}(\widehat{\beta}_{j|\tilde{R}_{j}}^{\top}\tilde{R}_{j,t}),
\label{eq9}
\end{equation}
\begin{equation}
\widehat{D}_{j|\tilde{R}_{j}}\overset{def}{=}\frac{\partial \widehat{\mbox{g}}(\widehat{\beta}_{j|R_{j}}^{\top}R_{j,t})}{\partial R_{j,t}}|_{R_{j,t}=\tilde{R}_{j,t}}=\widehat{\mbox{g}}^{\prime}(\widehat{\beta}_{j|\tilde{R}_{j}}^{\top}\tilde{R}_{j,t})\widehat{\beta}_{j|\tilde{R}_{j}}
\label{eq10}
\end{equation}

where $R_{j,t}\overset{def}{=}\{{X_{\_j,t}},\,M_{t-1},\,B_{j,t-1}\}$ denotes the set of information. Note that $X_{\_j,t}\overset{def}{=}\{X_{1,t}\,X_{2,t}\,...,X_{1,t}\}$ is a set of $\{k-1\}$ independent explanatory variables, such as the log returns of all the FIs expected the $FI_j$, and $k$ here denotes the number of FIs. The term $B_{j,t-1}$ characterizes the internal factors of the institution $j$.The parameters are defined through the term $\beta_{j|R_{j}}\overset{def}{=}\{\beta_{j|\_{j}},\,\beta_{j|M},\,\beta_{j|B_{j}}\}^{\top}$.
The $\widehat{CoVaR}^{TENET}$ stands for tail-event driven network risk by means of a single-index model (SIM) model and is estimated by plugging in $VaR$ of institution $i$ at level $j|Rj,t$, defined in Equation (\ref{eq6}) into Equation (\ref{eq8}).\\
Note that $\beta_{j|R_{j}}\overset{def}{=}\{\beta_{j|\_{j}},\,\beta_{j|M},\,\beta_{j|B_{j}}\}^{\top}$ and $R_{j,t}\overset{def}{=}\{\widehat{VaR}^{\tau}_{i,t},\,M_{t-1},\,B_{j,t-1}\}.\widehat{\mbox{g}}(\bullet)$ characterize the non-linear relationship among them. 
The parameter $D_{j|\tilde{R_j}}$ represents the gradient evaluating the marginal effect of covariates measured at $R_{j,t}=\tilde{R}_{j,t}$. The component wise expression is given by:\\ $\widehat{D}_{j|\tilde{R_j}}\equiv\{\widehat{D}_{j|\_{j}},\,\widehat{D}_{j|M},\,\widehat{D}_{j|B_{j}}\}^{\top}$.
Specifically, $\widehat{D}_{j|\_j}$ permits to quantity spillover effects through the FIs and to illustrate their evolution as a network system. 
The TENET network represents a set of FIs links. The estimation results of this interconnectedness can be summarized in weighted form of an adjacency matrix. 
Note $\widehat{D}^s_{j|i}$ an element in $\widehat{D}^s_{j|\_j}$  at estimation window $s$ for the  $FI_j$ given another $FI_i$ (where $i$ is an element in the other $FI_j$). 
Therefore, the adjacency matrix includes absolute values 
of $\widehat{D}^s_{j|i}$ (in upper triangular matrix) and $\widehat{D}^s_{i|j}$ (in lower triangular matrix), where $\widehat{D}^s_{j|i}$ denotes the influence from a $FI_i$ to a $FI_j$ and $\widehat{D}^s_{i|j}$ is the influence from $FI_j$ to $FI_i$. The adjacency matrix $A_s=(k*k)$ is represented in the following equation, where  for each time windows $s$ only one adjacency matrix is estimated.

\begin{equation}
A_{s} = 
 \begin{pNiceMatrix}[first-row,first-col,nullify-dots]
&I_1 & I_2 & I_3 & \cdots & I_k\\ 
I_1 & 0 & \lvert \widehat{D}^s_{1|2}\rvert &  \lvert\widehat{D}^s_{1|3}\rvert &\cdots &  \lvert\widehat{D}^s_{1|k}\rvert \\
I_2 & \lvert \widehat{D}^s_{2|1}\rvert &  0  &\lvert \widehat{D}^s_{2|3}\rvert & \cdots&\lvert \widehat{D}^s_{2|k}\rvert \\
 \vdots  & \vdots & \vdots & \vdots & \ddots & \vdots   \\
I_k & \lvert \widehat{D}^s_{k|1}\rvert & \lvert \widehat{D}^s_{k|2}\rvert &  \lvert \widehat{D}^s_{k|3}\rvert & \cdots & 0 
\end{pNiceMatrix}
\label{eq11}
\end{equation}

The matrix $A_s$ summarizes the overall connectedness between variables at time window $s$, and $I_i$ denotes the name of $FI_i$. 

\paragraph{Step 3: Systemic risk contributions}\

The objective here is to measure the systemic risk relevance of a specific FI by its total in- and out-connections, weighted by market capitalization. 
Hence, we define the Systemic Risk Receiver Index $(SRR)$ for a $FI_j$ at time windows $s$ as follow: 
\begin{equation}
  SSR_{j,s}\overset{def}{=} MC_{j,s}\Biggl\{\sum_{i\in {k_s^{IN}}}(\lvert \widehat{D}^s_{j|i}\rvert.MC_{i,s})\Biggl\}
  \label{eq12}
\end{equation}

The Systemic Risk Emitter Index $(SRE)$ for a $FI_j$ at time windows $s$ is given by the following equation: 

\begin{equation}
  SSE_{j,s}\overset{def}{=} MC_{j,s}\Biggl\{\sum_{i\in {k_s^{OUT}}}(\lvert \widehat{D}^s_{i|j}\rvert.MC_{i,s})\Biggl\}
  \label{eq13}
\end{equation}

Here $k_s^{IN}$ and $k_s^{OUT}$ denotes the set of FIs linked with the $FI_j$ at time windows $s$ by incoming and outgoing links respectively. $MC_{i,s}$ is the market capitalisation of FI $i$ at the starting point of time windows $s$. $\lvert \widehat{D}^s_{j|i}\rvert$ and $\lvert \widehat{D}^s_{i|j}\rvert$ are absolute partial derivatives derived from Equation (\ref{eq10}) which represents row (incoming) and column (outgoing) direction connection of $FI_j$. Thus both $SR{R_j,s}$ and $SRE_{j,s}$  take into consideration the $FI_j$ and its linked FI’ market capitalization as well as its connectedness within our network.
For more details see \cite{hardle2016tenet}.

\subsubsection{FRM Financial Risk Meter}
The FRM is a systemic risk measure based on the penalization parameter$(\lambda)$ of a linear quantile Lasso regression using moving-window approach.  In this section we present the methodology and algorithm that constitutes the FRM technology. Since the penalization parameters $(\lambda)$ are computed based on an $L_1$-norm (Lasso) quantile linear regression, this method is introduced first. 

\paragraph{Linear Quantile Lasso Regression Model}\

The FRM aims to simultaneously capture all interdependencies in one single number based on the log return of FIs and a set of macroeconomic variables that illustrate the state of the economy. Consider $J$ the number of FI under consideration, where $j\in\{1,...,J\}$. Therefore $p=J+M-1$ denotes the number of covariates. Let $t=\{1,...,T\}$ be the time point, where $T$ denotes the total number of daily observations. Suppose $s$ is the index of time windows, where $s\in\{1,...(T-(n-1))\}$ and $n$ is the length of windows size.

The Linear quantile Lasso regression for return series $X$ is defined as follow: 
\begin{equation}
    X^{s}_{j,t}=\alpha^{s}_{j,t}+A^{s\top}_{j,t}\beta^{s}_{j}+\varepsilon^{s}_{j,t}, \qquad 
    \label{eq14}
\end{equation}

Where $A^{s\top}_{j,t}\overset{def}{=}\begin{bmatrix}M^{s}_{t-1}\\
X^{s}_{-j,t}\end{bmatrix}$, $M^s_{t-1}$ is the $M$ dimensional vector of macro variables, $X^{s}_{-j,t}$ is the $p-M$ dimensional vector of log returns of all other FIs except $FI_j$ at time $t$ and in moving window $s$, $\beta^s_{j}$ is a $p\times 1$ vector defined for moving window $s$ and $\alpha^s_{j}$ is a constant term.

The regression is performed using an $L_1$-norm penalisation given a parameter $\lambda_j$, defined by \cite{tibshirani1996regression} as the least absolute shrinkage and selection operater (Lasso). According to \cite{bassett1978asymptotic}, the current company’s $\lambda_j$ are estimated by a modification of Lasso in a quantile regression [see \cite{belloni2011} and \cite{li20081} for more details), where the optimization is solved as follow:

\begin{equation}
\operatorname{\min_{\alpha^{s}_{j},\beta^{s}_{j}}}\Big\{n^{-1}\sum_{t=s}^{s+(n-1)}\rho_{\tau}\,(X_{j,t}^{s}-\alpha^{s}_{j}-A_{j,t}^{s\top}\beta_{j}^{s})+\lambda_{j}^s\parallel\beta_{j}^{s}\parallel_1\Big\}
\label{eq15}
\end{equation}

with the following check function:
\begin{equation}
\rho_\tau\,(u) = |u|^c \,|\tau - \mathrm{I}_{\{u<0\}}|
\label{eq16}
\end{equation}
where $c=1$ corresponds to quantile regression employed here and $c=2$ corresponds to expectile regression. 

\paragraph{Penalization Parameter $\lambda$}\

The formula of Lasso’s penalization parameter $\lambda$ can be estimated in a linear regression context, following the work of \cite{osborne2000lasso}
Treating $\lambda$ as a fixed value in the objective function of the penalized regression:
 \begin{equation}
     f(\beta,\lambda)=\Biggl\{\frac{1}{2}\sum_{i=1}^n (y_{i}-X_{i}^{\top}\beta)^2+\lambda\sum_{i=1}^p|\beta_{j}|\Biggl\}
     \label{eq17}
 \end{equation}
Here $ f(\beta,\lambda)$ is convex in the parameter $\lambda$. In addition, with diverging $\beta$ we note that $f(\beta,\lambda)\to\infty$. Consequently, the function $f(.,\lambda)$ admits at least one minimum, which is attained in $\widehat{\beta}(\lambda)$ [\cite{osborne1985finite}] if and only if the null vector $0\in R^p$ is an element of sub-differential:
 \begin{equation}
     \frac{\partial f(\beta,\lambda)}{\partial \beta}=-X^{\top}(Y-X\beta)+\lambda u(\beta)
     \label{eq18}
 \end{equation}
 
 Were:
 \begin{equation*}
u(\beta)=(u_1(\beta),...,u_p(\beta))^{\top} \text{is defined as} 
\begin{cases}
u_j(\beta)=1\; \; if\; \;\beta_j>0\\
u_j(\beta)=-1\; \; if\; \;\beta_j<0\\
u_j(\beta)\in[-1,\;1]; \; if\; \;\beta_j=0\\

\end{cases}
\end{equation*}
 
Therefore, considering that $ f(\beta,\lambda)$ admit a minimum in  $\widehat{\beta}$, the following equation has to be satisfied:
\begin{equation}
    0=-X^{\top}\{Y-X\widehat{\beta}(\lambda)\}+\lambda u(\widehat{\beta}(\lambda)
    \label{eq19}
\end{equation}

The estimator of the vector of parameters $\beta$ here is a function of the penalization parameter $\lambda$. This dependency follows from the formulation of the penalized regression method and its objective function Equation (\ref{eq17}). Following this method we select first the penalization parameter $\lambda$ and then search for $\beta_{\lambda}$ that minimizes  Equation (\ref{eq17}). Given that $u(\beta))^{\top} \beta=\sum_{j=1}^p\lvert\beta_j\rvert=\left \Vert \beta\right \|_1$, where $\left \Vert . \right \|_1$ represents here $L1$-norm of a $p$-dimensional vector, Equation (\ref{eq19}) can be rewritten in the following formula:

\begin{equation}
    \lambda = \displaystyle \frac{\{ Y - X\beta(\lambda)\}^{\top}X\beta\left(\lambda\right)}{\left\|\beta\right\|_{1}}
    \label{eq20}
\end{equation}

Looking to formula Equation (\ref{eq20}), we can define the elements that influence the value of $\lambda$ and its dynamics when treated in a time- varying framework. The following three elements affect the size of $\lambda$:\vspace{\baselineskip}

1. The size of residuals of the model;\vspace{\baselineskip}

2. The absolute size of the coefficients of the model, $\left \Vert\\ \beta\right \|_1$; \vspace{\baselineskip}

3. The singularity of a matrix $X^{\top}X$. \vspace{\baselineskip}

The formulae for the penalization parameter $\lambda$ in a quantile regression problem Equations (\ref{eq15} and \ref{eq16})can be derived following \cite{li20081}:

\begin{equation}
  \lambda = \displaystyle\frac{ \left( \theta \right)^{\top}X\widehat{\beta}\left(\lambda\right)}{\left\|\widehat{\beta}(\lambda)\right\|_{1}}  
  \label{eq21}
\end{equation}

Where $\theta=(\theta_1,...,\theta_n)^{\top}$ satisfies the following conditions:
\begin{equation*}
    \theta_i=\begin{cases}
    \tau\;\;\;\;\;\;\;\;\;\;\;\;\;\;\;\;\;\;\;\;\;\;\;\;\;\;\;\;\;if\;\;\;\;Y_i-X_i^{\top}\widehat\beta(\lambda)>0\\
    -(1-\tau)\;\;\;\;\;\;\;\;\;\;\;\;\;\;\;\;\;if\;\;\;\;Y_i-X_i^{\top}\widehat\beta(\lambda)<0\\
    \in(-(1-\:\tau)\;,\tau)\;\;\;\;if\;\;\;\;Y_i-X_i^{\top}\widehat\beta(\lambda)=0\\
    \end{cases}
\end{equation*}

Since Equation (\ref{eq15}) has an $L_1$ loss function and an $L_1$-norm penalty term, the optimization problem is an $L_1$-norm quantile regression estimation problem. The selection of the penalization parameter $\lambda^s_j$ is fundamental. There are different possibilities to choose $\lambda^s_j$, for example with the Generalized Approximate Cross- Validation criterion (GACV) or the Bayesian Information Criterion $(BIC)$. In this regards, \cite{yuan2006gacv} conducted simulations and concluded that GACV outperforms BIC in terms of statistical efficiency. Hence, we estimate $\lambda^s_j$ with the GACV criterion in the FRM model and set $\lambda^s_j$ as the solution of the following minimization problem: 

\begin{equation}
  \operatorname{min}GACV(\lambda_{j}^{s})=\operatorname{min}\frac{\sum^{s+(n-1)}_{t=s}\rho_{\tau}(X_{j,t}^{s}-\alpha^{s}_{j}-A_{j,t}^{s,\top}\beta_{j}^{s})}{n-df}  
  \label{eq22}
\end{equation}
where $df$ stands for the estimated effective dimension of the fitted model. The advantage of GACV is that it can be also adopted for $p > n$, which can be crucial for the FRM if the moving window size is small. For further details see \cite{mihoci2020frm}.
\vspace{\baselineskip}

\paragraph{Financial Risk Meter (FRM) }\

The FRM is estimated using a regression analysis as explained above and select the $\lambda_j$ for each FI $j$ using GACV. The distribution of the$\lambda^s_j$ in a moving window gives important information on the network dependencies among the financial nodes. The standard FRM is defined as the average of $\lambda_j$ over the set of $J$ FIs for all windows. It is formally written as follows:
\begin{equation}
    FRM=J^{-1}\sum_{j=1}^J\lambda_j
    \label{eq23}
\end{equation}

Note that the distribution of  $\lambda_j$ provides details regarding the overall market movement and gives information to macro-prudential decision makers about the network dynamics. Therefore, the FRM is adopted to identify the high joint tail event risks resulting from each FI. More precisely, the FI with high  $\lambda_j$ exhibits common high stress levels as the FI at the origin of the crisis. Therefore, this FI is considered as having high ”co-stress”. 
\vspace{\baselineskip}

\subsection {FRM Network and node centrality measures}
The FRM network illustrates the tail event interaction between the selected FIs based on the adjacency matrix $A_s$ (11). To describe the topology of the FRM networks, we focus on node centrality measures, specifically, eigenvalue centrality, degree centrality, indegree, outdegree, betweenness and closeness. The concept of centrality aims to measure the impact and the importance of a given node in the network. \vspace{\baselineskip}

\paragraph{Eigenvector centrality}\

\cite{bonacich1972factoring} proposed the so-called eigenvector centrality, which evolved to be a standard measure in network analysis. 

Consider a network $G = \{N, \omega\}$, constituted by a set of links $\omega$  that connecting pairs of nodes and a  set of nodes $N = {1, 2, ... , N}$. If there is a connection between two nodes $i$ and $j$, we denote it as $(i,j)\in\omega$. The network connections are defined by a simplified version of (11): the $(N\times N)$ adjacency matrix $A_{i,j}^s=[\beta_{ij}^s]$ whose element $\beta_{ij}\neq 0$ whenever $(i,j)\in \omega$, $s$ is the rolling window. 
 \begin{equation}
A^s = 
\begin{pmatrix}
0 & \beta_{1,2} & \cdots & \beta_{1,n} \\
\beta_{2,1} & 0 & \cdots & \beta_{2,n} \\
\vdots  & \vdots  & \ddots & \vdots  \\
\beta_{n,1} & \beta_{n,2} & \cdots & 0 
\end{pmatrix}
\label{eq24}
\end{equation}
 
Based on the \cite{bonacich1987power}; \citeyearpar{bonacich1972factoring} definition, the eigenvector centrality of node $i$, $v_i$, is expressed as the proportional sum of its neighbors' centralities. This concept has been extended by \cite{bonanno2004networks} suggesting that for the weighted networks,  $v_i$ is relative to the weighted sum of neighbors’ centralities of node $i$ with the adjencency matrix $A_{i,j}^s$ that illustrates the corresponding weighting factors. Therefore, the eigenvector centrality of node $v_i$ is calculated as follows:

\begin{equation}
    v_i=\delta^{-1}\sum_j\beta_{ij}^s v_j
    \label{eq25}
\end{equation}

where $\delta$ here denotes the eigenvalue. A large value of $v_i$ means that the node $i$ is highly central, implying that node $i$ is linked either to several other nodes or is linked to a few highly central nodes.
The Equation (\ref{eq25}) can be rewritten in matrix terms. We have $\delta v=A v$ specifying that the centrality vector $v$ is defined by the eigenvector of $A^s$ corresponding to the largest eigenvalue $\delta$.\vspace{\baselineskip}

In general, each eigenvector of $A^s$ is a solution to Equation (\ref{eq25}). Nevertheless, the centrality vector matching to the largest network component is specified using the eigenvector corresponding to the largest eigenvalue \cite{bonacich1972factoring} and \cite{peralta2016network}.

\paragraph{Closeness}\

Closeness is a centrality measure proposed by \cite{freeman1978centrality}. It highlights the distance of a given vertex to the rest of vertices in the network. It can be considered as duration of the information spread from one vertex to another. Closeness of a given vertex $j$ is defined as follow:
\begin{equation}
    Closeness_j=\sum{_{i=1}^N \frac{1}{d_{(i,j)}}}
\end{equation}
Where $d_{(i,j)}$ measure the distance between a vertix $i$ and another vertix $j$ in the network.
\paragraph{Betweenness}\

Betweenness is a centrality measure, which computes the number of times a vertex lies on the shortest path between other vertices in the network  [\cite{vyrost2019network}].\\ 
Suppose we need to compute this measure for a vertex $j\in\omega$, for any two distinct vertices other than $j$, named $l$ and $k$ $\in\omega$, the number of shortest paths between $l$ and $k$ is $n_{l,j,k}$ $\in N$ and the total number of shortest paths between $l$ and $k$ is $n_{l,k}\in N$, the betweenness for $j$ is then computed as follow:  
\begin{equation}
    Betweenness_j=\sum_{\substack{l\neq j\neq k{} \\ l,j,k\in \omega}} \frac{n_{l,j,k}}{n_{l,k}}
      \label{eq26} 
\end{equation}

\paragraph{Degree centrality}\ 

Degree centrality captures total connectedness in the network, it is given by the following equation:
\begin{equation}
    D=\sum_{j=1}^N \sum_{i=1}^N \textbf{1} (\beta_{j,i}^s)
       \label{eq27}
\end{equation}
where
\begin{equation*}
    \textbf{1} (\beta_{j,i}^s)=\begin{cases}
1\; \; if\; \;\beta_{j,i}^s\neq 0\\
0\; \; if\; \;\beta_{j,i}^s =0\\
\end{cases}
\end{equation*}

\paragraph{Indegree:} Indegree computes the number of inflows, or in this paper the number of other FIs influencing one node. Indegree of $FI_j$ is defined as:
\begin{equation}
    Ind_j=\sum_{i=1}^N\textbf{1} (\beta_{j,i}^s)
     \label{eq28}
\end{equation}
Here the $FI_j$ can be considered as a risk receiver. 
\paragraph{Outdegree:} Outdegree computes the number of out-going links, or in this paper the number of other FIs influenced by the node. Outdgree of $FI_i$ is defined as:

\begin{equation}
    Outd_i=\sum_{j=1}^N\textbf{1} (\beta_{j,i}^s)
     \label{eq29}
\end{equation}
Here the $FI_i$ can be considered as a "risk emitter".

\section{Portfolio optimisation}

\subsection{The minimum-variance}\

Consider $N$ risky assets with a vector of expected returns $\mu$, and covariance matrix, $\Sigma=[\sigma_{ii}]$. The problem is to define the optimal portfolio weights vector $w$, which minimizes the portfolio variance subject to $w^{\top}\textbf{1}=\textbf{1}$ where $\textbf{1}$ represents a column vector whose components are equivalent to one \cite{markowitz1959portfolio}. This approach is usually defined as minimum-variance or shortly $MinVar$. Formally the problem is stated as follow: 
\begin{equation}
    min_w\sigma^2_p=w^{\top}\Sigma{w};\;\;\;\;\;\text{subject to}\;\;\;\;w^{\top}\textbf{1}=\textbf{1}
    \label{minvareq}
\end{equation}

The solution of Equation (\ref{minvareq}) is:
\begin{equation}
    \widehat{w}^*_{minv}=\frac{\textbf{1}}{\textbf{1}^{\top}\Sigma^{-1}1}\Sigma^{-1}\textbf{1}
    \label{minvarweights}
\end{equation}

\subsection{A machine learning-based Hierarchical Risk Parity HRP}
\subsubsection{Classical HRP approach}
The HRP is a risk-based portfolio optimization approach that diversifies portfolios without imposing a positive-definite return covariance matrix [\cite{LopezdePrado59}]. The algorithm employs machine learning methods and graph theory to classify the underlying hierarchical correlation structure of the assets, which allow clusters of similar assets to compete for capital allocation. The algorithm of the HRP approach can be broken down into three main steps: tree clustering, quasi-diagonalization, and recursive bisection. In the following, we explain each step in detail. \vspace{\baselineskip}
 
\paragraph{Step 1- Hierarchical Tree Clustering}\
This step involves breaking down the assets of the considering portfolio into different hierarchical clusters exploiting a Hierarchical Tree Clustering algorithm. The clusters are formed as follows:\\
1.	For $N$ stock returns compute the correlation matrix [see \cite{hardle2019mva}(p 431-442)], which gives an $N\times N$ matrix $\Omega$ of these correlations $\rho$, \\
$\rho= \{ \rho_{i,j}\}_{i,j=1,...,N}$\\
Where $\rho_{i,j}$ is the correlation coefficient between a pair of assets $\{i,j\}$ $\rho[X_{i},X_{j}]$\\
2.	Transform the correlation matrix  to a correlation-distance matrix $D$, where for\\
$d:(X_{i},X_{j}) \subset B \rightarrow {R} \in [0,1];$
\begin{equation}
    d_{i,j} = d[X_{i},X_{j}]=\sqrt{\frac{1}{2}(1-\rho_{i,j})}
    \label{eq32}
\end{equation}
3. Compute a new distance matrix $\tilde{d}$ where, by taking the Euclidean distance among columns in a pair-wise manner, the augmented distance matrix is given as follow: 
\begin{equation}
    \tilde{d}= \tilde{d}[d_{i},d_{j}] = \sqrt{\sum_{n=1}^{N}(d_{n,i}-d_{n,j})^2}
    \label{eq33}
\end{equation}
where $\tilde{d}_{i,j}: (d_{i},d_{j}) \subset B \rightarrow {R} \in [0, \sqrt{N}]$.\\
Note that for two assets $i$ and $j$,  $ D_{i,j}$ represents the distance between them, however $\tilde{d}_{i,j}$ represents the closeness in similarity of $\{i,j\}$ with the rest of the portfolio. More precisely, a lower $\tilde{d}_{i,j}$ indicates that the assets $i$ and $j$ are similarly correlated to the rest of stocks in the portfolio.\\
4. Form the assets clusters in a recursive manner based in Equation(\ref{eq33}). The set of clusters is donated by $U$ and the first formed cluster$(i^*,j^*)$ is define as: 
\begin{equation}
    U[1]=(i^{*},j^{*}) =\text{argmin}(i,j)_{i \neq j}\{\tilde{d}_{i,j}\}
    \label{eq34}
\end{equation}
5.	Update the distance matrix $d$ by computing the distances of other assets from the newly formed cluster $U(1)$ using single linkage clustering. Therefore, for any asset $i$ outside of $U(1)$, the distance from $U(1)$ is updated as follows: 
\begin{equation}
    d_{i,U[1]}=\min\bigl[\{\tilde{d}_{i,j}\}_{j \in U[1]}\bigl]
    \label{eq35}
\end{equation}

Thereby, the algorithm recursively forms assets clusters and updates the distance matrix until we are left with one giant cluster of all stocks.
Finally, the clusters are visualised in a dendrogram. See \cite{hardle2019mva} (p. 363-393) for more details. \vspace{\baselineskip}

\paragraph{Step 2- Quasi Diagonalisation or Matrix Seriation}\

The Quasi-Diagonalisation of the covariance matrix or Matrix seriation is adopted to rearrange the data to clearly represent the inherent clusters. Using the hierarchical clusters from the previous step, the columns and rows of the covariance matrix are reorganized so that similar assets are placed together and dissimilar assets are placed far apart. More precisely, the larger covariances are positioned along the diagonal and smaller ones around this diagonal and since the off-diagonal elements are not completely zero. This is named a quasi-diagonal covariance matrix.\vspace{\baselineskip}

\paragraph{Step 3- Recursive Bisection}\

The final recursive bisection step implicates assigning actual portfolio weights to the assets in the portfolio.\\
1.	Assign a unit weight to all assets, 
\begin{equation*}
    W_i=1,\;\;\;\forall\;i=1,...N
\end{equation*}
2.	Bisect each cluster into two sub-clusters in a top-down manner, it means by starting with the topmost cluster, so, for each cluster, we obtain a left and right sub-cluster.\\
3.	Calculate the variance for each of these sub-cluster 
\begin{equation}
    V_{1,2}=w^{\top}\Sigma w\;\;\;\;; \Sigma \;\;\text{is the covariance matrix}
    \label{eq36}
\end{equation}
where, 
\begin{equation}
    w=  \frac{\mbox{diag}[\Sigma]^{-1}}{\mbox{tr[diag}[\Sigma]^{-1}]}
    \label{IVPweights}
\end{equation}
Since we are dealing with a quasi-diagonal matrix, the algorithm uses the property of the portfolio that the inverse-variance allocation is optimal for a diagonal covariance matrix. Hence, we adopt the inverse-variance allocation weights when computing the variance for sub-clusters.\\
4.	Calculate the weighting factor based on the quasi-diagonalised  covariance matrix
\begin{equation}
    \alpha_{1}=1-\frac{V_{1}}{V_{1}+V_{2}}, \text{so \ that} \ 0 \leq \alpha_{i} \leq 1\; \; ;\alpha_2=1-\alpha_1
    \label{eq38}
\end{equation}

5-	Update the weights $w_1$ and $w_2$ for both sub-clusters:
\begin{equation}
    w'_1=\alpha_1\cdot w_1;\;\;\;\
     w'_2=\alpha_2\cdot w_2  
     \label{eq39}
\end{equation}

6- Execute recursively steps 2-5. The algorithm stops when we have a single asset for each cluster and then the weights are assigned to all assets in the portfolio.\\
Since the weights are assigned in a top-down manner, only assets within each cluster compete for allocation for the final portfolio, rather than competing with all the assets in the portfolio, see \cite{vyrost2019network}.\vspace{\baselineskip}

\subsubsection{An uplifted HRP approach based on FRM}

The classical HRP approach focused on calculating the optimal portfolio weights based on the variance and the covariance matrix, in other words it is focused on estimating the spillover risk on the mean-variance levels, ignoring analysis of other quantile levels. To overcome this limitation, we extend the classical HRP based on FRM in order to take into consideration the hidden interdependency structures among the FIs tail quantile level when optimizing the EMs portfolio. 
The basic idea is to use the FRM output in order to compute the portfolio weights based on quantile level analysis. In the classical approach, these weights are computed using the variance (which is a measure of individual risk) and the covariance matrix that measures the relationship between a pair of assets in the mean-variance level. To improve on this, the FRM provides the penalty parameters $(\lambda_{j})$ as a measure of individual risk of each company or FI at the quantile level. In addition, the FRM provides also the adjacency matrix to estimate the interconnectedness between each pair of the studied FIs $(\beta_{i,j})$. By analogy, for the uplifted HRP approach based FRM, we replace the covariance matrix with the adjacency matrix. Recall that the diagonal elements of the adjacency matrix are zero (Equation (\ref{eq24})), which allow us to introduce the vector of penalty parameters $(\lambda_{j})$ in the diagonal of the adjacency matrix to replace the variance in the classical approach.  
Following this reasoning, the uplifted HRP steps will be as follows:\vspace{\baselineskip}

\paragraph{Step 1- Hierarchical Tree Clustering}:\\
1.	Calculate the the FRM adjacency matrix $A^s$  (Equation (\ref{eq24}))
Where $\beta_{i,j}$ is the degree of connectedness between a pair of assets $\{i,j\}$ \\
2. Introduce the vector of $(\lambda_{j})$ as diagonal elements in the adjacency matrix $A^s$, so we obtain the new adjacency matrix $\tilde{A}$
\begin{equation}
\tilde{A^s} = 
\begin{pmatrix}
\lambda_1 & \beta_{1,2} & \cdots & \beta_{1,n} \\
\beta_{2,1} & \lambda_2 & \cdots & \beta_{2,n} \\
\vdots  & \vdots  & \ddots & \vdots  \\
\beta_{n,1} & \beta_{n,2} & \cdots & \lambda_n
\end{pmatrix}
\label{eq40}
\end{equation}
3. Replace the covariance matrix in the HRP algorithm with the adjacency matrix $\tilde{A^s}$.\\
4.	Transform $\tilde{A^s}$ to an adjacency-distance matrix $D$, 
\begin{equation}
    D_{i,j} = D[X_i, X_j]=
    \begin{cases}
 \sqrt{\frac{1}{2}(1-\beta_{i,j})}\; \;\; \; if \; \;\; \; \vert \beta_{i,j}\vert <1\\
 \; \; \\
\mbox{max}D[X_i, X_j]\; \;\; \;\; \;\; \;\; \; otherwise.
\end{cases}
    \label{eq41}
\end{equation}
5. Compute a new distance matrix $\tilde{D}$ 
\begin{equation}
    \tilde{D}= \tilde{D}[D_{i},D_{j}] = \sqrt{\sum_{n=1}^{N}(D_{n,i}-D_{n,j})^2}
    \label{eq42}
\end{equation}
6. Form the assets clusters in a recursive manner based in Equation(\ref{eq42}). The set of clusters is donated by $\tilde{U}$ and the first formed cluster$(i^*,j^*)$ is define as: 
\begin{equation}
    \tilde{U}[1]=(i^{*},j^{*}) =\text{argmin}(i,j)_{i \neq j}\{\tilde{D}_{i,j}\}
    \label{eq43}
\end{equation}
7.	Update the distance matrix $D$ by computing the distances of other assets from the newly formed cluster $\tilde{U}(1)$ as follows: 
\begin{equation}
    D_{i,\tilde{U}[1]}=\min\bigl[\{\tilde{D}_{i,j}\}_{j \in U[1]}\bigl]
     \label{eq44}
\end{equation}

\paragraph{Step 2- Quasi Diagonalisation or Matrix Seriation}:\\
The Quasi-Diagonalisation of the adjacency matrix or Matrix seriation is adopted to rearrange the data to represent clearly the inherent clusters (as explained in the classical approach).

\paragraph{Step 3- Recursive Bisection}:\\ 
1.	Assign a unit weight to all assets, 
\begin{equation*}
    W_i=1,\;\;\;\forall\;i=1,...N
\end{equation*}
2.	Bisect each cluster into two sub-clusters.\\
3.	Calculate the variance for each sub-cluster 
\begin{equation}
    \tilde{V}_{1,2}=\tilde{w}^{\top}\tilde{A}  \tilde{w}\;\;\;\;; \tilde{A} \;\;\text{is the adjacency matrix}
     \label{eq45}
\end{equation}
where, 
\begin{equation}
    \tilde{w}=  \frac{\mbox{diag}[\tilde{A}]^{-1}}{\mbox{tr[diag}[\tilde{A}]^{-1}]}
     \label{eq46}
\end{equation}
4.	Calculate the weighting factor based on the quasi-diagonalised adjacency matrix.
\begin{equation}
    \tilde{\alpha}_{1}=1-\frac{\tilde{V}_{1}}{\tilde{V}_{1}+\tilde{V}_{2}}, \text{so \ that} \ 0 \leq \tilde{\alpha}_{i} \leq 1\; \; ;\tilde{\alpha}_2=1-\tilde{\alpha}_1
     \label{eq47}
\end{equation}

5-	Update the weights $\tilde{w}_1$ and $\tilde{w}_2$ for both sub-clusters:
\begin{equation}
    \tilde{w}'_1=\tilde{\alpha}_1\cdot w_1;\;\;\;\
    \tilde{w}'_2=\tilde{w}{\alpha}_2\cdot w_2  
     \label{eq48}
\end{equation}

6- Execute recursively steps 2-5, the algorithm stops when we have a single asset for each cluster and then the weights are assigned to all assets in the portfolio.\vspace{\baselineskip} 

\section{Financial Risk Meter in Emerging Markets}
\label{FRMEMinterp}
\subsection{FRM@EM data-set description}
In this paper we study the largest 25 Emerging Market financial institutions (FIs) by market capitalisation at any given point in time, with focus on the BRIMST FIs. We compile a database of daily price levels in as well as market capitalisations in U.S. Dollars from Bloomberg, and select the biggest financial institutions on a daily basis from the most liquid local EM equity market indices. On any given trading day in consideration, we take the price returns of those biggest $j=25$ FIs over an estimation window $s=63$ business days. 

As for the macroeconomic data, we follow \cite{tobias2016covar} concerning the developed market specific risk factors, to repeat, returns in US REITs, S\&P 500 index, U.S. three months treasury bill rates, the spread between 3 months and 10 year U.S. treasury rates, the spread between BAA rates corporate bonds by the rating agency Moody's to U.S. treasury bonds, and the implied volatility index VIX based on outstanding options on the S\&P 500 equity index. We add the following EM specific macroeconomic risk variables: The J.P. Morgan Emerging Bond Index Global Sovereign Spread index tracking the Emerging Bond Index yields over the benchmark U.S. Treasury bonds and thereby representing the risk compensation demanded from investors when investing in EM sovereign bonds, as well as the respective countries currency versus the U.S. dollar cross. 

\subsection{FRM@EM Interpretation and Network Analysis}
\begin{figure}[H]
  \centering
  \includegraphics[width=1\textwidth]{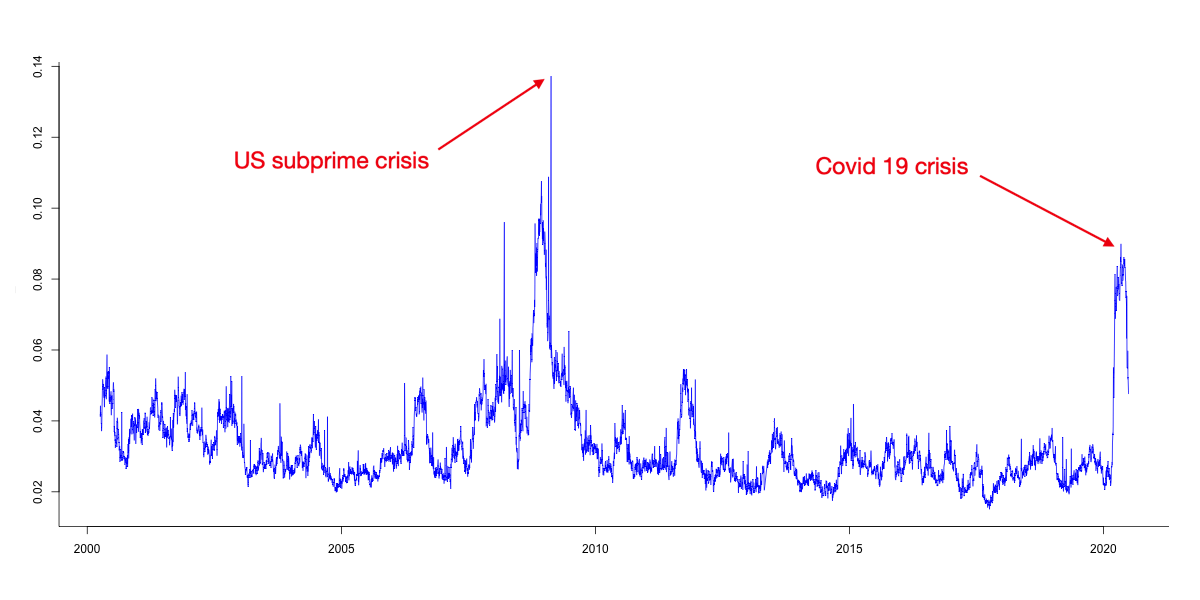}
  
  \caption{FRM@EMs Time series}
  \hspace*{\fill} \raisebox{-1pt}{\includegraphics[scale=0.008]{Graphics/Quantlets_Logo_Ring.jpeg}}
  \label{FRMEMtimeseriestotal}
\end{figure}

\begin{figure}[H]
  \centering
  \includegraphics[width=1\textwidth]{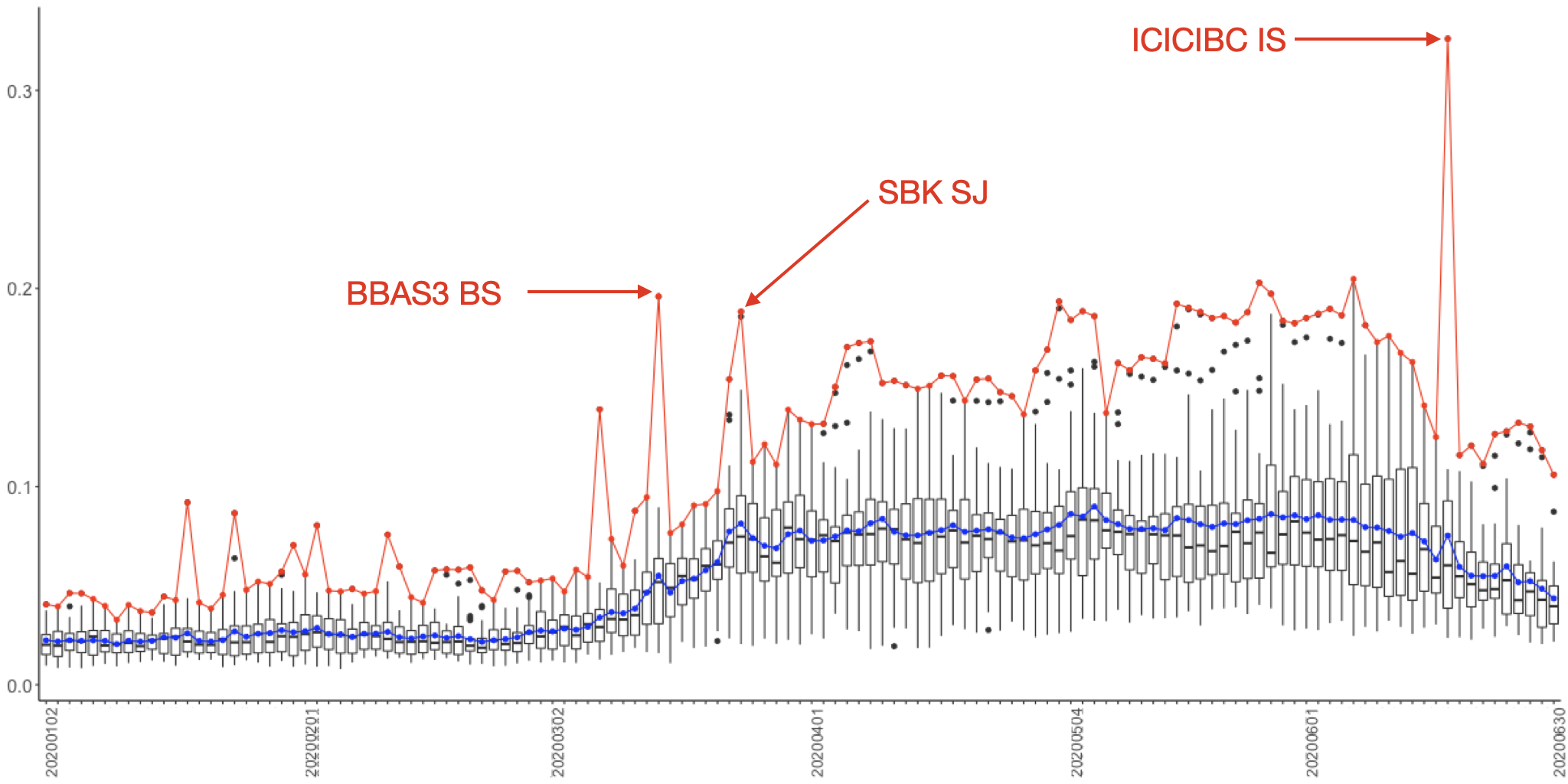}
  \caption{FRM@EM Boxplot, with \textcolor{blue}{mean} and \textcolor{red}{maximum} at $\tau=0.05$}
  \hspace*{\fill} \raisebox{-1pt}{\includegraphics[scale=0.008]{Graphics/Quantlets_Logo_Ring.jpeg}}
  \label{FRMEMboxplot}
\end{figure}

\begin{figure}[H]
 \centering
 \begin{subfigure}[b]{0.45\linewidth}
   \includegraphics[width=\linewidth]{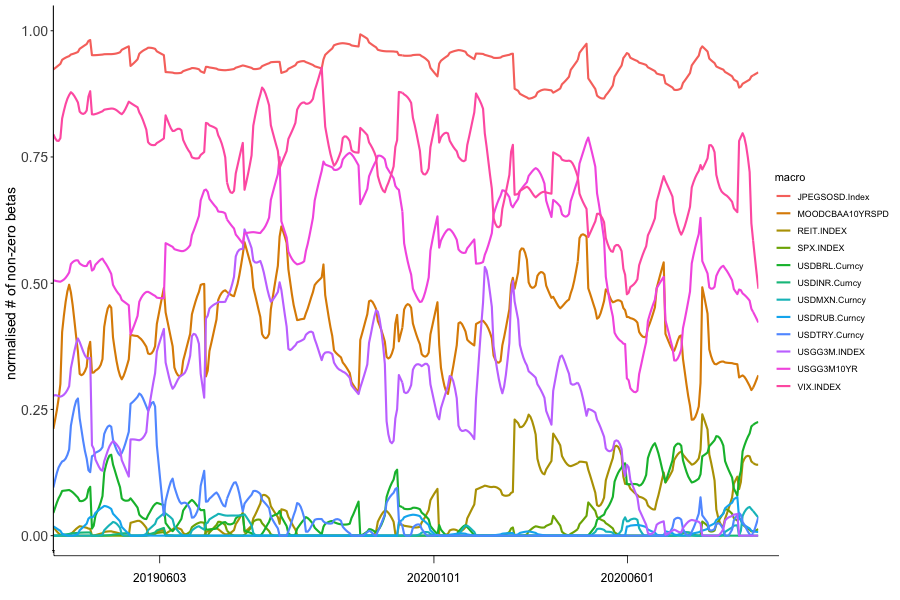}
   \caption{$\tau=0.05$}
 \end{subfigure}
 \begin{subfigure}[b]{0.45\linewidth}
   \includegraphics[width=\linewidth]{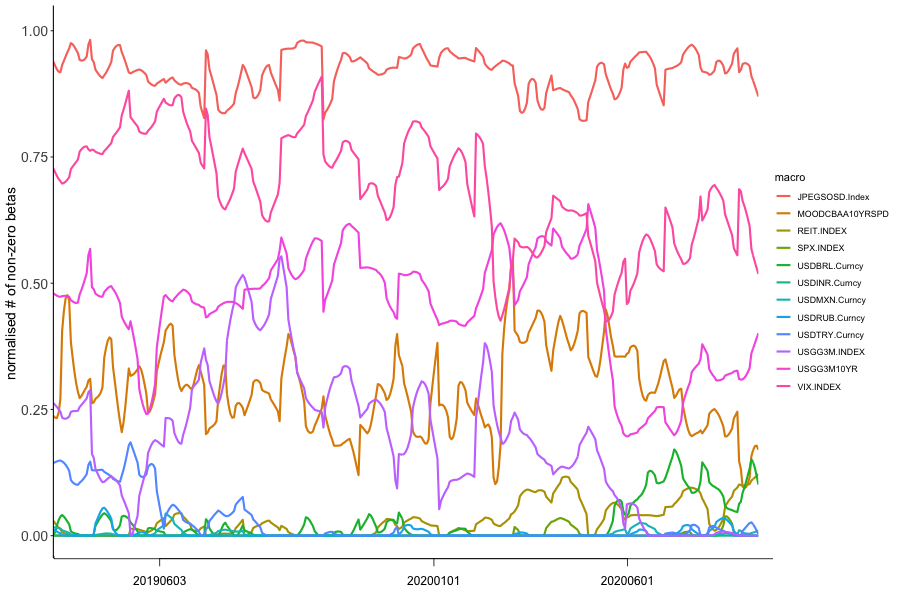}
   \caption{$\tau=0.1$}
   \end{subfigure}
   \begin{subfigure}[b]{0.45\linewidth}
   \includegraphics[width=\linewidth]{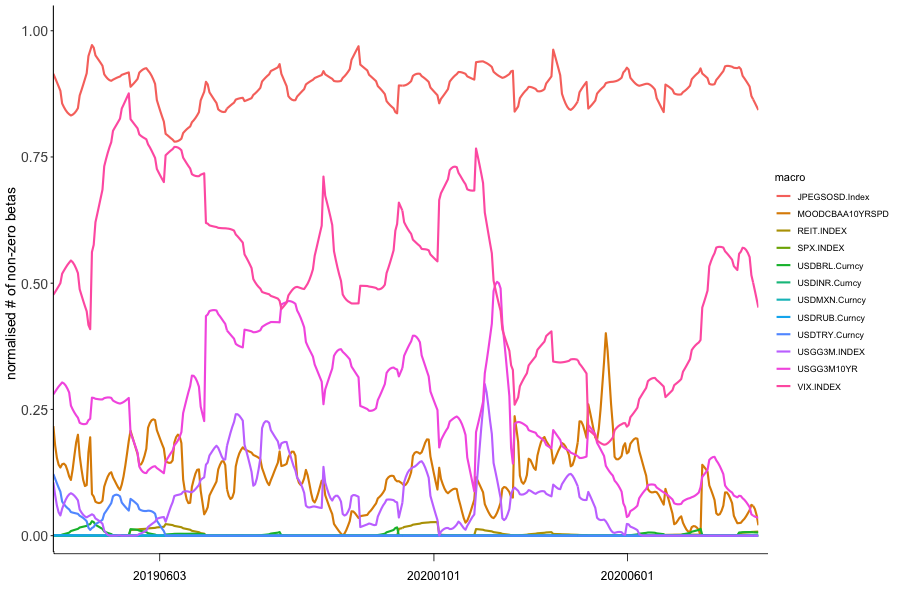}
   \caption{$\tau=0.25$}
 \end{subfigure}
 \begin{subfigure}[b]{0.45\linewidth}
   \includegraphics[width=\linewidth]{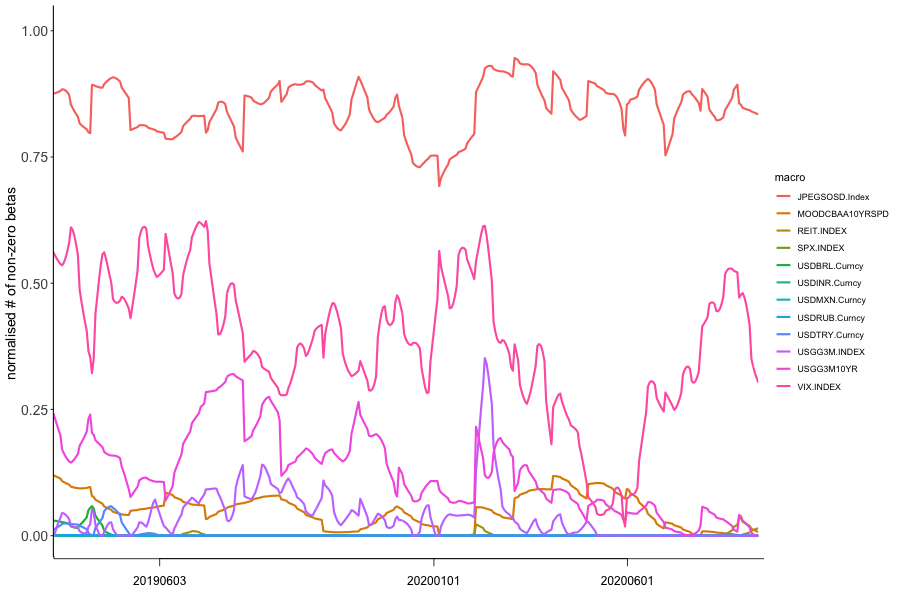}
   \caption{$\tau=0.5$}
 \end{subfigure}
 \caption{Macro variables' most frequent marginal return contribution across time with 7-day moving average: \textcolor{red}{EM Sovereign Spread}, \textcolor{pink}{VIX}, \textcolor{violet}{U.S. 3mth to 10yr yield spread},  \textcolor{brown}{Moody's BAA Corporate Yield Spread}, \textcolor{green}{S\&P 500 Index}}
 \hspace*{\fill} \raisebox{-1pt}{\includegraphics[scale=0.008]{Graphics/Quantlets_Logo_Ring.jpeg}}
 \label{fig:macro}
\end{figure}

\begin{sidewaysfigure}
  \centering
    \includegraphics[width=\textwidth, scale=2]{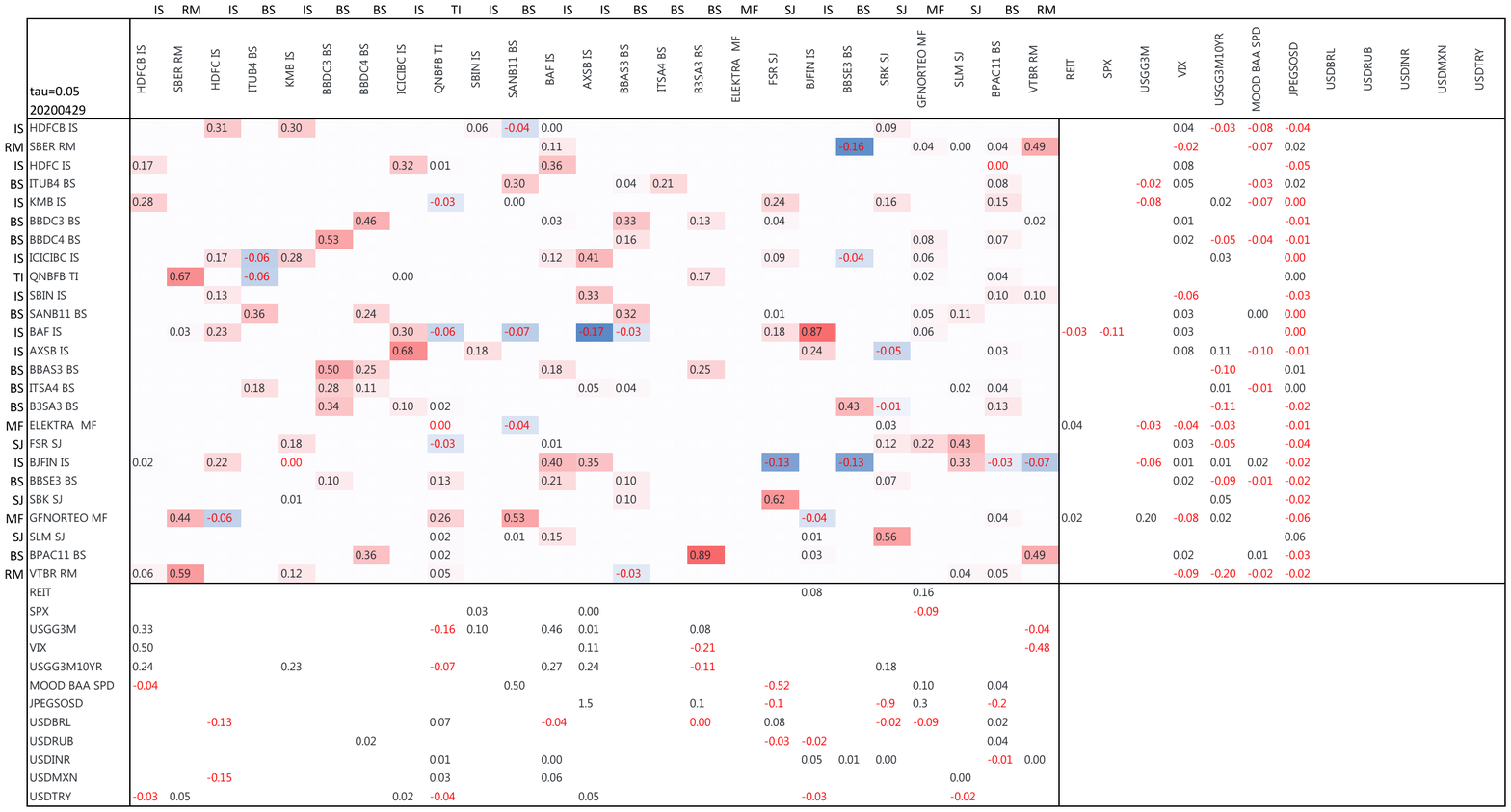}
    \caption{Adjacency matrix estimated at $\tau=0.05$ on 20200429}
    \hspace*{\fill} \raisebox{-1pt}{\includegraphics[scale=0.008]{Graphics/Quantlets_Logo_Ring.jpeg}}
    \label{ADJMAT20200429005}
\end{sidewaysfigure}

\begin{sidewaysfigure}
  \centering
    \includegraphics[width=\textwidth, scale=2]{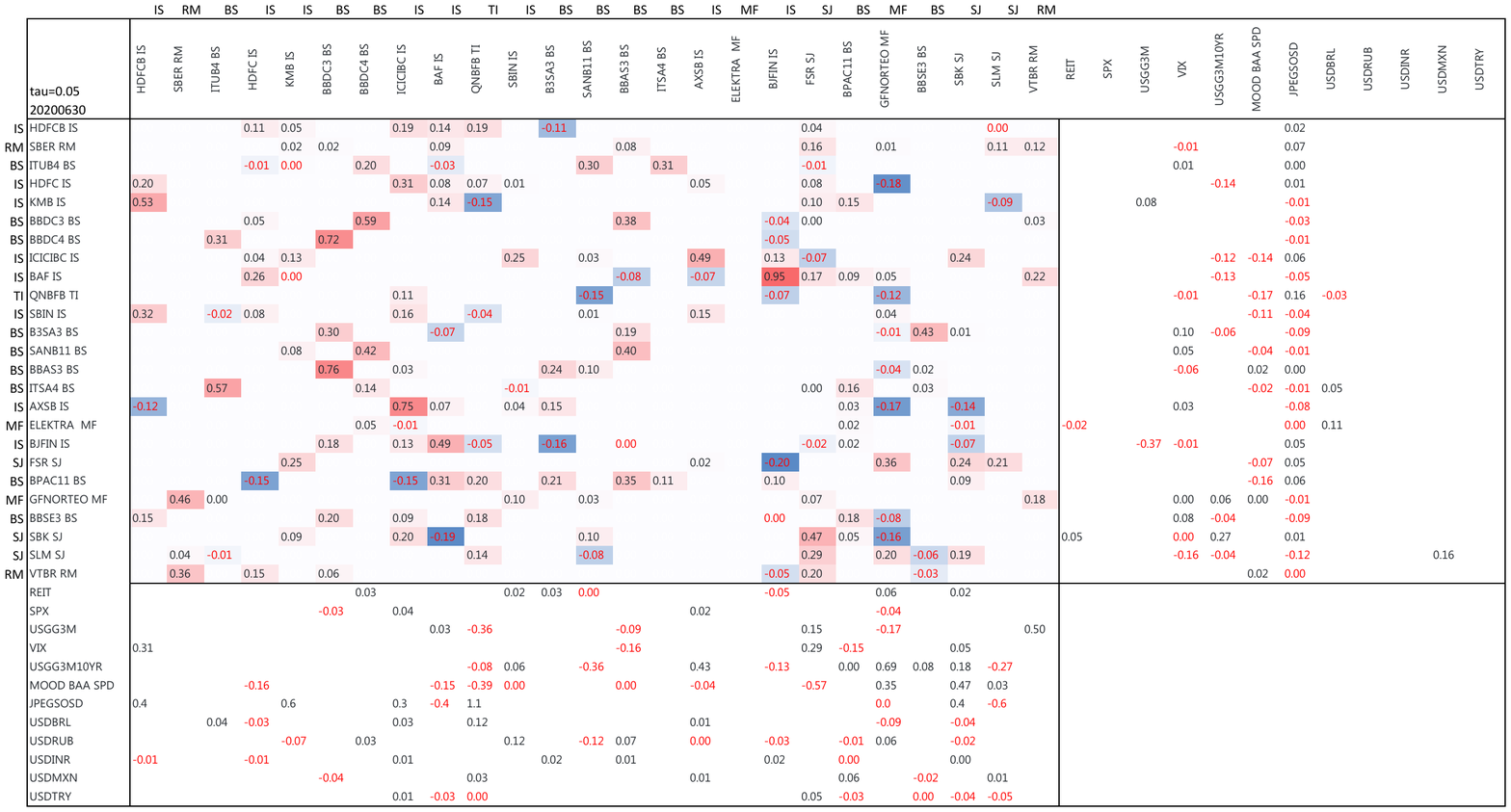}
    \caption{Adjacency matrix estimated at $\tau=0.05$ on 20200630}
    \hspace*{\fill} \raisebox{-1pt}{\includegraphics[scale=0.008]{Graphics/Quantlets_Logo_Ring.jpeg}}
    \label{ADJMAT20200630005}
\end{sidewaysfigure}

\begin{figure}[H]
  \centering
  \includegraphics[width=1\textwidth]{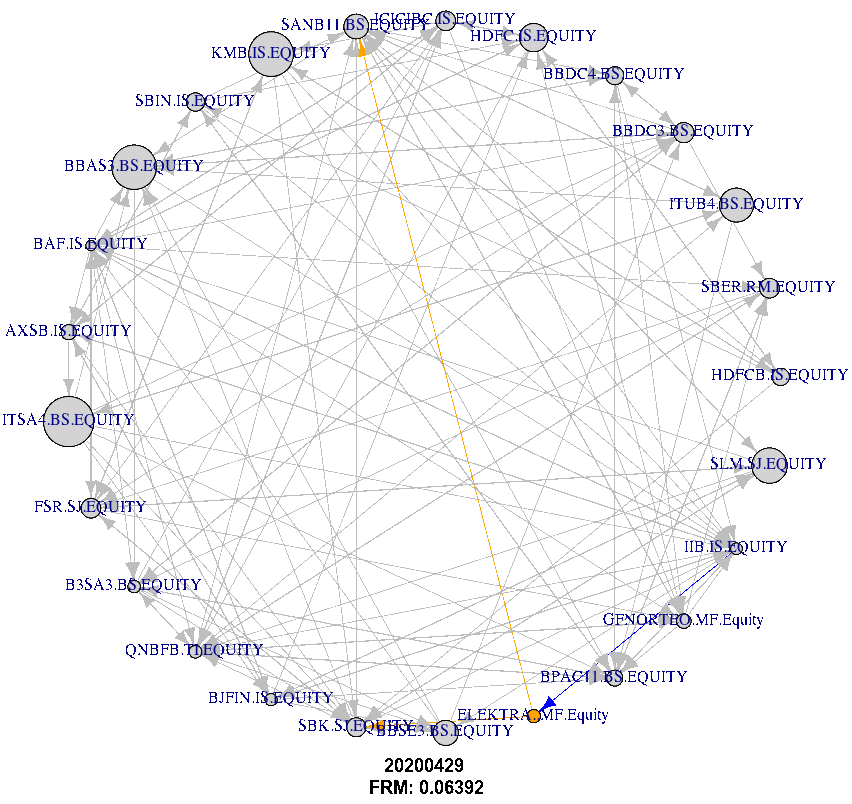}
  \caption{FRM@EM Network with one \textcolor{orange}{example node} and its \textcolor{blue}{in-degree} and \textcolor{orange}{out-degree} edges}
  \hspace*{\fill} \raisebox{-1pt}{\includegraphics[scale=0.008]{Graphics/Quantlets_Logo_Ring.jpeg}}
  \label{FRMEMnetwork}
\end{figure}

\begin{figure}[H]
 \centering
 \begin{subfigure}[b]{0.45\linewidth}
   \includegraphics[width=\linewidth]{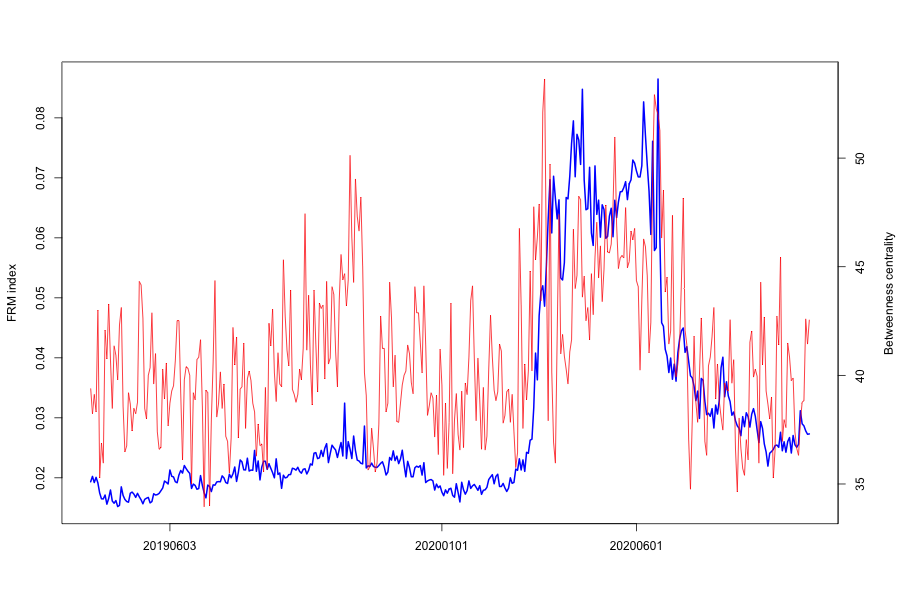}
   \caption{{\color{blue} FRM} and \color{red}{Betweenness}}
   \label{FRMbetweenness}
 \end{subfigure}
 \begin{subfigure}[b]{0.45\linewidth}
   \includegraphics[width=\linewidth]{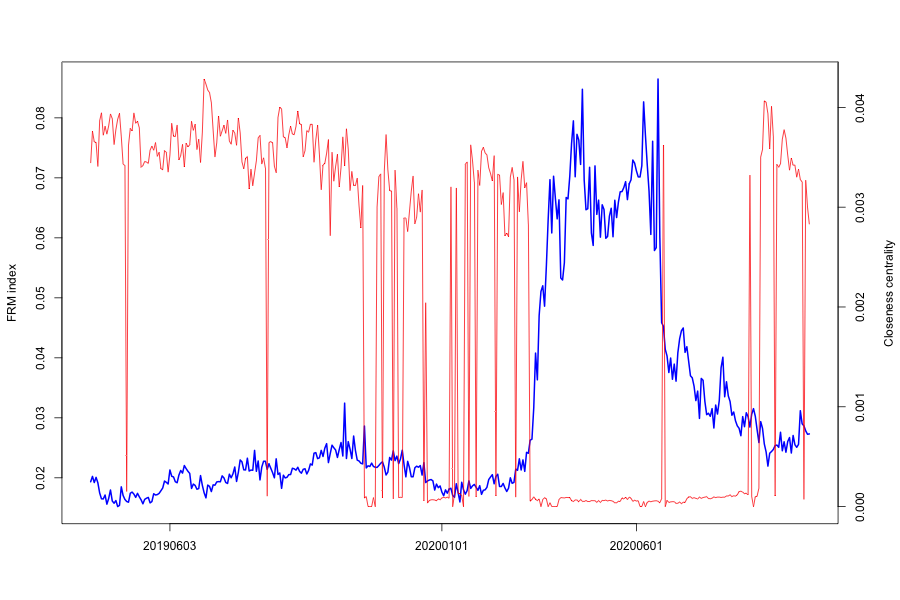}
   \caption{{\color{blue} FRM} and \color{red}{Closeness}}
   \label{FRMclosness}
   \end{subfigure}
   \begin{subfigure}[b]{0.45\linewidth}
   \includegraphics[width=\linewidth]{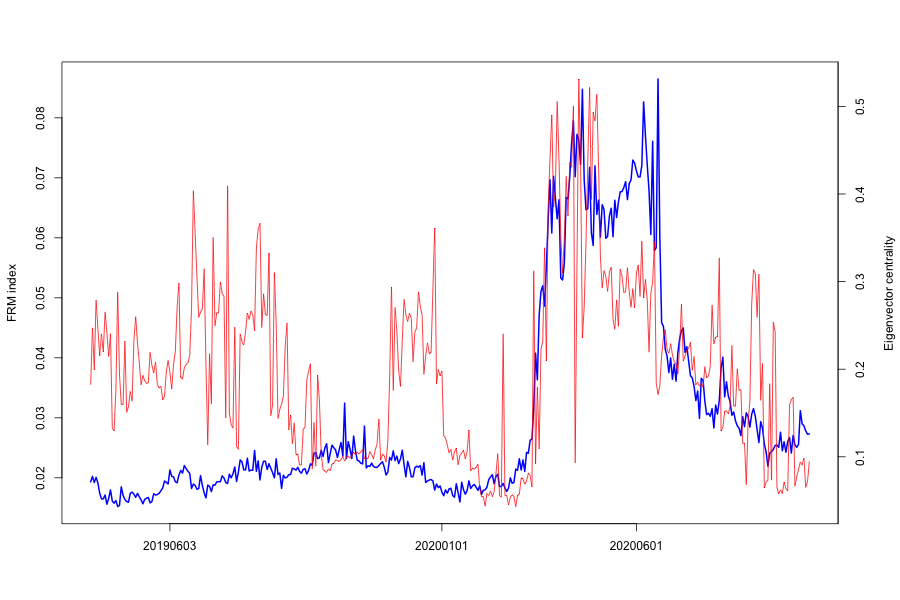}
   \caption{{\color{blue} FRM} and {\color{red} Eigenvector}}
   \label{FRMeigenvector}
 \end{subfigure}
 \begin{subfigure}[b]{0.45\linewidth}
   \includegraphics[width=\linewidth]{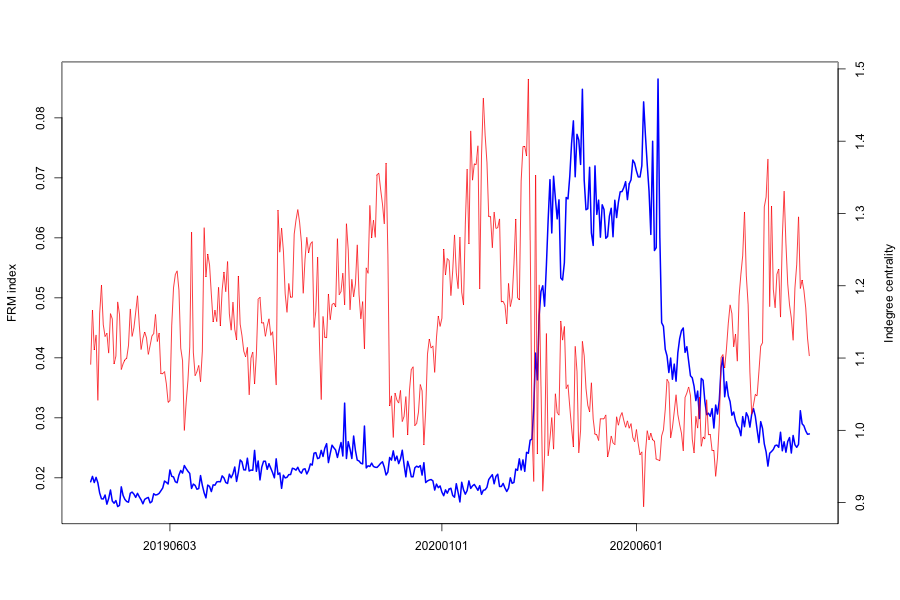}
   \caption{{\color{blue} FRM} and {\color{red} In-degree}}
   \label{FRMindegree}
 \end{subfigure}
 \begin{subfigure}[b]{0.45\linewidth}
   \includegraphics[width=\linewidth]{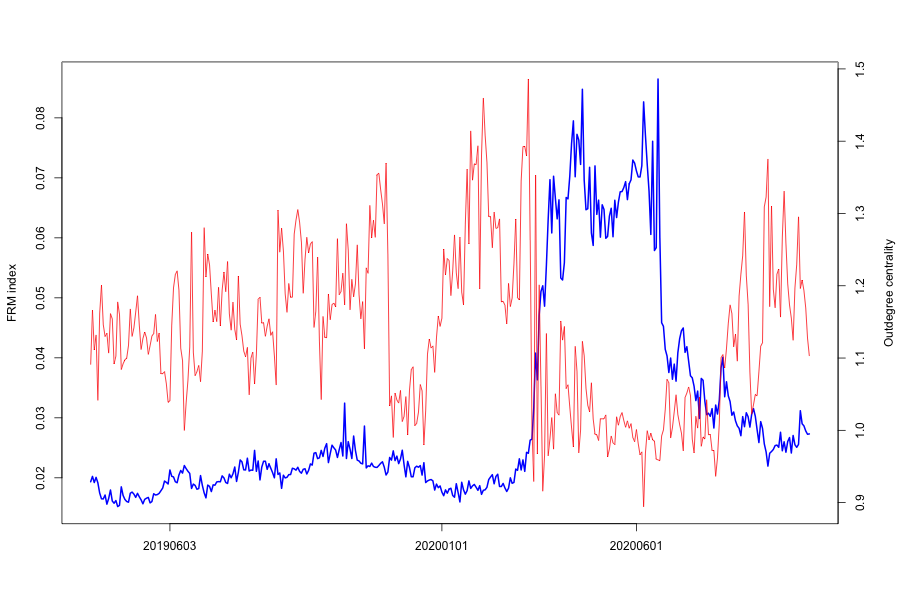}
   \caption{{\color{blue} FRM} and {\color{red} Out-degree}}
   \label{FRMoutdgree}
 \end{subfigure}
 \caption{{\color{blue} FRM} at $\tau=0.05$ and {\color{red} Centrality measures}}
 \hspace*{\fill} \raisebox{-1pt}{\includegraphics[scale=0.008]{Graphics/Quantlets_Logo_Ring.jpeg}}
 \label{FRMcentmeasures}
\end{figure}

Figure \ref{FRMEMtimeseriestotal} depicts the FRM@EM from April 2000 to June 2020. Clearly observable are the periods of distress in the global financial system around 2008, 2012, more recently in 2020 but also during the EM specific market distress periods such as 2002 (Argentina) and 2013 (following the Federal Reserve Board's Open Market Committee forward guidance). \\
Figure \ref{FRMEMboxplot} introduces one of the FRM's tool: FRM is not just the mean but actually a distribution of $\lambda^s_j$ as well as the adjacency matrix containing all (Lasso penalised) $\beta^{s}_{j}$ between FIs, and also between FIs and macroeconomic risk variables. As an example of the information contained, we mark some of the more extreme maxima in Figure \ref{FRMEMboxplot}. For example, into the crisis, Standard Bank Group (SBK SJ) had a very high $\lambda^s_j$ reading, indicating the bank was a "risk receiver", thus at risk to be impacted by spill over effects. \\
But we can also have a more detailed look at the adjacency matrix containing the $\beta^{s}_{j}$ between FIs and macroeconomic risk variables, with two examples shown for 20200429 in Figure \ref{ADJMAT20200429005} as well as post crisis 20200630 in Figure \ref{ADJMAT20200630005}, both estimated at $\tau=0.05$, and colour scaled in blue (low/negative) to high (red/positive). Observable are the high dependencies between countries of the same region (\textbf{B}S for Brazil, \textbf{R}M for Russia, \textbf{I}S for India, \textbf{M}F for Mexico, \textbf{S}J for South Africa and \textbf{T}I for Turkey), and often negative relationships (adjusted for co-movements with macroeconomic risk variables) in between regions. However, there are also detectable co-dependencies, which necessitate closer inspection from both investors as well as regulators. For example in Figure \ref{ADJMAT20200429005} South Africa's Sanlam Ltd. (SLM SJ)'s returns are explained to a significant extend by Bajaj Finance Ltd. of India (BAF IS), as both financial services companies, providing an assortment of financial services. Similarly, Banco BTG Pactual (BPAC11 BS)'s returns are explained not only by other Brazilian FIs, but also significantly by Russia's VTB Bank (VTBR RM), both operating in more banking and investment banking related markets. Clearly, sub-sector dependencies across EM FIs are to be considered to prevent risk clusters. \\
Another component to consider is the impact from macroeconomic risk variables' changes on FIs returns. As can be seen on both adjacency matrices, the "classic" macroeconomic risk variables do have an impact, however, to a large extend, the Emerging Market Sovereign Yield Spread to U.S. Treasuries (JPEGSOSD) is the main influence. In fact, this is true across from $\tau=0.05$ to $\tau=0.50$. In Figure \ref{fig:macro} we show the smoothed (rolling seven day mean) of the share of FIs impacted by the respective macroeconomic risk variable. The Emerging Market Yield spread is the dominant driver, followed by more general market risk measures such as the VIX, Moody's Baa corporate yield spreads, and the shape of the U.S. Treasury yield curve. Emerging market currencies as a cross versus the U.S. Dollar have a lesser impact. It is mostly one or two EM currencies having an overall impact, and not only on domestic banks. In Figure \ref{ADJMAT20200630005} for example, the Brazilian Real (USDBRL) has marginal negative return contribution to Bajaj Finance Ltd in India (BAF IS), and positive return contribution to Brazilian Itausa SA (ITSA4 BS) and Mexican Grupo Elektra SAB (ELEKTRA MF). \\
In Figure \ref{FRMcentmeasures}, we show the time series of FRM against various centrality measures. We observe that Betweenness, Eigenvector have similar trends as the FRM EM series, especially during the crisis period of March to May 2020. On the other hand, Closeness centrality drops sharply into the crisis period. 
When the FRM rises, the number of $\beta^{s}_{ij}$ equal to zero increase, increasing with it the distance between the vertices. The average length of one node (one FI in our case) and all other FIs increases, thereby sharply reducing Closeness centrality. 
Betweeness as a centrality measure of a vertex within a network rises when the FRM rises, as information flow has a very high probability to pass between some central FIs, indicating a concentration of risk around certain FIs which we call "risk emitters". 
Similarly, Eigenvector centrality is a measure of an FIs influence in an observed financial system network. Central FIs Eigencentrality rises sharply around a crisis period, as the FRM increases in value.
In- and Out-degree centrality drop when FRM rises, since the edges or connections between FIs have reduced sharply, to mostly the network's risk emitters.
We can conclude therefore, that a close inspection if the distribution of $\lambda^s_j$ as well as the detailed information within the adjacency matrix across a range of $\tau$ are particularly important. In Section \ref{portfolioconstruction} we aim to make use of this richness of information for the construction of more robust, tail-event network behaviour attentive portfolios. As an indication, Figure \ref{FRMEMnetwork} shows a network graph with edges between the 25 largest FIs, estimated at $\tau=0.05$ on 20200429. We highlight one exemplary bank, and its edges stemming from the adjacency matrix. Condensing such information into clusters of risk as outlined above is the focus of the following portfolio construction discussion. 
In Figure \ref{fig:macro} we depict the macroeconomic risk variables influence over time. Clearly, the EM Sovereign Yield Spread to U.S. Treasury Bonds is the prominent macroeconomic risk variable not only during tail events but also at "normal" times estimated at $\tau=0.50$. We can conclude that most EM risk is rapidly priced into yield spreads, and then consequently impacts financial institutions. This link between the sovereign and banks needs to be considered for investors and policy makers as well. Other strongly influencing risk variables are more expected such as the VIX, the US yield curve shape, and Corporate Bond yield spreads. However, EM currency fluctuations only have a marginal effect in tail events, on some of the FIs. With the increase of debt issuance in local currency denominated debt, the risk of mismatches versus developed market currencies has diminished overall, and has lower influence on EM FIs.

\section{Portfolio Construction}
\label{portfolioconstruction}
The portfolio is constructed using daily FIs prices for the period between 20200101 and 20200630 (which yields 130 observations of the 25 biggest EM FIs).\\
The performance of the uplifted portfolio approaches (Inv$\lambda$ and upHRP) are benchmarked against the minimum-variance portfolio (MinVar), the inverse variance (IVP), and the classical HRP approaches. The MinVar weights are simply computed based on Equation (\ref{minvarweights}), and often lead to concentration in low-volatility assets. The IVP strategy can be considered as a naive risk parity strategy (Equation (\ref{IVPweights})), since it is agnostic with respect to asset correlations. Thereafter, an overlapping region between portfolio optimisation startegies and FRM network centrality is developed.\\
As shown in the FRM network (Figure \ref{FRMEMnetwork}), all FIs are treated as potential substitutes without specifying any hierarchical structure among them. Therefore, tree structures that integrate hierarchical relationships are needed. For that purpose, the upHRP algorithm aims to build and make use of a clustered adjacency matrix.
\paragraph{\textbf{Matrix seriation}}\

For the classical approach, the steps of the HRP are applied using the price time series of the 25 biggest FIs selected during the studied period. Figure \ref{fig:Classical HRP: Matrix Seriation}  presents the estimation results:

\begin{figure}[H]
  \centering
  \begin{subfigure}[b]{0.45\linewidth}
    \includegraphics[width=\linewidth]{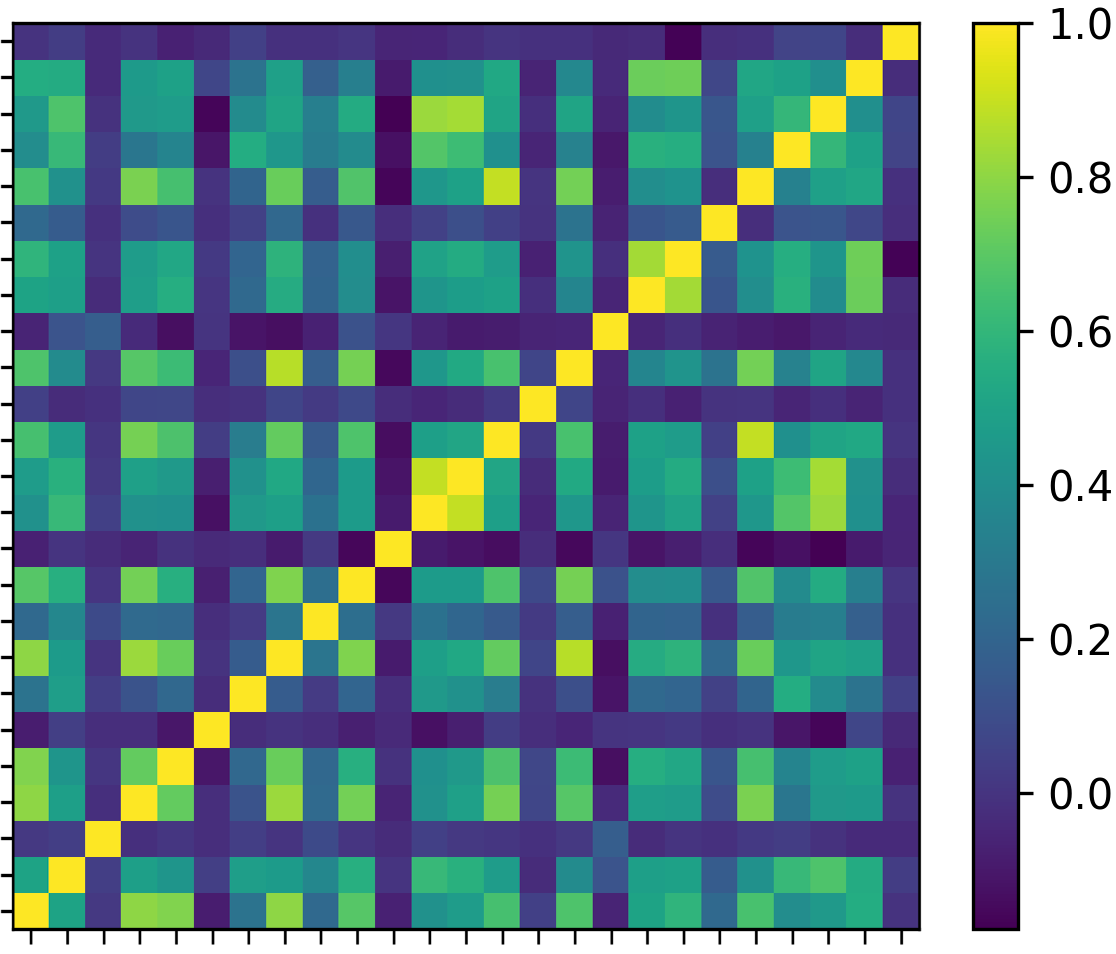}
    \caption{Unclustered Correlations}
    \label{UnclusteredCorrelations}
  \end{subfigure}
  \begin{subfigure}[b]{0.45\linewidth}
    \includegraphics[width=\linewidth]{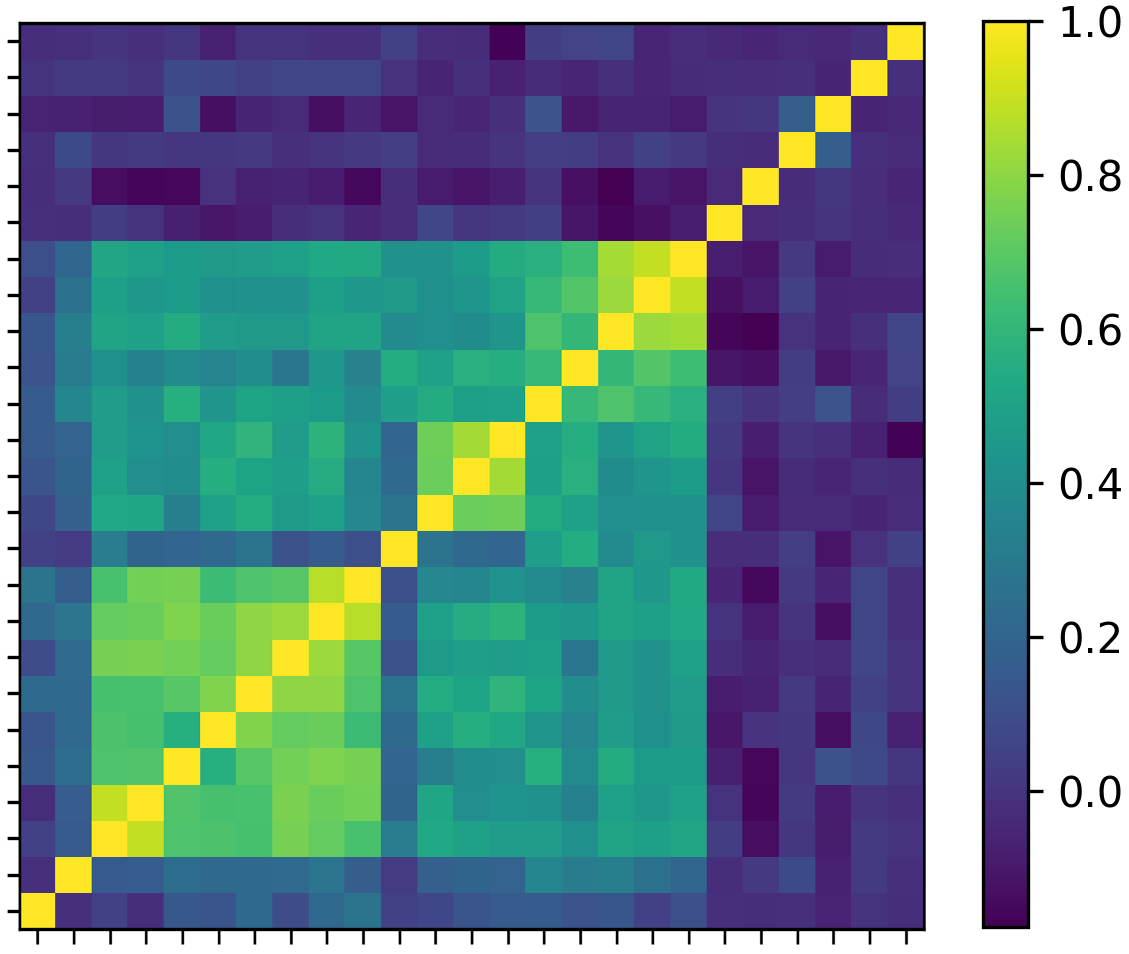}
    \caption{clustered Correlations}
    \label{ClusteredCorrelations}
  \end{subfigure}
  \caption{Classical HRP: Matrix Seriation}
  \hspace*{\fill} \href{https://github.com/QuantLet/FRM-EM-paper}{\includegraphics[scale=0.008]{Graphics/Quantlets_Logo_Ring.jpeg}}
  \label{fig:Classical HRP: Matrix Seriation}
\end{figure}
Figure \ref{UnclusteredCorrelations} shows the original or the unclustered correlation matrix. Figure \ref{ClusteredCorrelations} illustrates the correlation matrix after reordering in clusters using the hierarchical tree clustering, also called clustered correlation matrix. This matrix then serves as the input for the asset allocation procedure. As shown in Figure \ref{ClusteredCorrelations} the reorganizing results group similar FIs together and the dissimilar further away in a new correlation matrix, which helps to construct more meaningful asset allocation decisions and building more risk diversified portfolios. More precisely, from Figure \ref{ClusteredCorrelations} the lighter-colored squares (indicating a higher correlation coefficient) are all concentrated around the diagonal matrix.\vspace{\baselineskip}
\begin{figure}[H]
  \centering
  \begin{subfigure}[b]{0.45\linewidth}
    \includegraphics[width=\linewidth]{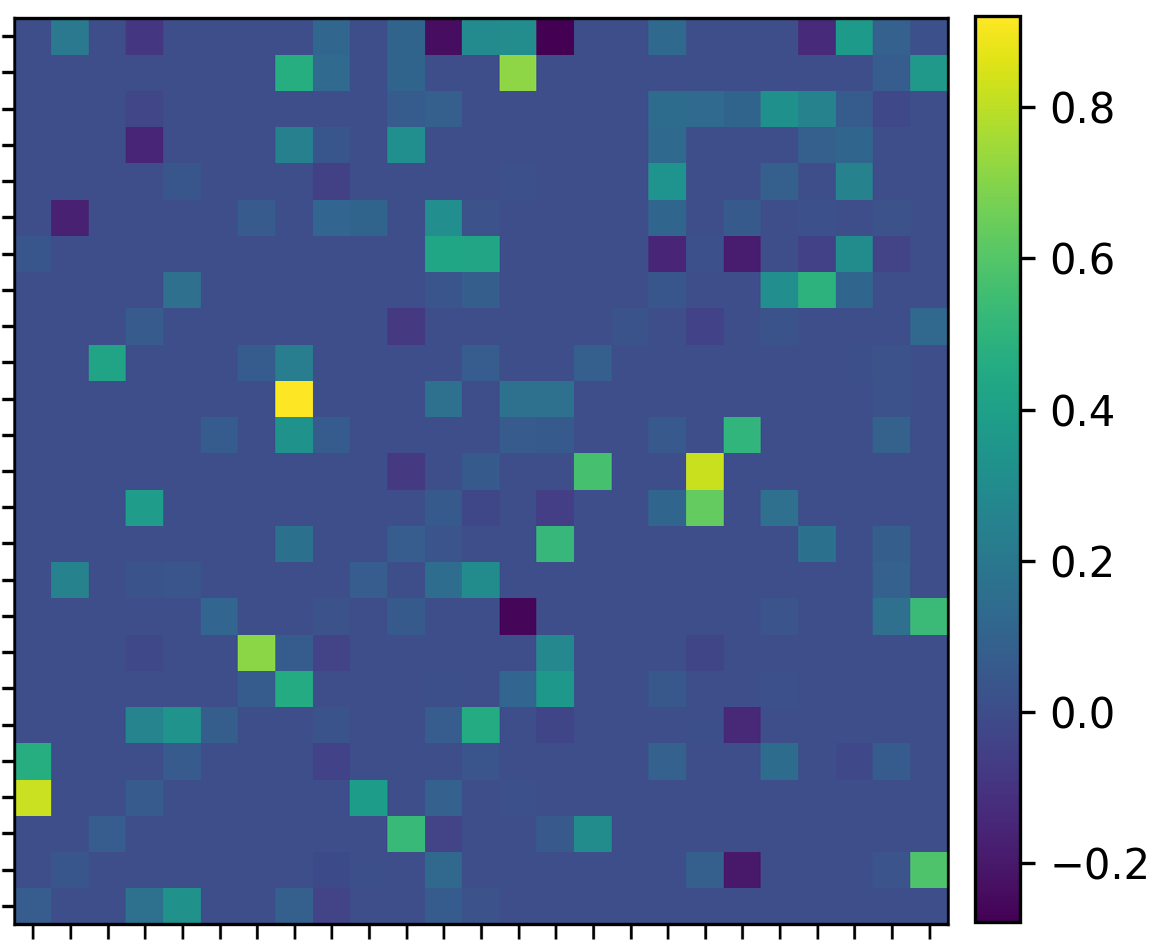}
    \caption{Unclustered Adjacency matrix}
    \label{UnclusteredAdjmatx}
  \end{subfigure}
  \begin{subfigure}[b]{0.45\linewidth}
    \includegraphics[width=\linewidth]{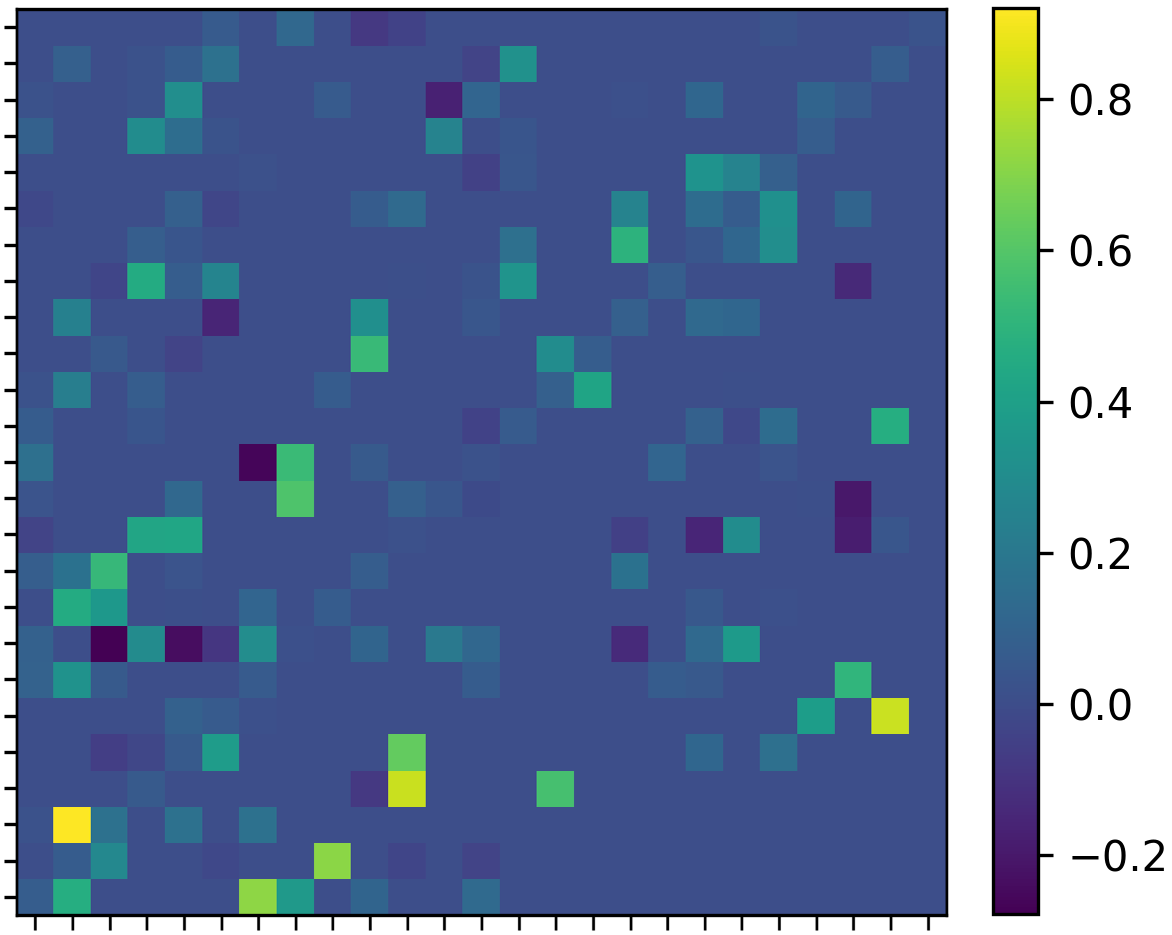}
    \caption{clustered Adjacency matrix}
    \label{ClusteredAdjmatx}
  \end{subfigure}
  \caption{Uplifted HRP: Matrix Seriation}
  \hspace*{\fill} \href{https://github.com/QuantLet/FRM-EM-paper}{\includegraphics[scale=0.008]{Graphics/Quantlets_Logo_Ring.jpeg}}
  \label{fig:Uplifted HRP: Matrix Seriation}
\end{figure}

For the uplifted approach, the quasi-diagonalization step is applied to the adjacency matrix dated on 20200630 with $\tau=0.05$. In this stage, the algorithm aims at reordering the adjacency matrix by placing similar assets together. As we can notice from Figure \ref{ClusteredAdjmatx} the closer assets (similar given ($\beta$)) are placed together forming assets clusters.\vspace{\baselineskip}

\textbf{Tree clustering or Dendrogram}\vspace{\baselineskip}

\begin{figure}[H]
  \centering
  \includegraphics[width=1\textwidth]{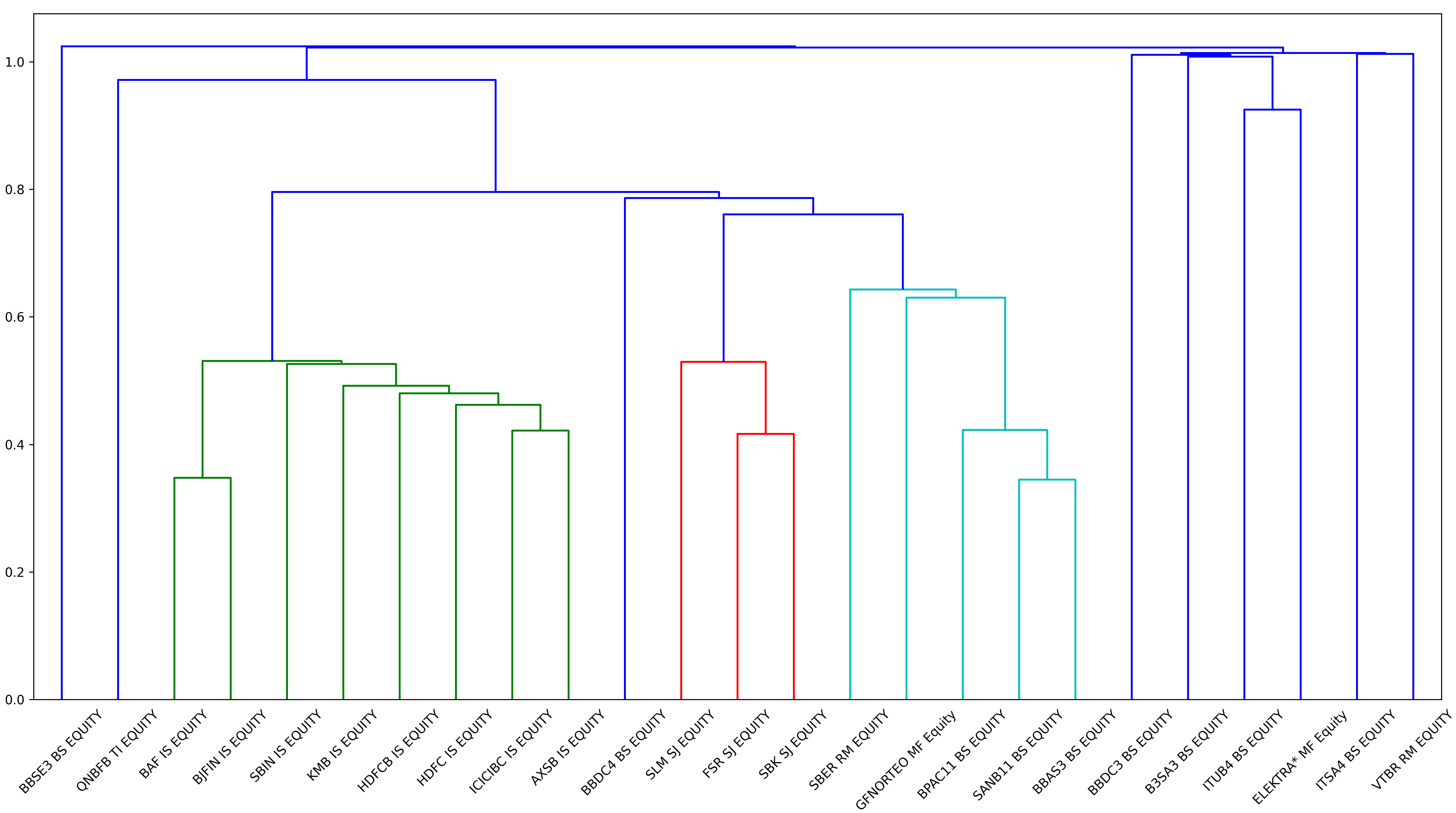}
  \caption{Classical HRP: EM@Dendrogram}
  \hspace*{\fill} \href{https://github.com/QuantLet/FRM-EM-paper}{\includegraphics[scale=0.008]{Graphics/Quantlets_Logo_Ring.jpeg}}
  \label{ClassicalHRPEM@Dendrogram}
\end{figure}

The clusters are visualised in the form of a cluster diagram called a dendrogram. Figure \ref{ClassicalHRPEM@Dendrogram} illustrates the hierarchical clusters for our FIs data, where the x-axis indicates the name of the FIs in the studied portfolio and the y-axis measures the distance between the two merging FIs. 
The key to interpreting a dendrogram is to focus on the height at which any two FIs are joined together. In Figure \ref{ClassicalHRPEM@Dendrogram}, we can see that Bajaj Finance (BAF IS) and Bajaj Finserv Ltd (BJFIN IS) (plotted with green) are expectedly most similar, as the height of the link that joins them together is the lowest. Note that both of them are same ultimate parent Indian financial services companies focused on insurance. The dendrogram also detects that the most similar FIs most often belong to the same market (FirstRand Ltd (FSR SJ) and Standard Bank Group Ltd (SBK SJ) plotted with red), or a similar financial sector. Finally, the highest cluster represents the giant cluster that joins all the FIs formed clusters together.

\begin{figure}[H]
  
  \centering
  \includegraphics[width=1\textwidth]{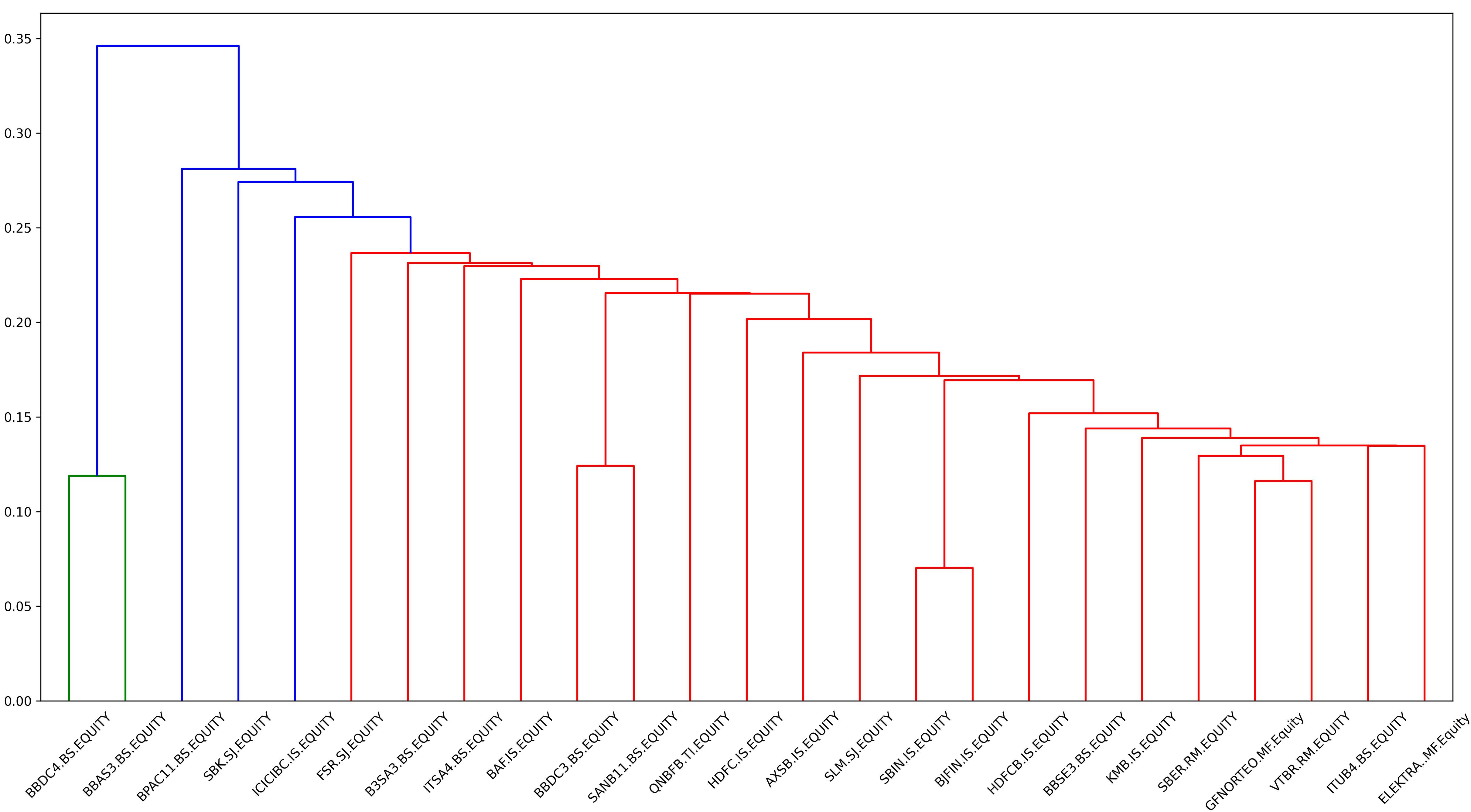}
  \caption{Uplifted HRP: EM@Dendrogram}
  \hspace*{\fill} \href{https://github.com/QuantLet/FRM-EM-paper}{\includegraphics[scale=0.008]{Graphics/Quantlets_Logo_Ring.jpeg}}
  \label{upHRPEM@Dendrogram}
\end{figure}
Based on the FRM clustered adjacency matrix, Figure \ref{upHRPEM@Dendrogram} indicates that the tree clustering architecture differs from the previous one (Figure \ref{ClassicalHRPEM@Dendrogram}). Indeed, here State Bank of India (SBIN IS) is a perturbation of BJFIN IS, and consequently the two assets present the first formed cluster since they are the most similar FIs. Also, Grupo Financiero Banorte (GFNORTEO MF) and VTR Bank (VTRB RM) are perturbations of Sberbank of Russia (SBER RM), hence these three FIs are clustered together, forming a cluster of two different markets, contrary to the previous dendrogram, Figure \ref{ClassicalHRPEM@Dendrogram}, where the most similar FIs belong the same market. \vspace{\baselineskip}

\paragraph{\textbf{Calculate allocation through recursive bisection}}\

\begin{figure}[H]
  \centering
  \begin{subfigure}[b]{1\linewidth}
    \includegraphics[width=\linewidth]{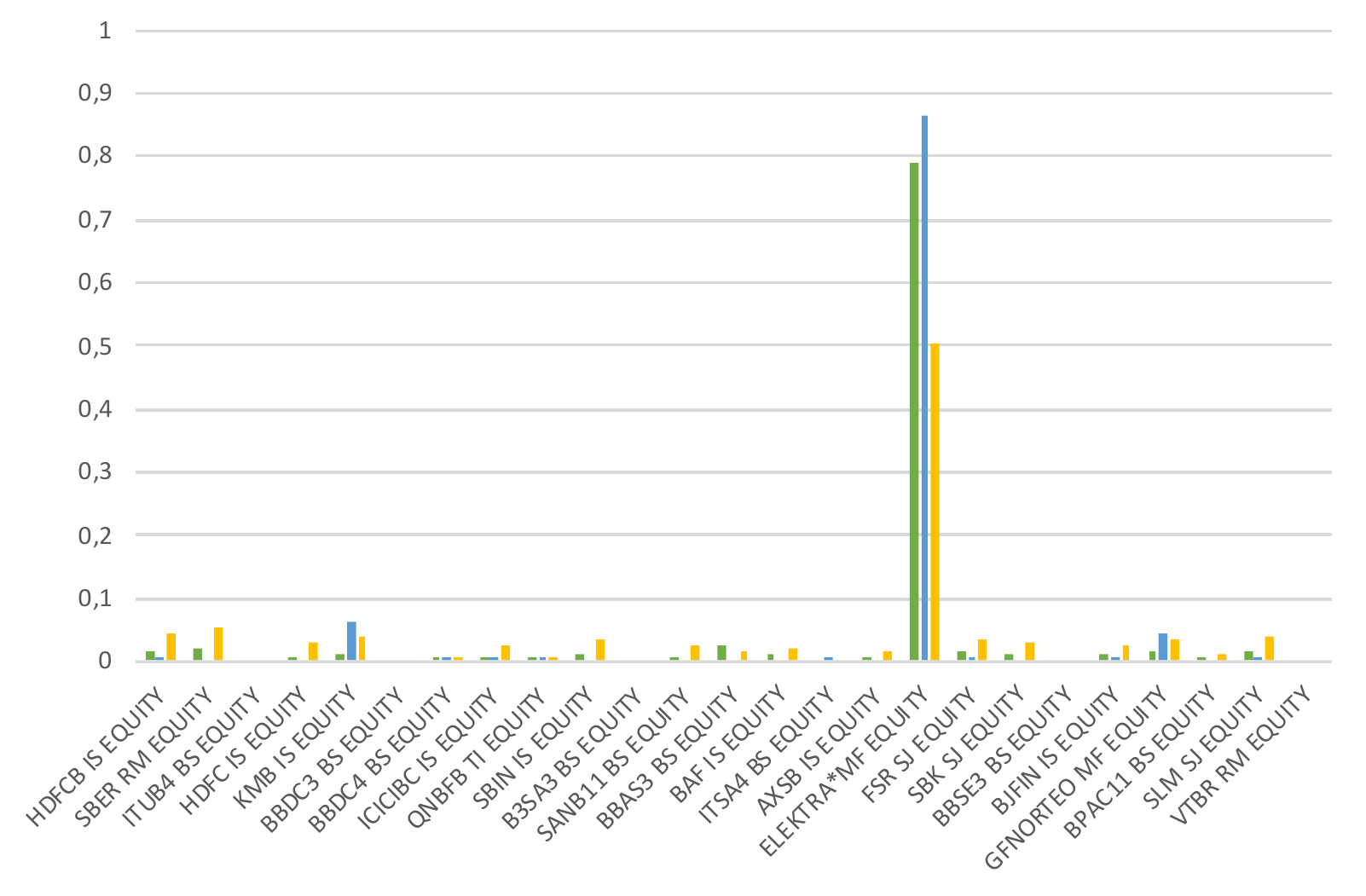}
    \caption{Optimal weights {\color{cyan} MinVar}, {\color{yellow} IVP}, {\color{green} HRP}}
    \label{classicalWeightsallocations}
  \end{subfigure}
  \begin{subfigure}[b]{1\linewidth}
    \includegraphics[width=\linewidth]{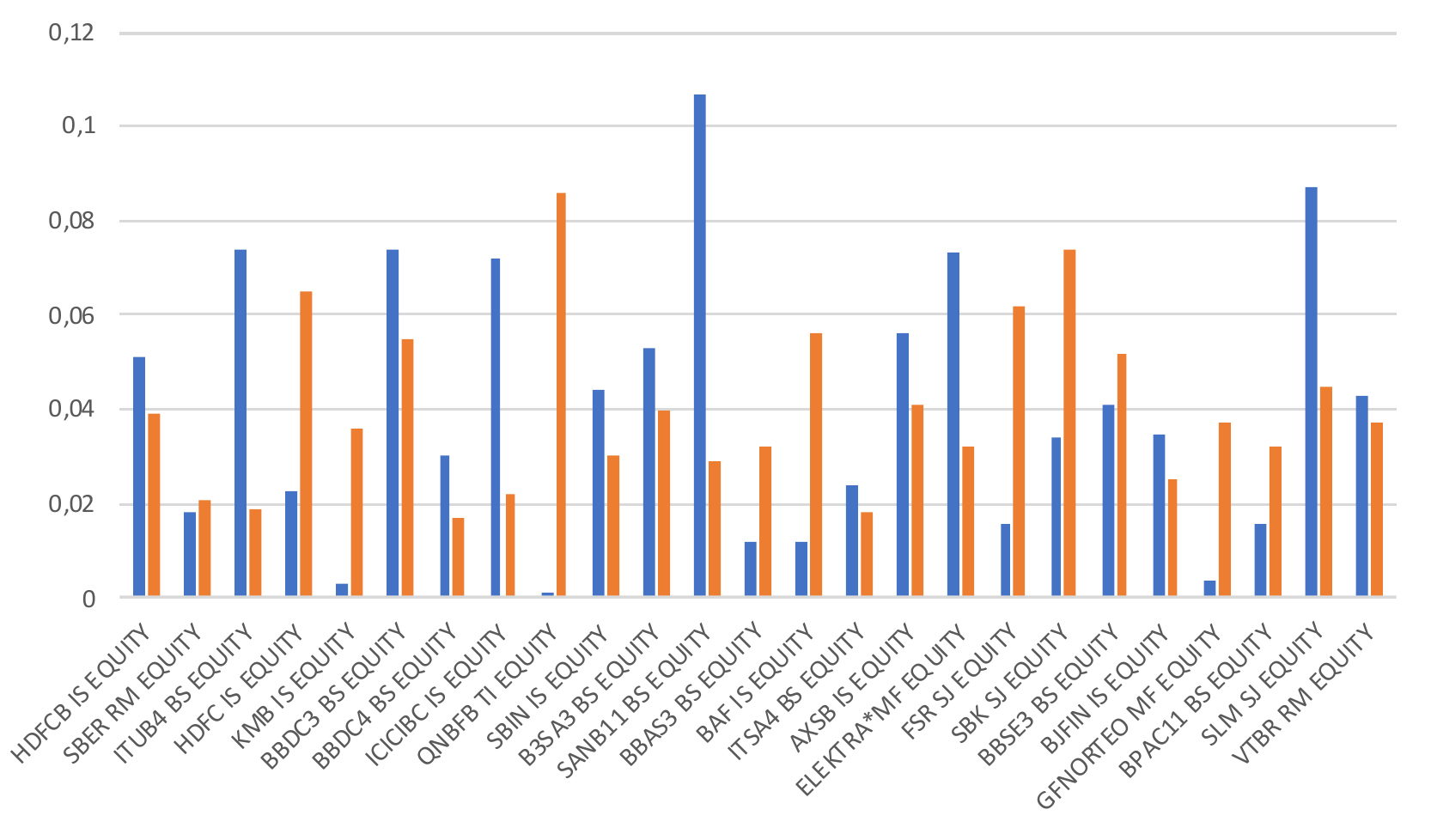}
    \caption{Optimal weights {\color{orange} Inv$\lambda$}, {\color{blue} upHRP}
    \bigskip
    }
    \label{UpliftedWeightsallocations}
  \end{subfigure}
  \caption{Weights allocations of FIs: classical approach vs uplifted approach}
  \hspace*{\fill} \href{https://github.com/QuantLet/FRM-EM-paper}{\includegraphics[scale=0.008]{Graphics/Quantlets_Logo_Ring.jpeg}}
  \label{figWeights allocationsFIs}
\end{figure}

Table \ref{WeightallocationsofFIsClassicalvsUplifted} and Figure \ref{figWeights allocationsFIs} specify further explanation regarding the weight’s allocation of the studied FIs according to the adopted strategies. 

\begin{table}[H]
\begin{center}
    \caption{Weight allocations of FIs: Classical approach vs {\color{blue} uplifted approach}}
    \label{WeightallocationsofFIsClassicalvsUplifted}
\begin{tabular}{lccccc}
\hline\\
\multicolumn{1}{c}{\textbf{EM@ FIs}}      & \textbf{MinVar}                 & \textbf{IVP}                    & \textbf{HRP}                   & \textbf{{\color{blue}Inv$\lambda$}}                             & \textbf{{\color{blue}upHRP}}                                  \\
\hline
{\color[HTML]{212121} HDFCB.IS.EQUITY}    & {\color[HTML]{212121} 0.002}    & {\color[HTML]{212121} 0.045}    & {\color[HTML]{212121} 0.017}   & {\color[HTML]{212121} 0.039}                   & {\color[HTML]{212121} 0.051}                   \\ \hline
{\color[HTML]{212121} SBER.RM.EQUITY}     & {\color[HTML]{212121} 0}        & {\color[HTML]{212121} 0.054}    & {\color[HTML]{212121} 0.020}    & {\color[HTML]{212121} 0.021}                   & {\color[HTML]{212121} 0.018}                   \\ \hline
{\color[HTML]{212121} ITUB4.BS.EQUITY}    & {\color[HTML]{212121} 0}        & {\color[HTML]{212121} 0}        & {\color[HTML]{212121} 0}       & {\color[HTML]{212121} 0.019}                   & {\color[HTML]{212121} 0.074}                   \\ \hline
{\color[HTML]{212121} HDFC.IS.EQUITY}     & {\color[HTML]{212121} 0}        & {\color[HTML]{212121} 0.032}    & {\color[HTML]{212121} 0.007}   & {\color[HTML]{212121} 0.065}                   & {\color[HTML]{212121} 0.023}                   \\ \hline
{\color[HTML]{212121} KMB.IS.EQUITY}      & {\color[HTML]{212121} 0.062}    & {\color[HTML]{212121} 0.038}    & {\color[HTML]{212121} 0.011}   & {\color[HTML]{212121} 0.036}                   & {\color[HTML]{212121} 0.003}                   \\ \hline
{\color[HTML]{212121} BBDC3.BS.EQUITY}    & {\color[HTML]{212121} 0}        & {\color[HTML]{212121} 0}        & {\color[HTML]{212121} 0}       & {\color[HTML]{212121} 0.055}                   & {\color[HTML]{212121} 0.074}                   \\ \hline
{\color[HTML]{212121} BBDC4.BS.EQUITY}    & {\color[HTML]{212121} 0.002}    & {\color[HTML]{212121} 0.005}    & {\color[HTML]{212121} 0.002}   & {\color[HTML]{212121} 0.017}                   & {\color[HTML]{212121} 0.030}                    \\ \hline
{\color[HTML]{212121} ICICIBC.IS.EQUITY}  & {\color[HTML]{212121} 0.002}    & {\color[HTML]{212121} 0.027}    & {\color[HTML]{212121} 0.005}   & {\color[HTML]{212121} 0.022}                   & {\color[HTML]{212121} 0.072}                   \\ \hline
{\color[HTML]{212121} QNBFB.TI.EQUITY}    & {\color[HTML]{212121} 0.008}    & {\color[HTML]{212121} 0.009}    & {\color[HTML]{212121} 0.006}   & {\color[HTML]{212121} 0.084}                   & {\color[HTML]{212121} 0.001}                   \\ \hline
{\color[HTML]{212121} SBIN.IS.EQUITY}     & {\color[HTML]{212121} 0}        & {\color[HTML]{212121} 0.035}    & {\color[HTML]{212121} 0.010}    & {\color[HTML]{212121} 0.030}                    & {\color[HTML]{212121} 0.044}                   \\ \hline
{\color[HTML]{212121} B3SA3.BS.EQUITY}    & {\color[HTML]{212121} 0}        & {\color[HTML]{212121} 0}        & {\color[HTML]{212121} 0}       & {\color[HTML]{212121} 0.040}                    & {\color[HTML]{212121} 0.053}                   \\ \hline
{\color[HTML]{212121} SANB11.BS.EQUITY}   & {\color[HTML]{212121} 0}        & {\color[HTML]{212121} 0.024}    & {\color[HTML]{212121} 0.007}   & {\color[HTML]{212121} 0.029}                   & {\color[HTML]{212121} 0.107}                   \\ \hline
{\color[HTML]{212121} BBAS3.BS.EQUITY}    & {\color[HTML]{212121} 0}        & {\color[HTML]{212121} 0.016}    & {\color[HTML]{212121} 0.025}   & {\color[HTML]{212121} 0.032}                   & {\color[HTML]{212121} 0.012}                   \\ \hline
{\color[HTML]{212121} BAF.IS.EQUITY}      & {\color[HTML]{212121} 0}        & {\color[HTML]{212121} 0.021}    & {\color[HTML]{212121} 0.014}   & {\color[HTML]{212121} 0.056}                   & {\color[HTML]{212121} 0.012}                   \\ \hline
{\color[HTML]{212121} ITSA4.BS.EQUITY}    & {\color[HTML]{212121} 0.001}    & {\color[HTML]{212121} 0}        & {\color[HTML]{212121} 0}       & {\color[HTML]{212121} 0.018}                   & {\color[HTML]{212121} 0.024}                   \\ \hline
{\color[HTML]{212121} AXSB.IS.EQUITY}     & {\color[HTML]{212121} 0}        & {\color[HTML]{212121} 0.018}    & {\color[HTML]{212121} 0.009}   & {\color[HTML]{212121} 0.041}                   & {\color[HTML]{212121} 0.056}                   \\ \hline
{\color[HTML]{212121} ELEKTRA.MF.Equity}  & {\color[HTML]{212121} 0.866}    & {\color[HTML]{212121} 0.503}    & {\color[HTML]{212121} 0.787}   & {\color[HTML]{212121} 0.032}                   & {\color[HTML]{212121} 0.073}                   \\ \hline
{\color[HTML]{212121} FSR.SJ.EQUITY}      & {\color[HTML]{212121} 0.001}    & {\color[HTML]{212121} 0.033}    & {\color[HTML]{212121} 0.018}   & {\color[HTML]{212121} 0.062}                   & {\color[HTML]{212121} 0.016}                   \\ \hline
{\color[HTML]{212121} SBK.SJ.EQUITY}      & {\color[HTML]{212121} 0}        & {\color[HTML]{212121} 0.029}    & {\color[HTML]{212121} 0.011}   & {\color[HTML]{212121} 0.074}                   & {\color[HTML]{212121} 0.034}                   \\ \hline
{\color[HTML]{212121} BBSE3.BS.EQUITY}    & {\color[HTML]{212121} 0}        & {\color[HTML]{212121} 0}        & {\color[HTML]{212121} 0}       & {\color[HTML]{212121} 0.052}                   & {\color[HTML]{212121} 0.041}                   \\ \hline
{\color[HTML]{212121} BJFIN.IS.EQUITY}    & {\color[HTML]{212121} 0.004}    & {\color[HTML]{212121} 0.026}    & {\color[HTML]{212121} 0.012}   & {\color[HTML]{212121} 0.025}                   & {\color[HTML]{212121} 0.035}                   \\ \hline
{\color[HTML]{212121} GFNORTEO.MF.Equity} & {\color[HTML]{212121} 0.044}    & {\color[HTML]{212121} 0.035}    & {\color[HTML]{212121} 0.017}   & {\color[HTML]{212121} 0.037}                   & {\color[HTML]{212121} 0.004}                   \\ \hline
{\color[HTML]{212121} BPAC11.BS.EQUITY}   & {\color[HTML]{212121} 0}        & {\color[HTML]{212121} 0.010}     & {\color[HTML]{212121} 0.003}   & {\color[HTML]{212121} 0.032}                   & {\color[HTML]{212121} 0.016}                   \\ \hline
{\color[HTML]{212121} SLM.SJ.EQUITY}      & {\color[HTML]{212121} 0.007}    & {\color[HTML]{212121} 0.039}    & {\color[HTML]{212121} 0.018}   & {\color[HTML]{212121} 0.045}                   & {\color[HTML]{212121} 0.084}                   \\ \hline
{\color[HTML]{212121} VTBR.RM.EQUITY}     & {\color[HTML]{212121} 0}        & {\color[HTML]{212121} 0}        & {\color[HTML]{212121} 0}       & {\color[HTML]{212121} 0.037}                   & {\color[HTML]{212121} 0.043}           \\ \hline          
\end{tabular}
  \end{center}
\end{table}

\
Table \ref{WeightallocationsofFIsClassicalvsUplifted} and Figure \ref{figWeights allocationsFIs} specify further explanation regarding the weight’s allocation of the studied FIs according to the adopted strategies. 
Several points are worth noting. First, for both classical and uplifted approaches, IVP, HRP, Inv$\lambda$ and upHRP and similarly allocate their assets. This result can be explained by the fact that they are based on the inverse variance method to compute their weights. However, while IVP and Inv$\lambda$ show more stable weight distributions, HRP and upHRP adjust their weight allocation more frequently, since it also takes into consideration the correlation (for HRP)/ tail connectedness (for upHRP) and the asset clusters when assigning weights. Second, results also show that the MinVar allocations have the most concentrated weight allocation. While it accorded zero allocation to certain FIs (i.e SBIN IS, B3SA3 BS, etc.), it significantly overweights others (ELEKTRA MF). 
Third, the MinVar, IVP and HRP classical approaches over-weighted the Mexican ELEKTRA MF (0.866, 0.503, 0.787, respectively). In this case, any distress situation affecting this FI will have a great impact on those concentrated portfolios. Contrary, these extreme weight concentrations disappear with the Inv$\lambda$ and upHRP uplifted approaches, where the weights are wider distributed among FIs, providing a well-diversified portfolio.  

\begin{table}[H]
\begin{center}
    \caption{ Weight's allocation of EMs: Classical approach vs {\color{blue}Uplifted approach}}
    \label{WeightsallocationEMs}
\begin{tabular}{lrrrrrr}
\hline
{\color[HTML]{212121} \textbf{EM@ FIs}}       & \multicolumn{1}{l}{{\color[HTML]{212121} \textbf{Nbr of FIs}}} & \multicolumn{1}{l}{{\color[HTML]{212121} \textbf{MinVar}}} & \multicolumn{1}{l}{{\color[HTML]{212121} \textbf{IVP}}} & \multicolumn{1}{l}{{\color[HTML]{330001} \textbf{HRP}}} & \multicolumn{1}{l}{{\color[HTML]{212121} \textbf{{\color{blue}Inv$\lambda$}}}} & \multicolumn{1}{l}{{\color[HTML]{212121} \textbf{{\color{blue}upHRP}}}} \\ \hline
{\color[HTML]{212121} \textbf{India (IS)}}    & {\color[HTML]{212121} 8}                                       & {\color[HTML]{212121} 7 \%}                                & {\color[HTML]{212121} 24.2\%}                           & {\color[HTML]{212121} 8.5\%}                            & {\color[HTML]{212121} 29.2\%}                                 & {\color[HTML]{212121} 22.4\%}                            \\ \hline
{\color[HTML]{212121} \textbf{Brasil (BS)}}   & {\color[HTML]{212121} 9}                                       & {\color[HTML]{212121} 0.3\%}                               & {\color[HTML]{212121} 5.5\%}                            & {\color[HTML]{212121} 3.7\%}                            & {\color[HTML]{212121} 29.4\%}                                 & {\color[HTML]{212121} 43.1\%}                            \\ \hline
{\color[HTML]{212121} \textbf{Mexico (MF)}}   & {\color[HTML]{212121} 2}                                       & {\color[HTML]{212121} 91 \%}                               & {\color[HTML]{212121} 53.8\%}                           & {\color[HTML]{212121} 80.4\%}                           & {\color[HTML]{212121} 6.9\%}                                  & {\color[HTML]{212121} 7.7\%}                             \\ \hline
{\color[HTML]{212121} \textbf{Russia (RM)}}   & {\color[HTML]{212121} 2}                                       & {\color[HTML]{212121} 0 \%}                                & {\color[HTML]{212121} 5.4\%}                            & {\color[HTML]{212121} 2 \%}                             & {\color[HTML]{212121} 5.8\%}                                  & {\color[HTML]{212121} 6.1\%}                             \\ \hline
{\color[HTML]{212121} \textbf{Turkey (TI)}}   & {\color[HTML]{212121} 1}                                       & {\color[HTML]{212121} 0.8\%}                               & {\color[HTML]{212121} 0.1\%}                           & {\color[HTML]{212121} 0.6\%}                            & {\color[HTML]{212121} 8.6\%}                                  & {\color[HTML]{212121} 0.1\%}                             \\ \hline
{\color[HTML]{212121} \textbf{S.Africa (SJ)}} & {\color[HTML]{212121} 3}                                       & {\color[HTML]{212121} 0.8\%}                               & {\color[HTML]{212121} 10.1\%}                           & {\color[HTML]{212121} 4.7\%}                            & {\color[HTML]{212121} 18.1\%}                                 & {\color[HTML]{212121} 13.7\%}                            \\ \hline
\end{tabular}
\end{center}
\end{table}

 Table \ref{WeightsallocationEMs} reports the percentage of weight allocation for the different approaches, giving further intuition on how the portfolio optimization approaches allocate their weights through EMs. From these allocation results, we can understand a few stylized features: First, the MinVar concentrates 91\% of the allocation on the two existing Mexican FIs and it assigns zero weights for the Russian FIs and only 0.3\% for the 9 Brazilian FIs. Similarly, the IVP and the HRP concentrate allocation on the same market with 53.8\% and 80.4\% weights respectively. However, they provide some diversification across the other markets compared to MinVar strategy. The classical approaches concentrate their weights on the Mexican market that contributed only by two FIs to the basket of the 25 biggest FIs in the EMs, which poses significant idiosyncratic risk.
Nevertheless, for the uplifted portfolio approaches, the weights are well distributed across the EMs. For example, the Inv$\lambda$ allocates 29\% of weights to the Brazilian and Indian FIs, while, the upHRP allocates 29.4\% to Indian FIs and 43.1\% to the Brazilian FIs since this last market has the highest number of FIs in the selection of the 25 biggest EM@FIs. Moreover, the upHRP allocates only 0.1\% for the Turkish FIs, since this market contributes only by one FI. 
To recapitulate, the upHRP appears to find a compromise between classical concentrated weight approaches and the Inv$\lambda$ strategy. The classical strategies can concentrate weights on a few FIs, leading to vulnerabilities. The Inv$\lambda$ evenly assigns weights across all FIs, ignoring the correlation structure. This makes it exposed to systemic shocks. However, the upHRP finds a compromise between diversifying across all FIs and diversifying across clusters, which makes it more resistant against both types of shocks.\vspace{\baselineskip}

\paragraph{\textbf{Backtesting results}}

\begin{table}[H]
\begin{center}
    \caption{Backtesting on EM FIs data: Classical approach vs {\color{blue} uplifted approach}}
    \label{BacktestingEMFIsdata}
\begin{tabular}{ccccc}
\hline
\multicolumn{1}{l}{}                      & {\color[HTML]{212121} \textbf{Mean}} & {\color[HTML]{212121} \textbf{Std}} & {\color[HTML]{212121} \textbf{Sharpe ratio}} & {\color[HTML]{212121} \textbf{Effective n}} \\ \hline
{\color[HTML]{212121} \textbf{MinVar}}    & {\color[HTML]{212121} 0.0043}        & {\color[HTML]{212121} 0.0818}       & {\color[HTML]{212121} 0.0528}                & {\color[HTML]{212121} 1.3225}               \\ \hline
{\color[HTML]{212121} \textbf{IVP}}       & {\color[HTML]{212121} 0.0039}        & {\color[HTML]{212121} 0.0712}       & {\color[HTML]{212121} 0.0560}                & {\color[HTML]{212121} 3.7121}               \\ \hline
{\color[HTML]{212121} \textbf{HRP}}       & {\color[HTML]{212121} 0.0035}        & {\color[HTML]{212121} 0.0490}       & {\color[HTML]{212121} 0.0721}                & {\color[HTML]{212121} 1.6063}               \\ \hline
\multicolumn{5}{l}{$\tau=5\%$}                                                                                                                                                                                          \\ \hline
{\color[HTML]{212121} \textbf{{\color{blue}Inv$\lambda$}}} & {\color[HTML]{212121} 0.0767}        & {\color[HTML]{212121} 0.1859}       & {\color[HTML]{212121} 0.4128}                & {\color[HTML]{212121} 20.986}               \\ \hline
{\color[HTML]{212121} \textbf{{\color{blue}upHRP}}}      & {\color[HTML]{212121} 0.0522}        & {\color[HTML]{212121} 0.1886}       & {\color[HTML]{212121} 0.2768}                & {\color[HTML]{212121} 16.801}               \\ \hline
\multicolumn{5}{l}{$\tau=10\%$}                                                                                                                                                                                          \\ \hline
{\color[HTML]{212121} \textbf{{\color{blue}Inv$\lambda$}}} & {\color[HTML]{212121} 0.0894}        & {\color[HTML]{212121} 0.1868}       & {\color[HTML]{212121} 0.4788}                & {\color[HTML]{212121} 21.643}               \\ \hline
{\color[HTML]{212121} \textbf{{\color{blue}upHRP}}}      & {\color[HTML]{212121} 0.0807}        & {\color[HTML]{212121} 0.1967}       & {\color[HTML]{212121} 0.4104}                & {\color[HTML]{212121} 16.296}               \\ \hline
\multicolumn{5}{l}{$\tau=25\%$}                                                                                                                                                                                          \\ \hline
{\color[HTML]{212121} \textbf{{\color{blue}Inv$\lambda$}}} & {\color[HTML]{212121} 0.0894}        & {\color[HTML]{212121} 0.2195}       & {\color[HTML]{212121} 0.4073}                & {\color[HTML]{212121} 23.429}               \\ \hline
{\color[HTML]{212121} \textbf{{\color{blue}upHRP}}}      & {\color[HTML]{212121} 0.0792}        & {\color[HTML]{212121} 0.2280}       & {\color[HTML]{212121} 0.3476}                & {\color[HTML]{212121} 20.199}               \\ \hline
\multicolumn{5}{l}{$\tau=50\%$}                                                                                                                                                                                          \\ \hline
{\color[HTML]{212121} \textbf{{\color{blue}Inv$\lambda$}}} & {\color[HTML]{212121} 0.0883}        & {\color[HTML]{212121} 0.2093}       & {\color[HTML]{212121} 0.4219}                & {\color[HTML]{212121} 21.117}               \\ \hline
{\color[HTML]{212121} \textbf{{\color{blue}upHRP}}}      & {\color[HTML]{212121} 0.0766}        & {\color[HTML]{212121} 0.2145}       & {\color[HTML]{212121} 0.3571}                & {\color[HTML]{212121} 21.508}               \\ \hline
\end{tabular}
\end{center}
\end{table}

\begin{figure}[H]
  \centering
  \includegraphics[width=0.8\textwidth]{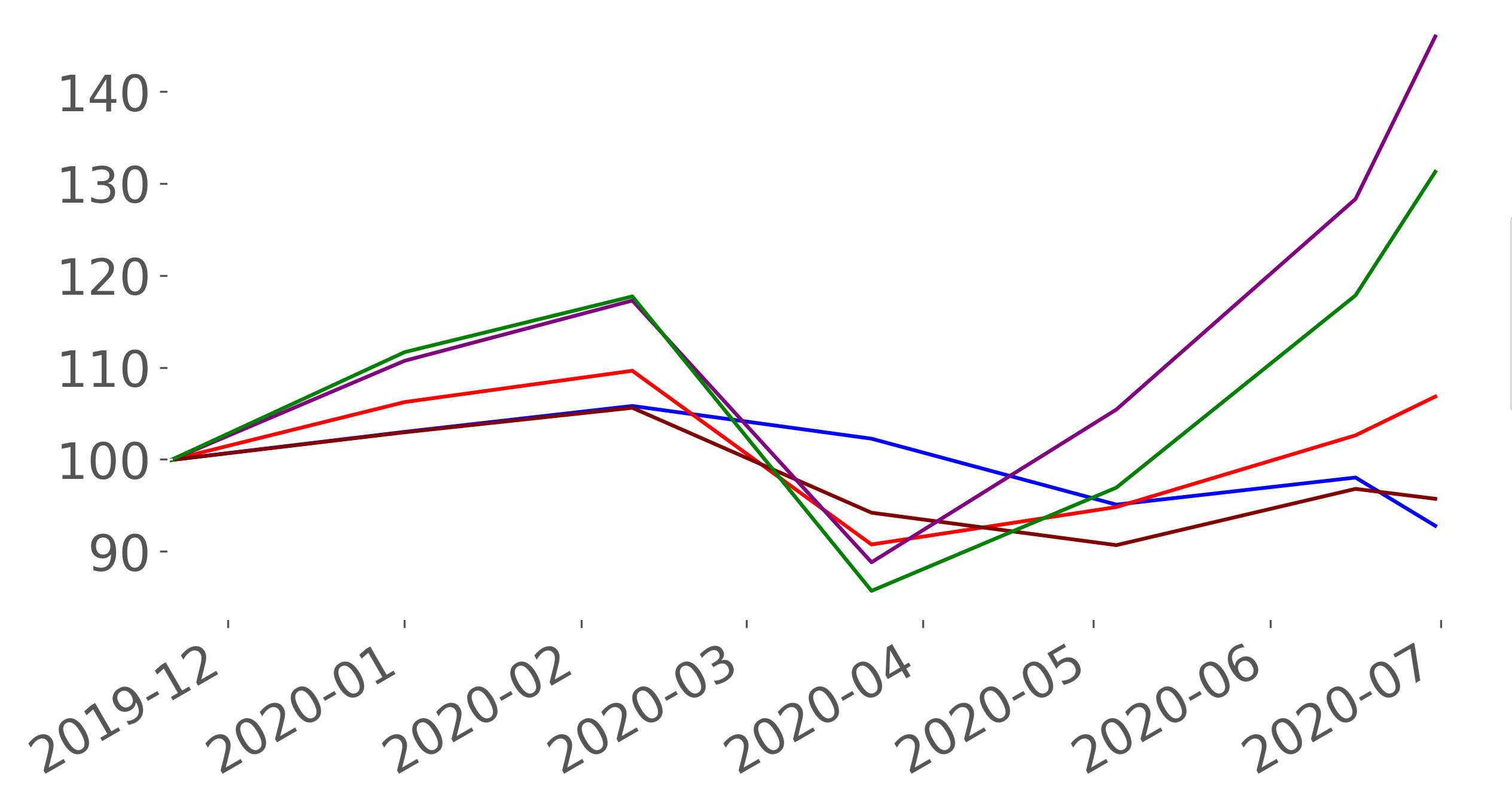}
  \caption{Backtesting on EM FIs data: classical approaches
   ({\color{blue}MinVar},{\color{red}IVP}, {\color{brown}HRP}) vs uplifted approaches ({\color{purple}Inv$\lambda$}, {\color{green}upHRP})}
   \hspace*{\fill} \href{https://github.com/QuantLet/FRM-EM-paper}{\includegraphics[scale=0.008]{Graphics/Quantlets_Logo_Ring.jpeg}}
   \label{figBacktestingEMFIsdataclassicalvsUplifted}
\end{figure}

\begin{figure}[H]
  \centering
  \begin{subfigure}[b]{0.45\linewidth}
    \includegraphics[width=\linewidth]{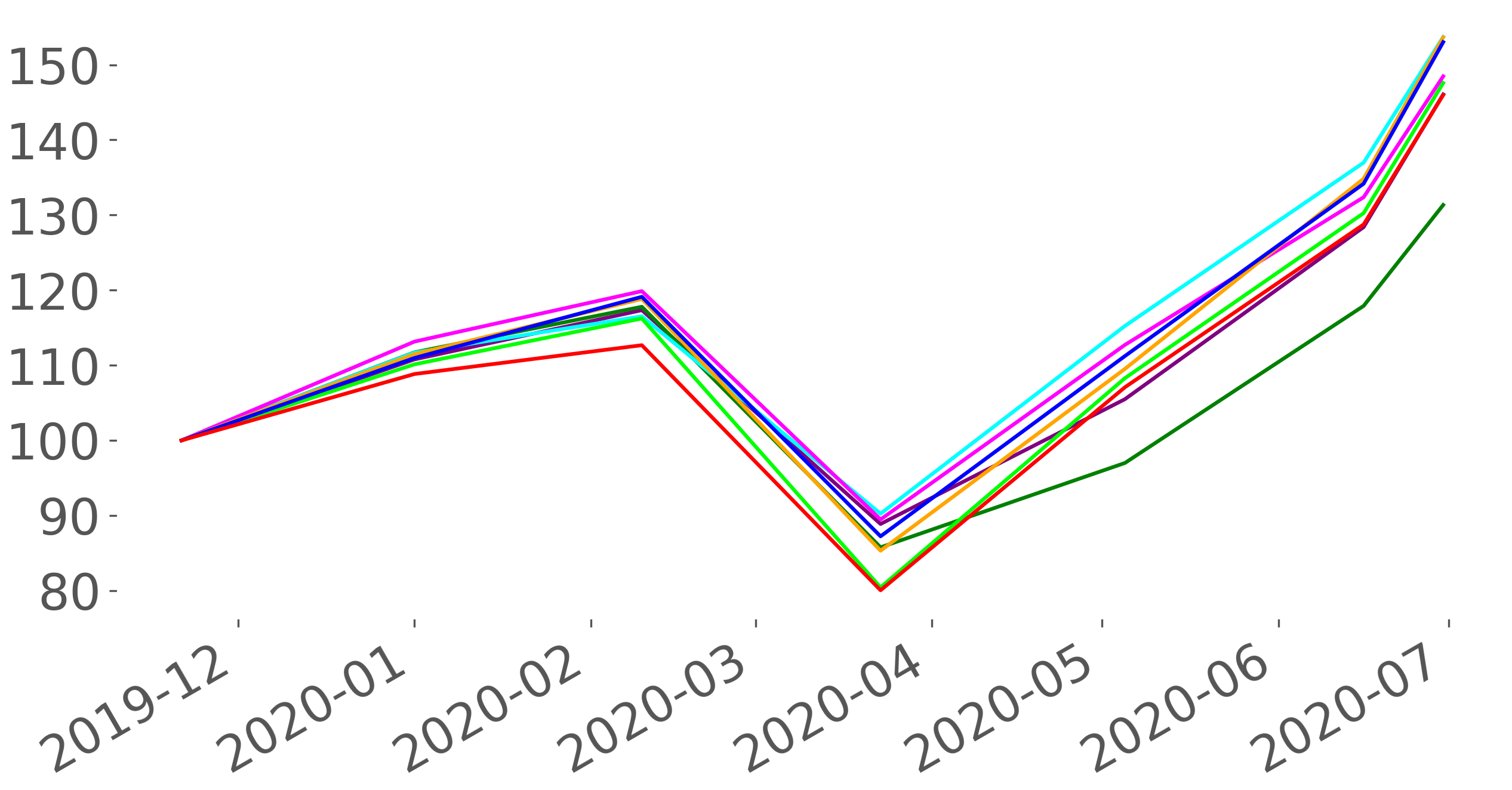}
    \caption{Backtesting on EM asset data: uplifted approach ({\color{purple}Inv$\lambda$ 5\%}, {\color{green}upHRP 5\%}, {\color{cyan}Inv$\lambda$ 10\%}, {\color{magenta}upHRP 10\%}, {\color{orange}Inv$\lambda$ 25\%}, {\color{lime}upHRP 25\%}, {\color{blue}Inv$\lambda$ 50\%}, {\color{red}upHRP 50\%})}
    \label{BacktestingUpliftedapproachwithdifferenttau}
  \end{subfigure}
  \begin{subfigure}[b]{0.5\linewidth}
    \includegraphics[width=\linewidth]{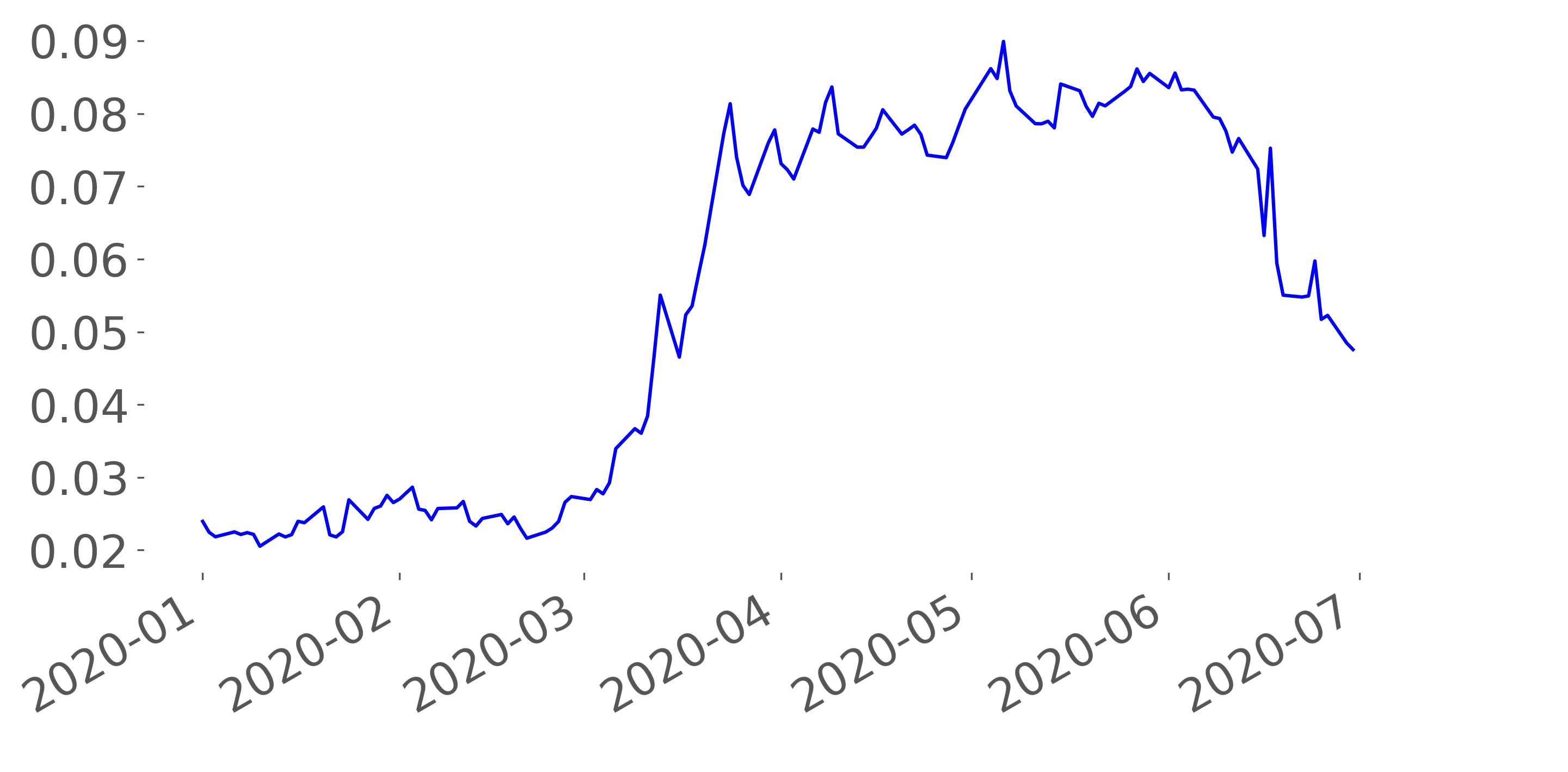}
    \caption{Financial Risk Meter (FRM)
   \bigskip
  \bigskip
   \smallskip
   }
   \label{FinancialRiskMeter2020}
  \end{subfigure}
  \begin{subfigure}[b]{0.8\linewidth}
    \includegraphics[width=\linewidth]{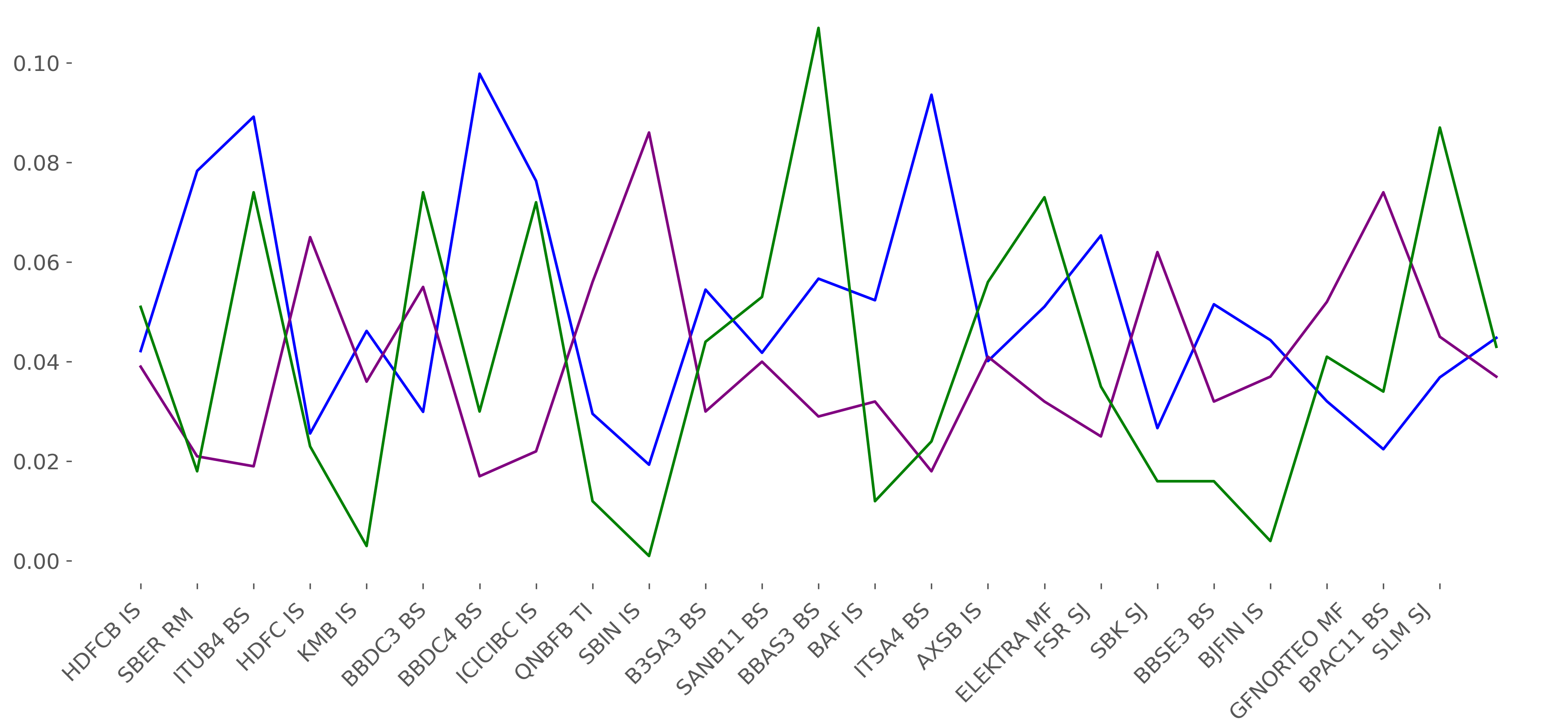}
    \caption{Penalisation parameter ($\lambda$) vs Optimal weights ({\color{blue}Penalisation parameter ($\lambda$}, {\color{purple}Inv$\lambda$ 5\% weights}, {\color{green}upHRP 5\% weights})}
    \label{Penalisationparametervsoptimalweights}
  \end{subfigure}
  \caption{Impact of Penalisation parameter ($\lambda$) and FRM on optimal weights and portfolio returns}
  \hspace*{\fill} \href{https://github.com/QuantLet/FRM-EM-paper}{\includegraphics[scale=0.008]{Graphics/Quantlets_Logo_Ring.jpeg}}
  \label{figImpactofPenalisationparameterandFRMonoptimalweightsand portfolio returns}
\end{figure}

For the backtesting, we adopt a 30 days rebalancing period. Table  \ref{BacktestingEMFIsdata} recapitulates the out-of-sample performance of the studied approaches.\\
For the classical approaches, the mean and the volatility of the HRP are 0.003 and 0.049, respectively. While the MinVar delivers the highest return value (0.004) and the highest volatility (0.081), the IVP provides approximately the same portfolio return (0.003) with higher volatility (0.0712). Therefore, the HRP balances out both return and volatility most efficiently providing the best approach in terms of Sharpe ratio (0.072 compared to 0.052 and 0.056 of the MinVar and IVP portfolios, respectively).\\ 
For the uplifted approach (with $\tau=5\%$), the return and volatility of the upHRP approach are 0.052 and 0.188, respectively. While the Inv$\lambda$ offers a higher return (0.076), and approximately the same volatility, the Inv$\lambda$ provides a risk-return balance registering the best Sharpe ratio value (0.412 compared to 0.276 of the upHRP portfolio). \\
Besides the sharp ratio, we measured the portfolio diversification effects using the Effective $N$ $Effect(N)$ measurement, proposed by \cite{strongin2000beating} as one of allocation concentration and defined as follow:
\begin{equation}
    Effect(N)=\frac{1}{\sum_{j=1}^N \widehat{\omega}_{j,t}^2}
    \label{eq49}
\end{equation}
The portfolio has high concentration risk if $Effect(N)$ is close to one.\vspace{\baselineskip}

Comparing the classical and the uplifted approaches(with $\tau=5\%$), it seems that the upHRP improves the Sharpe ratio of the HRP approach (0.072 and 0.276, respectively). Moreover, the HRP has discarded five EMs in favor of one single market, with ($Effect(N)=1.6)$, see Table \ref{WeightsallocationEMs}). Therefore, the HRP’s portfolio is deceitfully diversified, since any distress situation affecting this market will have a greater negative impact on HRP’s portfolio than the upHRP, which allocates only 6.9\% to the Mexican market and providing more diversification across EMs ($Effect(N)=16.8$, see Table \ref{BacktestingEMFIsdata}). \vspace{\baselineskip}

In fact, the main innovation of upHRP strategy is to apply the HRP algorithm based on tail dependence clustering instead of the standard correlation-based clustering. First, this strategy is motivated by the fact that the correlation coefficients can change drastically during financial crises due to contagion effects, and such a crisis can spill over quickly. Consequently, diversifying a portfolio based on correlation clusters may be a failing strategy without attention to tail events. Second, while the correlation matrix illustrates the relationship based on the mean-variance of the distribution, the FRM measures the co-movements in extreme events based on both tail of the distribution. To exemplify, lower tail dependence is associated with the capacity to diversify during crises, which can improve the tail risk management of a given strategy. More precisely, the main idea is to cluster the studied FIs illustrating a high probability to experience extremely negative events contemporaneously. Finally, empirical results show that the proposed uplifted approach has the potential to compete with the classical portfolio optimization approach by providing desirable diversification properties, especially if this hierarchical risk parity strategy is based on the tail dependence coefficient (provided by the FRM adjacency matrix), which is a benefit to tail risk management.\vspace{\baselineskip}

Besides, for the uplifted approach, we take into consideration four tail risk levels (5\%, 10\%, 25\%, 50\%). The results in Table \ref{BacktestingEMFIsdata} indicate that:
The $Effect N$ is an increasing function of the tail risk level, therefore for a higher tail risk level, more diversification is needed to guarantee stable portfolio returns.
Comparing the Inv$\lambda$ strategy with the upHRP, we noticed that the Inv$\lambda$ strategy needs to include a higher number of assets compared to the upHRP (i.e for $\tau=10\%$, $Effect(N)= 21.64$ and 16.24, respectively) to provide the same portfolio performance level ( approximately the same sharp ratio=0.4). On the one hand, this result can be explained by the fact that for the upHRP, only FIs within the same cluster compete for portfolio allocation rather than competing with all the FIs in the portfolio, which avoids the redundancy in the FIs included in the optimal portfolio. On the other hand, the larger number of FIs increases estimated parameters, which also can increase the risk of estimation error and thus biased results. For that purpose, for relatively similar values of sharp ratio (portfolio efficiency), the upHRP is preferred over the Inv$\lambda$.\vspace{\baselineskip}

Figure \ref{figBacktestingEMFIsdataclassicalvsUplifted} illustrates the backtesting on EM’s FIs with 30 days rebalancing period. It plots the portfolio turnover for the first half of 2020 according to the classical strategies (MinVar, IVP, and HRP) and the uplifted strategies (Inv$\lambda$ and upHRP), where the classical approaches show a relative stability in returns compared to the uplifted approaches since, the classical approaches concentrate their weights on one Mexican FI (see Table \ref{WeightsallocationEMs}), which limited their profit (increase in portfolio returns in mid-February and in June) and also limit their loss (end of March ) during the COVID 19 crisis. However, this is a trade-off between historical stability and very high concentration risk. Any idiosyncratic risk impact would lead to substantial portfolio losses. 

Figure \ref{BacktestingUpliftedapproachwithdifferenttau} plots the backtesting on EM’s FIs with 30 day rebalancing period for the uplifted strategies, taking into consideration the adjacency matrices with different tail risk levels (5\%, 10\%, 25\%, and 50\%). The Figure indicates that all portfolio returns follow the same trend, despite the difference in portfolio composition, since the biggest 25 FIs that compose the adjacency matrices vary over time and the risk levels.\\
Indeed, by comparing this portfolio return’s trend with the FRM plot (Figure \ref{FinancialRiskMeter2020}) during the same period, we noticed a negative relationship, where the high-risk period plotted by the FRM is translated by a decrease in the portfolio returns (March). Moreover, Figure \ref{Penalisationparametervsoptimalweights} indicates that the uplifted approach underweights the high risky FIs (with high penalization parameter $\lambda$) and overweight the less risky FIs, specially for the Inv$\lambda$, since the upHRP takes into consideration the tail co-movement between the FIs. 

\subsection{Bridging optimal portfolio weights and network centrality}
\begin{figure}[H]
  \centering
  \begin{subfigure}[b]{0.45\linewidth}
    \includegraphics[width=\linewidth]{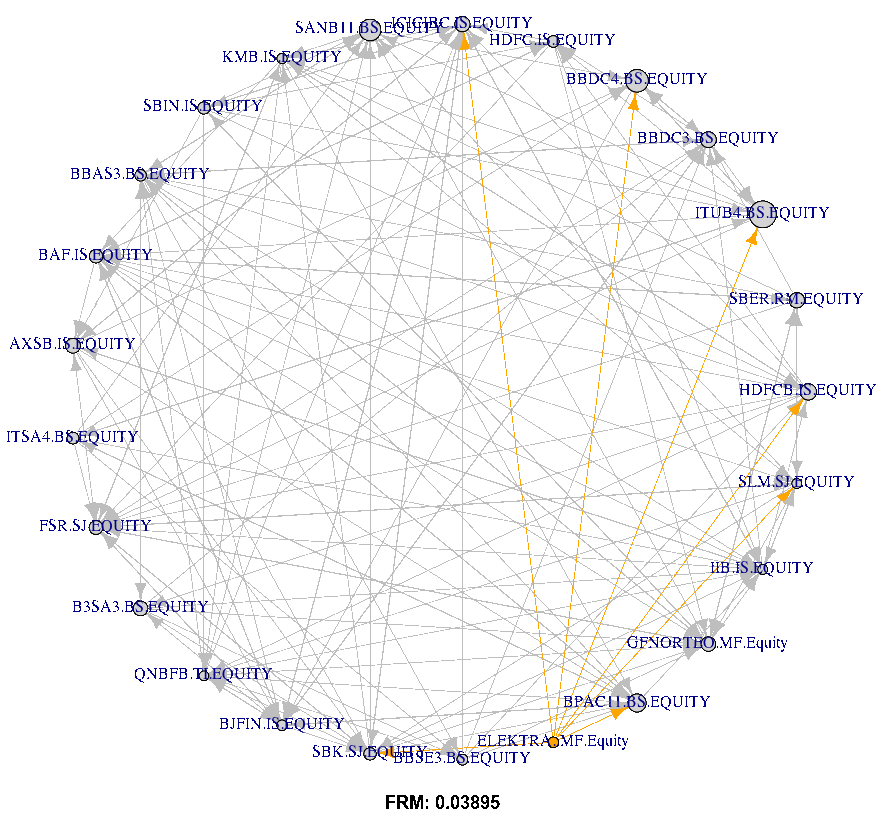}
    \caption{$\tau=0.05$}
    \label{tau005}
  \end{subfigure}
  \begin{subfigure}[b]{0.45\linewidth}
    \includegraphics[width=\linewidth]{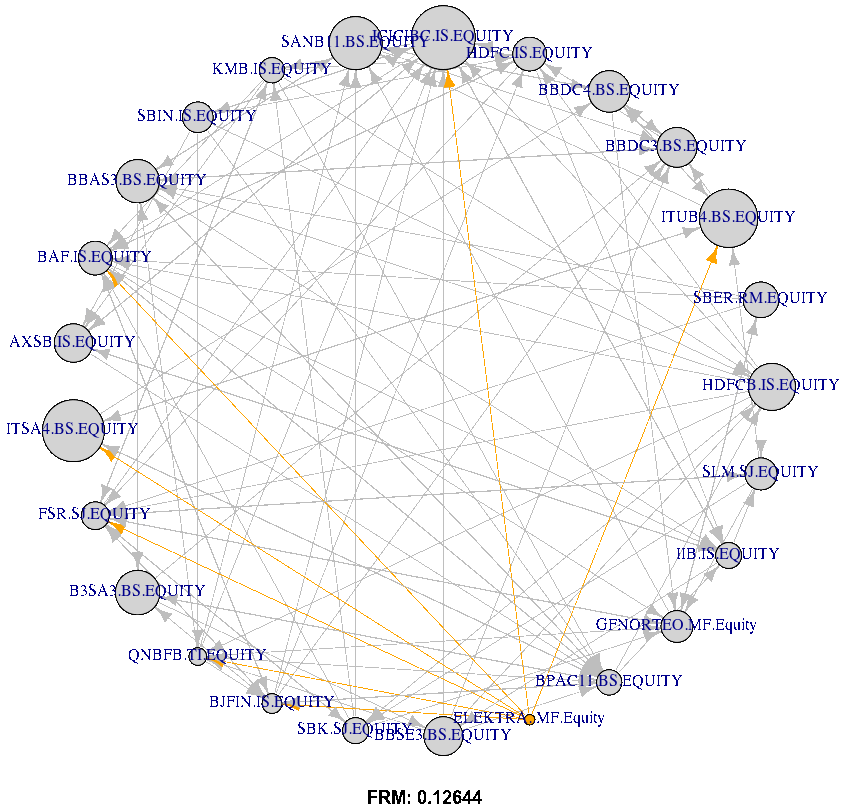}
    \caption{$\tau=0.1$}
    \label{tau01}
    \end{subfigure}
    \begin{subfigure}[b]{0.45\linewidth}
    \includegraphics[width=\linewidth]{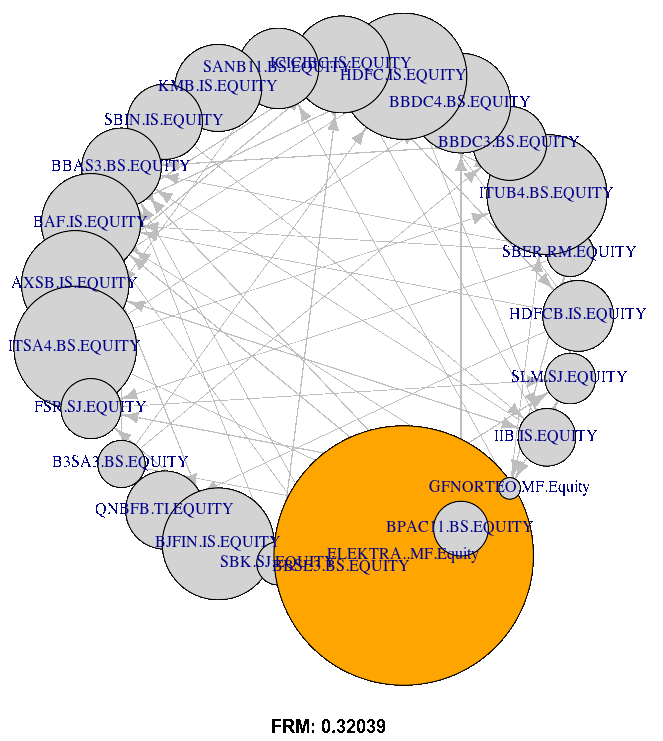}
    \caption{$\tau=0.25$}
    \label{tau025}
  \end{subfigure}
  \begin{subfigure}[b]{0.45\linewidth}
    \includegraphics[width=\linewidth]{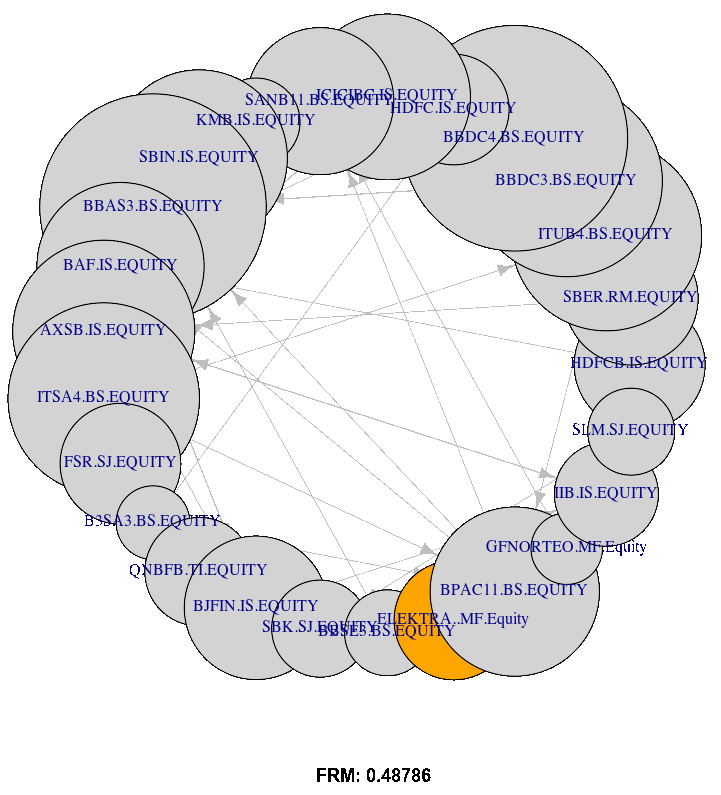}
    \caption{$\tau=0.5$}
    \label{tau05}
  \end{subfigure}
  \caption{Network graphs with the size of nodes representing total degree centrality on 20200630}
  \hspace*{\fill} \raisebox{-1pt}{\includegraphics[scale=0.008]{Graphics/Quantlets_Logo_Ring.jpeg}}
  \label{fig:NETWORKtau}
\end{figure}

\begin{figure}[H]
  \centering
  \begin{subfigure}[b]{0.8\linewidth}
    \includegraphics[width=\linewidth]{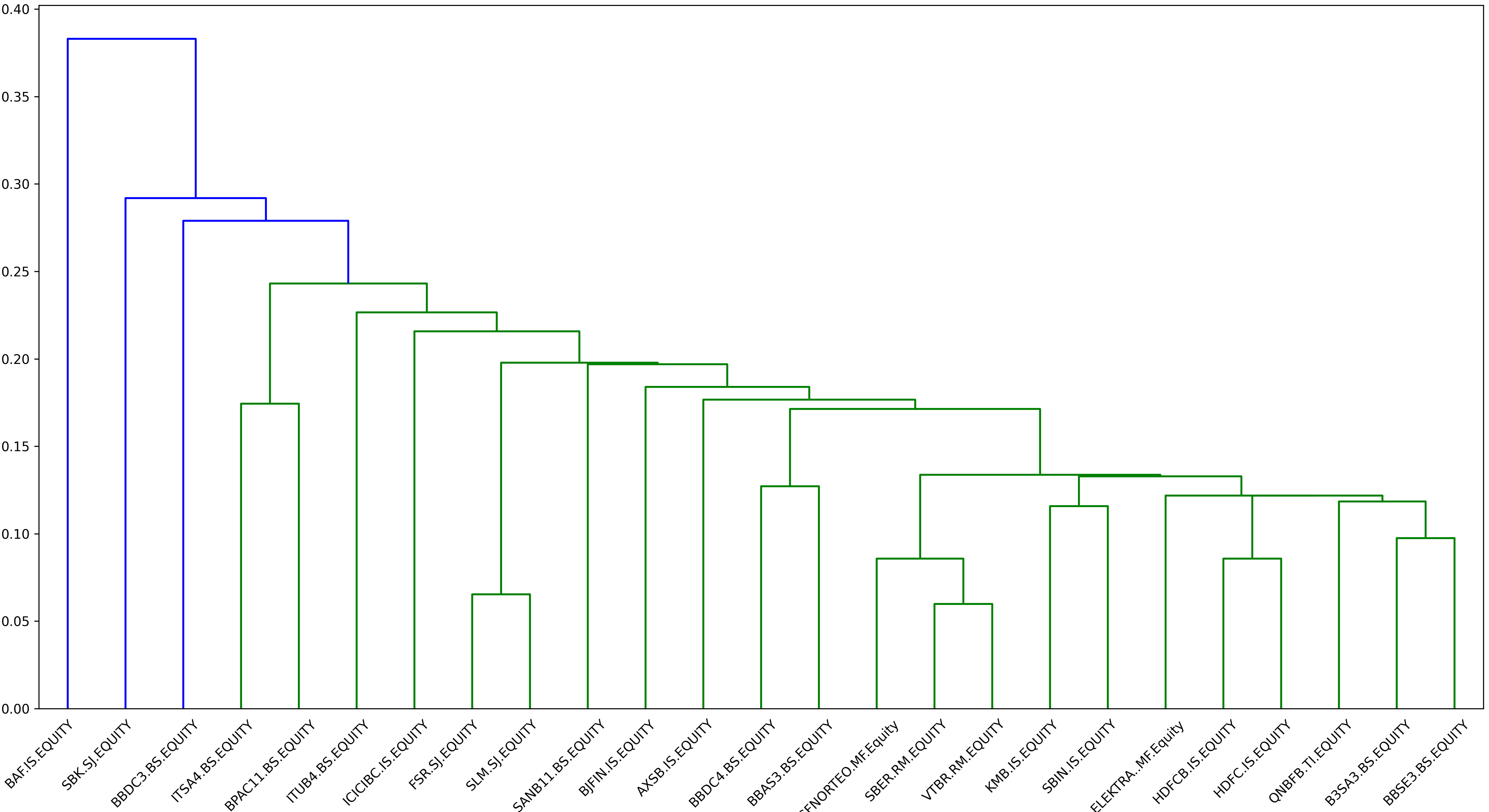}
    \caption{$\tau=0.1$}
    \end{subfigure}
    \begin{subfigure}[b]{0.8\linewidth}
    \includegraphics[width=\linewidth]{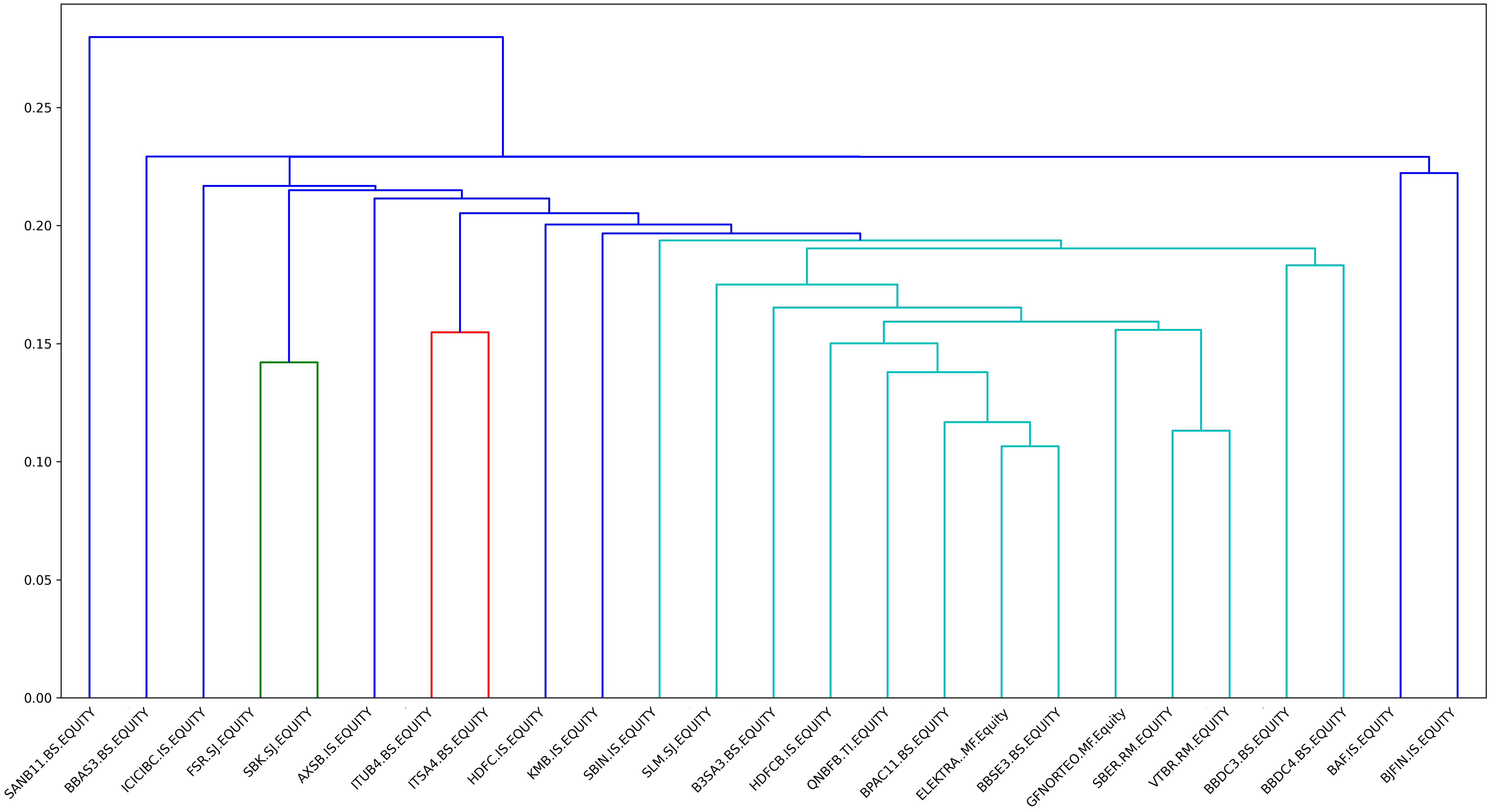}
    \caption{$\tau=0.25$}
  \end{subfigure}
  \begin{subfigure}[b]{0.8\linewidth}
    \includegraphics[width=\linewidth]{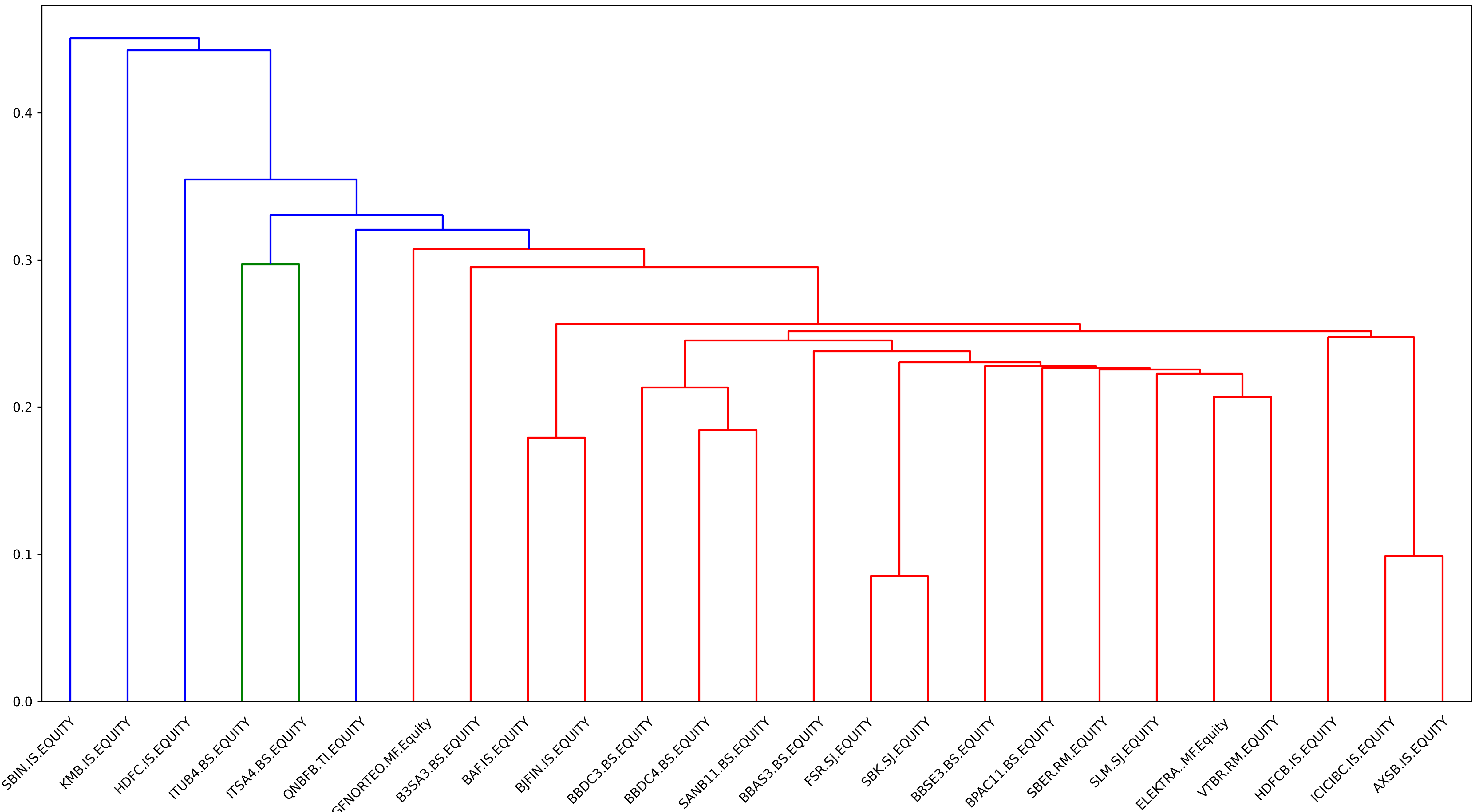}
    \caption{$\tau=0.5$}
  \end{subfigure}
  \caption{Sensitivity of the upHRP' dendrogram of  to different tail risk levels $\tau$}
  \hspace*{\fill} \href{https://github.com/QuantLet/FRM-EM-paper}
  {\includegraphics[scale=0.008]{Graphics/Quantlets_Logo_Ring.jpeg}}
  \label{figdendrogramofdifftau}
  
\end{figure}

Figure \ref{fig:NETWORKtau} plots the FRM Network with the size of nodes representing total degree centrality  (indegree and outdegree). The network information is provided by the adjacency matrices estimated at $\tau$ equal to 0.05, 0.10, 0.25, and 0.50, with dates 20200630. The Figure shows that the total degree centrality (both in and out) increase and become more visible with the increasing of tail risk level $\tau$.
By interpreting the adjacency matrix of the network, an overlapping region between the portfolio theory and network theory can be established.\\
Indeed, according to Figure \ref{tau005}, Itau Unibanco (ITUB4 BS) is a highly central FI followed by Banco Santander Brasil (SANB11 BS) and Banco Bradesco SA (BBDC4 BS). Therefore, the Brazilian market illustrates high-CoStress compared to the other markets (see also the boxplot, Figure \ref{FRMEMboxplot}. Contrary to that, QNB Finansbank (QNBF TI) and Housing Development Finance Corp (HDFC IS) are low-central FIs.\\
Regarding ELEKTRA MF, the networks (Figure \ref{fig:NETWORKtau}) indicate that this FI is only a risk emitter (high out-degrees with zero in-degree), which makes this FI insensitive to the other FIs shocks. Moreover,  the classical portfolio approaches overweight ELEKTRA MF (see Table \ref{WeightallocationsofFIsClassicalvsUplifted}). Therefore, at least historically, ELEKTRA MF had more stable return patterns, but as a risk emitter, any change in risk perception concerning this FI would rapidly spill over into the entire market.\\
Moreover, referring to Table \ref{Optimalweightsfordifftau}, the results of the optimal weight indicate that the Inv$\lambda$ portfolio is insensitive to the centrality degree. It often allocates similar weights for the high and low central FIs (i.e. BBDC4 BS and GFNORTEO MF, see Figure (\ref{tau05}). This result is due to the fact that this portfolio approach allocates weights based only on the individual tail risk level $\Lambda$, varying by tail risk levels $\tau$ but is insensitive to the network centrality degree.\\
Referring to Table \ref{Optimalweightsfordifftau} and Figure  \ref{fig:NETWORKtau}, results from different $\tau$ levels indicate that the upHRP approach underweights the high central FIs (i.e. BBDC4 BS (for $\tau$=5\% and $\tau$=50\%), ICICIBC IS (for $\tau$=10\%), ITUB BS and HDFC IS (for $\tau$=25\%)). Contrary to that, the upHRP strategy overweights the low central FIs (i.e. ELEKTRA MF and SLM SJ (for $\tau$=5\%), ELEKTRA MF (for $\tau$=10\%), GFNORTEO MF and B3SA3 BS (for $\tau$=25\% and $\tau$=50\%). These findings are in line with previous research, studying the relationship between the optimal portfolio weights and network centrality in the mean-variance level. In fact, our findings are consistent with \cite{peralta2016network} and \cite{pozzi2013spread} in establishing that "optimal portfolio strategies should overweigh low-central assets and underweight high-central ones". Nonetheless, they contradict the findings of \cite{vyrost2019network}, who argue that “asset weights must be ordered in the same way as the reciprocal of asset centrality in a given network”. \\
Indeed, the sensibility of the Adjacency matrix to different tail risk levels $\tau$ results in centrality degree variation as well as FIs clustering (see Figures \ref{upHRPEM@Dendrogram} and \ref{figdendrogramofdifftau}), which leads to upHRP optimal weights variation through different  $\tau$ for all FIs.\\  
In fact, the low concentration of weights for the uplifted strategies (Inv$\Lambda$ and upHRP) as well as the sensitivity of weights to different tail risk levels, don’t allow us to conclude about the best FI that offers greater diversification benefits in EMs (contrary to the classical approach that concentrates weights on ELEKTRA MF). However, uplifted strategies overweight the Brasilian FIs (for different $\tau$). Moreover, by contributing with the highest number of FIs to the basket of the 25 biggest FIs in EMs (see Table \ref{WeightsallocationEMs}), the Brazilian market seems more interesting to international investors, besides the need of diversification advantages provided by the other EMs.

\section{Policy recommendation}
Besides the outlined portfolio construction in Section \ref{portfolioconstruction}, we want to derive some recommendations for both policy makers as well as investors. \\
In Section \ref{FRMEMinterp} we have shown that particular EM FIs are not only influenced by FIs and macroeconomic risk variables from the same region, but also have tail-event co-movement dependencies to other EM geographic regions and currencies, the latter all the more important at lower levels of $\tau=0.10$ or $\tau=0.05$. Secondly, a dominant driver of EM FIs returns is the fiscal and economical stance of EM Sovereign bond issuers. If the risk perception for sovereign issuers increases, the return of EM FIs is impacted across the board and across the multivariate distribution. Lastly, co-movements can be detected within similar economic sub-sectors across geographical regions, which is also apparent in the portfolio construction exercises in Section \ref{portfolioconstruction}. We recommend investors to analyse concentrations across economic sub-sectors across regions, as well as to analyse dependencies between equity investments and bond investments in EM, given the clear linkages between the sovereign and the domestic EM FIs. Lastly, currency fluctuations have marginal return contributions in tail-events especially. In so far as EM FI investments are unhedged into developed market currencies, there is risk of a compound effect on returns. \\
As for the portfolio construction, the "classical" approaches show a relative stability in portfolio returns compared to the uplifted approaches, which limited their profit and also limit their loss during the COVID 19 crisis. But, this often comes at the cost of high concentration risks, exposing the portfolio to idiosyncratic tail events. The uplifted portfolio approaches show more volatile turnover with higher loss and profit, but are much better diversified, preventing sizeable risk clusters. Even though a risk averse investor might opt for the classical approach at first, closer inspection of tail risk behaviour and concentration risk should let the investor prefer the uplifted portfolio.\\
EM policy makers can derive important recommendations from our analysis. Firstly, coordination between EM regulatory bodies is of importance in order to mute EM FI fluctuations. Here, particular attention should be put on linkages in same sub-sector operating FIs across regions, which will be increasingly more important as globalisation continues. Secondly, fluctuations in risk perception of the sovereign issuer has an immediate impact on EM FI returns. Regulatory bodies are therefore advised to preemptively verify sound capitalisation of their domestic banks, even if the sovereign issuers distress is stemming from another geographic region. This is further amplified by a tail-event leading to a weakening of an EM's currency versus for example the U.S. Dollar. Global financial linkages, we show, between EM FIs lead to spill-over effects. Overall, in order to protect the domestic as well as global EM economies and their FIs, EM regulatory bodies should continuously work towards closer coordination between Emerging Market economies. This will help increase robustness versus developed market economies distress, of which more is likely to come, as developed markets fight with difficult fiscal situations and low growth patterns. But also, closer coordination between countries will prevent spill-over effects from one geographical region's FIs onto others, thereby increasing the attractiveness of the asset class further. Our approach via dendograms rapidly indicates any such risk clusters, and can be updated at ease and frequently during financial crises.

\section{Conclusion}

In this study, we examine the co-movements of EM FIs across six geographical regions, with aim to analyse within EM country co-dependencies in tail events but importantly also between regions as well as FIs of the same sub-sector across regions. We also analyse the important macro-economical risk variables impacting EM FIs and conclude that in addition to the developed market variables suggested by \cite{tobias2016covar}, EM specific macroeconomic risk variables have significant explanatory power. This can be used to construct more robust total asset class portfolio allocation and supplies EM regulatory bodies with detailed information on co-dependencies for better and faster stabilisation measures during periods of distress.\\
We also propose a novel asset allocation method – Hierarchical Risk Parity based the tail event information from the FRM technology, allowing us to extend this approach to the quantile level and replace the covariance matrix with the rich information contained in the FRM adjacency matrix. We applied this proposed approach to a portfolio of the biggest 25 FIs in EM, and our results show that uplifted strategies provide appropriate diversification properties. In comparison, the Inv$\lambda$ portfolios tend to be too static and the classical approaches result in too concentrated portfolios. Bridging optimal portfolio weights and network centrality, we conclude that the Inv$\lambda$ insensitive to the network centrality degree. However the upHRP portfolio underweight high-central FIs and overweight low-central ones, therefore the Inv$\lambda$ is less at risk of spill-over effects across EM regions, FIs, and financial sub-sectors. 

\bibliography{bibliography.bib}

\section{Appendix}

\begin{table}[H]
\begin{center}
    \caption{Sensitivity of the optimal weights of the uplifted strategies to different tail risk levels $\tau$}
    \label{Optimalweightsfordifftau}
\resizebox{\textwidth}{!} {%
\begin{tabular}{ccccccccc}
\hline
\multirow{2}{*}{\textbf{FIs}} & \multicolumn{2}{c}{5\%} & \multicolumn{2}{c}{10\%} & \multicolumn{2}{c}{25\%} & \multicolumn{2}{c}{50\%} \\ \cline{2-9} 
                              & Inv$(\lambda)$    & upHRP     & Inv$(\lambda)$     & upHRP     & Inv$(\lambda)$     & upHRP     & Inv$(\lambda)$     & upHRP     \\ \hline
HDFCB.IS.EQUITY               & 0.039        & 0.051    & 0.030         & 0.017    & 0.021         & 0.320    & 0.038         & 0.027    \\ \hline
SBER.RM.EQUITY                & 0.021        & 0.018    & 0.021         & 0.086    & 0.053         & 0.052    & 0.037         & 0.031    \\ \hline
ITUB4.BS.EQUITY               & 0.019        & 0.074    & 0.023         & 0.069    & 0.020          & 0.026    & 0.026         & 0.066    \\ \hline
HDFC.IS.EQUITY                & 0.065        & 0.023    & 0.038         & 0.034    & 0.022         & 0.047    & 0.026         & 0.027    \\ \hline
KMB.IS.EQUITY                 & 0.036        & 0.003    & 0.026         & 0        & 0.027         & 0.052    & 0.022         & 0.031    \\ \hline
BBDC3.BS.EQUITY               & 0.055        & 0.074    & 0.041         & 0.050     & 0.032         & 0.022    & 0.045         & 0.069    \\ \hline
BBDC4.BS.EQUITY               & 0.017        & 0.030     & 0.027         & 0.043    & 0.020         & 0.039    & 0.030          & 0.027    \\ \hline
ICICIBC.IS.EQUITY             & 0.022        & 0.072    & 0.022         & 0.026    & 0.024         & 0.023    & 0.034         & 0.052    \\ \hline
BAF.IS.EQUITY                 & 0.056        & 0.012    & 0.032         & 0.049    & 0.044         & 0.014    & 0.057         & 0.033    \\ \hline
QNBFB.TI.EQUITY               & 0.086        & 0.001    & 0.023         & 0.001    & 0.029         & 0.026    & 0.028         & 0.046    \\ \hline
SBIN.IS.EQUITY                & 0.030        & 0.044    & 0.036         & 0.049    & 0.025         & 0.024    & 0.022         & 0.066    \\ \hline
B3SA3.BS.EQUITY               & 0.040        & 0.053    & 0.060         & 0.030    & 0.030         & 0.114    & 0.030         & 0.054    \\ \hline
SANB11.BS.EQUITY              & 0.029        & 0.107    & 0.033         & 0.059    & 0.025         & 0.050    & 0.028         & 0.053    \\ \hline
BBAS3.BS.EQUITY               & 0.032        & 0.012    & 0.026         & 0.039    & 0.019         & 0.018    & 0.026         & 0.039    \\ \hline
AXSB.IS.EQUITY                & 0.018        & 0.024    & 0.029         & 0.083    & 0.044         & 0.024    & 0.042         & 0.019    \\ \hline
ITSA4.BS.EQUITY               & 0.041        & 0.056    & 0.128         & 0.033    & 0.091         & 0.042    & 0.068         & 0.045    \\ \hline
ELEKTRA..MF.Equity            & 0.032        & 0.073    & 0.031         & 0.043    & 0.058         & 0.016    & 0.046         & 0.044    \\ \hline
BJFIN.IS.EQUITY               & 0.025        & 0.035    & 0.043         & 0.052    & 0.039         & 0.056    & 0.035         & 0.024    \\ \hline
FSR.SJ.EQUITY                 & 0.062        & 0.016    & 0.062         & 0.051    & 0.021         & 0.057    & 0.052         & 0.023    \\ \hline
BPAC11.BS.EQUITY              & 0.032        & 0.016    & 0.051         & 0.030     & 0.060         & 0.018    & 0.058         & 0.033    \\ \hline
BBSE3.BS.EQUITY               & 0.037        & 0.004    & 0.032         & 0.007    & 0.041         & 0.010    & 0.042         & 0.029    \\ \hline
GFNORTEO.MF.Equity            & 0.052        & 0.041    & 0.044         & 0.024    & 0.090         & 0.076    & 0.030         & 0.070    \\ \hline
SBK.SJ.EQUITY                 & 0.074        & 0.034    & 0.052         & 0.050    & 0.046         & 0.073    & 0.070         & 0.039    \\ \hline
SLM.SJ.EQUITY                 & 0.045        & 0.087    & 0.049         & 0.029    & 0.050         & 0.078    & 0.048         & 0.018    \\ \hline
VTBR.RM.EQUITY                & 0.037        & 0.043    & 0.042         & 0.044    & 0.058         & 0.021    & 0.058         & 0.038    \\ \hline
\end{tabular}%
}
\end{center}
\end{table}

\begin{table}[H]
\begin{center}
    \caption{List of Abbreviations}
    \label{abbreviations}
\resizebox{\textwidth}{!} {%
\begin{tabular}{lll}
\hline
\multicolumn{2}{c}{\textbf{Financial Institutions}}     & \multicolumn{1}{c}{\textbf{Emerging Market}}     \\ \hline
ITUB4 BS        & Itau Unibanco Holding SA              & BRAZIL                                           \\ \hline
BBDC3 BS        & Banco Bradesco SA                     & BRAZIL                                           \\ \hline
BBDC4 BS        & Banco Bradesco SA                     & BRAZIL                                           \\ \hline
B3SA3 BS        & B3 SA - Brasil Bolsa Balcao           & BRAZIL                                           \\ \hline
SANB11 BS       & Banco Santander Brasil SA             & BRAZIL                                           \\ \hline
BBAS3 BS        & Banco do Brasil SA                    & BRAZIL                                           \\ \hline
ITSA4 BS        & Itausa SA                             & BRAZIL                                           \\ \hline
BBSE3 BS        & BB Seguridade Participacoes SA        & BRAZIL                                           \\ \hline
BPAC11 BS       & Banco BTG Pactual SA                  & BRAZIL                                           \\ \hline
HDFCB IS        & HDFC Bank Ltd                         & INDIA                                            \\ \hline
HDFC IS         & Housing Development Finance Co        & INDIA                                            \\ \hline
KMB IS          & Kotak Mahindra Bank Ltd               & INDIA                                            \\ \hline
ICICIBC IS      & ICICI Bank Ltd                        & INDIA                                            \\ \hline
SBIN IS         & State Bank of India                   & INDIA                                            \\ \hline
BAF IS          & Bajaj Finance Ltd                     & INDIA                                            \\ \hline
AXSB IS         & Axis Bank Ltd                         & INDIA                                            \\ \hline
BJFIN IS        & Bajaj Finserv Ltd                     & INDIA                                            \\ \hline
ELEKTRA MF      & Grupo Elektra SAB DE CV               & MEXICO                                           \\ \hline
GFNORTEO MF     & Grupo Financiero Banorte SAB d        & MEXICO                                           \\ \hline
SBER RM         & Sberbank of Russia PJSC               & RUSSIA                                           \\ \hline
VTBR RM         & VTB Bank PJSC                         & RUSSIA                                           \\ \hline
FSR SJ          & FirstRand Ltd                         & SOUTH AFRICA                                     \\ \hline
SBK SJ          & Standard Bank Group Ltd               & SOUTH AFRICA                                     \\ \hline
SLM SJ          & Sanlam Ltd                            & SOUTH AFRICA                                     \\ \hline
QNBFB TI        & QNB Finansbank AS                     & TURKEY                                           \\ \hline
\multicolumn{3}{l}{}                                                                                       \\ \hline
\multicolumn{3}{c}{\textbf{Macroeconomic Risk Factors}}                                                    \\ \hline
REIT            & \multicolumn{2}{l}{Dow Jones Equity REIT Total Re}                                       \\ \hline
SPX             & \multicolumn{2}{l}{S\&P 500 INDEX}                                                       \\ \hline
USGG3M          & \multicolumn{2}{l}{US Generic Govt 3 Mth}                                                \\ \hline
VIX             & \multicolumn{2}{l}{Cboe Volatility Index}                                                \\ \hline
USGG3M10YR      & \multicolumn{2}{l}{US Generic Govt 3 Mth to 10yr yield spread}                           \\ \hline
MOOD BAA SPD    & \multicolumn{2}{l}{Moody's Bond Indices Corporate BAA Yield Spread to US Treasury Bonds} \\ \hline
JPEGSOSD        & \multicolumn{2}{l}{JPMorgan EMBIG  Sovereign Bond Yield Spread to U.S. Treasury}         \\ \hline
USDBRL          & \multicolumn{2}{l}{Brazilian Real Spot vs USD}                                           \\ \hline
USDRUB          & \multicolumn{2}{l}{Russian Ruble Spot vs USD}                                            \\ \hline
USDINR          & \multicolumn{2}{l}{Indian Rupee Spot vs USD}                                             \\ \hline
USDMXN          & \multicolumn{2}{l}{Mexican Peso Spot vs USD}                                             \\ \hline
USDTRY          & \multicolumn{2}{l}{Turkish Lira Spot vs USD}                                             \\ \hline
USDZAR          & \multicolumn{2}{l}{South African Rand Spot vs USD}                                       \\ \hline

\end{tabular}%
}
\end{center}
\end{table}

\end{document}